\documentclass{aa}

\usepackage{graphicx}
\usepackage{txfonts}
\usepackage{natbib}
\bibpunct{(}{)}{;}{a}{}{,} 

\usepackage{color}
\usepackage{esvect}
\usepackage{makecell}

\newcommand{\code}[1]{\texttt{#1}}
\newcommand{\saber}{{\small astro}\textsc{Saber}}
\usepackage{hyperref}
\hypersetup{breaklinks=True,
    colorlinks=true,
    linkcolor=blue,
    citecolor=blue,
    filecolor=cyan,      
    urlcolor=blue,
    pdftitle={CoAt clouds toward giant molecular filaments},
    pdfauthor={J. Syed}}
    
\begin{document}

   \title{Cold atomic gas identified by \ion{H}{i} self-absorption}

   \subtitle{Cold atomic clouds toward giant molecular filaments}

   \author{J.~Syed\inst{\ref{inst_mpia}}
          \and H.~Beuther\inst{\ref{inst_mpia}}
          \and P.~F.~Goldsmith\inst{\ref{inst_jpl}}
          \and Th.~Henning\inst{\ref{inst_mpia}}
          \and M.~Heyer\inst{\ref{inst_umass}}
          \and R.~S.~Klessen\inst{\ref{inst_ita},\ref{inst_unihd}}
          \and J.~M.~Stil\inst{\ref{inst_calgary}}
          \and J.~D.~Soler\inst{\ref{inst_inaf}}
          \and L.~D.~Anderson\inst{\ref{inst_westvir},\ref{inst_gbt},\ref{inst_grav_westvir}}
          \and J.~S.~Urquhart\inst{\ref{inst_kent}}
          \and M.~R.~Rugel\inst{\ref{inst_cfa},\ref{inst_nrao}}
          \and K.~G.~Johnston\inst{\ref{inst_leeds}}
          \and A.~Brunthaler\inst{\ref{inst_mpifr}}
          }

   \institute{Max-Planck-Institut f\"ur Astronomie, K\"onigstuhl 17, 69117 Heidelberg, Germany\\
              \email{syed@mpia.de}\label{inst_mpia}
              \and Jet Propulsion Laboratory, California Institute of Technology, 4800 Oak Grove Drive, Pasadena, CA 91109, USA\label{inst_jpl}
              \and Astronomy Department, University of Massachusetts, Amherst, MA, 01003 USA\label{inst_umass}
              \and Universit\"at Heidelberg, Zentrum für Astronomie, Institut für Theoretische Astrophysik, Albert-Ueberle-Str. 2, 69120 Heidelberg, Germany\label{inst_ita}
              \and Universit\"at Heidelberg, Interdisziplin\"ares Zentrum für Wissenschaftliches Rechnen, INF 205, 69120 Heidelberg, Germany\label{inst_unihd}
              \and Department of Physics and Astronomy, The University of Calgary, 2500 University Drive NW, Calgary AB T2N 1N4, Canada\label{inst_calgary}
              \and Istituto di Astrofisica e Planetologia Spaziali (IAPS). INAF. Via Fosso del Cavaliere 100, 00133 Roma, Italy\label{inst_inaf}
              \and Department of Physics and Astronomy, West Virginia University, Morgantown, WV 26506, USA\label{inst_westvir}
              \and Adjunct Astronomer at the Green Bank Observatory, P.O. Box 2, Green Bank, WV 24944, USA\label{inst_gbt}
              \and Center for Gravitational Waves and Cosmology, West Virginia University, Chestnut Ridge Research Building, Morgantown, WV 26505, USA\label{inst_grav_westvir}
              \and Centre for Astrophysics and Planetary Science, University of Kent, Canterbury CT2 7NH, UK\label{inst_kent}
              \and Harvard Smithsonian Center for Astrophysics, 60 Garden Street, Cambridge, MA, 02138, USA\label{inst_cfa}
              \and National Radio Astronomy Observatory, 1003 Lopezville Rd, Socorro, NM 87801, USA\label{inst_nrao}
              \and School of Physics \& Astronomy, Sir William Henry Bragg Building, The University of Leeds, Leeds, LS2 9JT, UK\label{inst_leeds}
              \and Max-Planck-Institut f\"ur Radioastronomie, Auf dem H\"ugel 69, 53121 Bonn, Germany\label{inst_mpifr}
             }

   \date{Received XX March XXXX; accepted XX October XXXX}

 
  \abstract
   {Stars form in the dense interiors of molecular clouds. The dynamics and physical properties of the atomic interstellar medium (ISM) set the conditions under which molecular clouds and eventually stars will form. It is, therefore, critical to investigate the relationship between the atomic and molecular gas phase to understand the global star formation process.}
   {Using the high angular resolution data from The \ion{H}{i}/OH/Recombination line survey of the Milky Way (THOR), we aim to constrain the kinematic and physical properties of the cold atomic hydrogen gas phase toward the inner Galactic plane.}
   {\ion{H}{i} self-absorption (HISA) has proven to be a viable method to detect cold atomic hydrogen clouds in the Galactic plane. With the help of a newly developed self-absorption extraction routine (\saber{}), we build upon previous case studies to identify \ion{H}{i} self-absorption toward a sample of Giant Molecular Filaments (GMFs).}
   {We find the cold atomic gas to be spatially correlated with the molecular gas on a global scale. The column densities of the cold atomic gas traced by HISA are usually of the order of $10^{20}\rm\,cm^{-2}$ whereas those of molecular hydrogen traced by \element[][13]{CO} are at least an order of magnitude higher. The HISA column densities are attributed to a cold gas component that accounts for a fraction of $\sim$5\% of the total atomic gas budget within the clouds. The HISA column density distributions show pronounced log-normal shapes that are broader than those traced by \ion{H}{i} emission. The cold atomic gas is found to be moderately supersonic with Mach numbers of a $\sim$few. In contrast, highly supersonic dynamics drive the molecular gas within most filaments.}
   {While \ion{H}{i} self-absorption is likely to trace just a small fraction of the total cold neutral medium within a cloud, probing the cold atomic ISM by the means of self-absorption significantly improves our understanding of the dynamical and physical interaction between the atomic and molecular gas phase during cloud formation.}

   \keywords{ISM: clouds --
                ISM: atoms --
                ISM: molecules --
                radio lines: ISM --
                stars: formation
               }

   \maketitle
%

\section{Introduction}
Atomic hydrogen provides the raw material to form molecular clouds, the sites of star formation. The dynamical and physical conditions under which molecular clouds form are therefore critical to understand the global star formation process. On a large scale ($\gtrsim10^2\rm\,pc$), molecular clouds form out of the diffuse atomic interstellar medium \citep[ISM; for a review see][]{2001RvMP...73.1031F,Draine2011,2016SAAS...43...85K} and are shaped by galactic dynamics and turbulence, stellar feedback, and magnetic fields. One major constituent of the ISM is the neutral atomic gas that provides the raw material to molecular clouds out of which stars eventually form and this gas phase takes in most of the energy and momentum feedback from its environment. The kinematic and physical relationship between the atomic and molecular gas phase of the ISM is then of central interest in the understanding of cloud formation processes.
   
According to the classical photodissociation region (PDR) model, layers of cold atomic hydrogen can effectively shield the cloud from photo-dissociating UV radiation at sufficiently high densities, allowing a transition of atomic hydrogen to its molecular form. In this idealized picture, pockets of high-density molecular hydrogen are embedded in an envelope of cold atomic hydrogen.

Using high angular resolution data of \ion{H}{i} emission, the structure of the atomic ISM can be studied in great detail. However, probing the physical properties of the atomic gas from \ion{H}{i} emission studies alone is not straightforward. In thermal pressure equilibrium, theoretical considerations based on ISM heating and cooling processes predict two stable phases of atomic hydrogen at the observed pressures in the ISM, namely the cold neutral medium (CNM) and warm neutral medium \citep[WNM;][]{1969ApJ...155L.149F,1977ApJ...218..148M,2003ApJ...587..278W,2019ApJ...881..160B}. Observations of \ion{H}{i} emission are thus generally attributed to both CNM and WNM, which have significantly different physical properties (see below). In an attempt to observationally isolate the CNM from the bistable emission, \ion{H}{i} self-absorption \citep[HISA;][]{2000ApJ...540..851G,2003ApJ...585..823L,2020A&A...634A.139W,2020A&A...642A..68S} is a viable method to identify cold atomic gas and study \ion{H}{i} clouds in the inner Milky Way but it heavily depends on the presence of sufficient background emission. In the following, we refer to cold atomic gas traced by HISA as ``CoAt'' gas, to make a distinction between the CNM as a whole and HISA-traced cold gas, which is a subset of the CNM. Due to the Galactic rotation, any positive (negative) line-of-sight velocity in the first (fourth) Galactic quadrant generally corresponds to a near kinematic and far kinematic distance within the solar circle \citep[see e.g.,][]{1988gera.book..295B}. HISA has been used to resolve this kinematic distance ambiguity for molecular clouds or Galactic \ion{H}{ii} regions. Sources of interest at the far distance are less likely to show HISA as there is less background to absorb. Any detection of corresponding HISA would then place molecular clouds or \ion{H}{ii} regions at the near kinematic distance \citep[e.g.,][]{2002ApJ...566L..81J,2009ApJ...690..706A,2021MNRAS.500.3027D}.

Since the warm component of atomic hydrogen is more diffuse and has a lower density, it fills up a larger volume than the cold component \citep{1977ApJ...218..148M,2005fost.book.....S,2009ARA&A..47...27K}.
\ion{H}{i} self-absorption occurs when a cold \ion{H}{i} cloud is located in front of a warmer \ion{H}{i} emitting cloud. Self-absorption can occur within the same cloud but can also be induced by an emitting cloud in the far background that has the same velocity as the absorbing medium with respect to the local standard of rest $\varv_{\rm LSR}$. Therefore, the clouds do not have to be physically associated for HISA to be observable.
    
The CNM is observed to have temperatures $\lesssim$300$\rm\,K$ and number densities of ${\gtrsim} n_{\mathrm{min,CNM}}{=}10\rm\,cm^{-3}$ while the thermally stable WNM exceeds temperatures of $\sim$5000$\rm\,K$ with number densities ${\lesssim} n_{\mathrm{max,WNM}}{=}0.1\rm\,cm^{-3}$ \citep{2003ApJ...586.1067H,2009ARA&A..47...27K}. In contrast to the properties of the CNM, the atomic gas traced by HISA has typical spin temperatures below 100\,K, in most cases even below 50\,K \citep[e.g.,][]{2000ApJ...540..851G,2003ApJ...585..823L,2008ApJ...689..276K,2020A&A...634A.139W}, thus highlighting the limited sensitivity of the HISA method for higher-temperature gas. For densities between $n_{\mathrm{min,CNM}}$ and $n_{\mathrm{max,WNM}}$, the gas is thermally unstable (denoted by UNM: unstable neutral medium) and it will move toward a stable CNM or WNM branch under isobaric density perturbations \citep{1965ApJ...142..531F}.

\ion{H}{i} self-absorption is found throughout the Milky Way in various environments. Many studies have focused on the detection of HISA, first measured in 1954 \citep{1954AJ.....59..324H,1955ApJ...121..569H}, in known sources, but statistical treatments of the kinematic properties and densities of the HISA-traced cold gas in large-scale high-resolution maps are still scarce.

Extensive investigations of the HISA properties are limited to individual observational case studies \citep[e.g.,][]{2000ApJ...540..851G,2003ApJ...585..823L,2003ApJ...598.1048K,2008ApJ...689..276K,2020A&A...634A.139W,2020A&A...642A..68S} and few simulations \citep[e.g.,][]{2022MNRAS.512.4765S}. In this paper, we aim to build upon these case studies and investigate the HISA properties toward a sample of Giant Molecular Filaments \citep[GMFs;][]{2014A&A...568A..73R} as they are likely to be at an early evolutionary stage of giant molecular clouds forming out of the atomic phase of the interstellar medium \citep{2018ApJ...864..153Z}. These giant filaments potentially trace the Galactic structure, such as spiral arms and spurs, and large concentrations of molecular gas.

\citet{2022MNRAS.512.4765S} present synthetic \ion{H}{i} observations including \ion{H}{i} self-absorption toward molecular clouds and investigate the observational effects and limitations of HISA. Their synthetic observations are based on 3D-MHD simulations within the scope of the SILCC-Zoom project that include out-of-equilibrium \ion{H}{i}-to-$\rm H_2$ chemistry, detailed radiative transfer calculations, as well as observational effects like noise and limited spatial and spectral resolution that are similar to those of the THOR survey (see Sect.~\ref{sec:methods_observation}).
Using commonly employed methods to derive the HISA properties, their results show that the \ion{H}{i} column densities inferred from self-absorption tend to underestimate the real column densities of cold \ion{H}{i} in a systematic way.

Traditionally, HISA features are obtained through various methods. To quantify the absorption depth, the strength of the warm emission background inducing \ion{H}{i} self-absorption has to be determined first. Since the warm atomic gas component is often assumed to have less spatial variation in intensity, the emission background is commonly estimated by taking an ``off'' position spectrum at a different line of sight close to the location where HISA is expected to occur \citep[e.g.,][]{2000ApJ...540..851G,2003ApJ...598.1048K}. However, it is challenging to select a position that is close enough, such that the off position can serve as a good proxy for the true HISA background, but far enough not to be interfered with by (partially) self-absorbing medium. Therefore, the location and spatial distribution of self-absorbing gas has to be known prior to estimating the background, which may work for single isolated cases. But particularly toward the Galactic plane, where multiple emission components and the Galactic rotation can add to the confusion along the line of sight, finding a clean off position has proven to be difficult to accomplish since the off spectrum can vary significantly over the angular size of the background cloud \citep[see][]{2020A&A...634A.139W}.

Another approach is to recover self-absorption in the spectral domain of \ion{H}{i} observations. If the location of HISA in the \ion{H}{i} emission spectrum is known, the spectral baseline can be determined by fitting the emission range around a HISA feature with a simple polynomial or Gaussian function \citep{2003ApJ...585..823L,2003ApJ...598.1048K,2020A&A...634A.139W,2020A&A...642A..68S}. However, the assumption of a velocity range where HISA is located introduces an additional source of bias, together with the specific fitting function that is used to derive the background. To address this issue, \citet{2008ApJ...689..276K} and \citet{2018MNRAS.479.1465D} have employed second and higher derivatives of the emission spectra to search for narrow HISA features \citep[HINSA;][]{2003ApJ...585..823L,2005ApJ...622..938G,2007ApJ...654..273G} over the entire spectral range in a more unbiased way. Sharp kinks and dips in the spectra that are due to self-absorption are therefore expected to become readily apparent when investigating the derivatives. This technique allows HISA features to be filtered out without prior knowledge of their central velocities but it relies on high sensitivity, a well-sampled HISA line width, and HISA features that are much narrower than the average emission component. However, the spectral baselines of these identified absorption features would then still need to be obtained using, for example, polynomial fits or making other physical assumptions of the HISA properties \citep[e.g.,][]{2008ApJ...689..276K}.

In this paper, we address the lack of a versatile self-absorption reconstruction algorithm that can be applied to any data set, at any spectral resolution, and self-absorption line width, and without the prior assumption that the cold \ion{H}{i} gas is tightly correlated with molecular gas. We present the algorithm \saber{} (\textbf{S}elf-\textbf{A}bsorption \textbf{B}aseline \textbf{E}xtracto\textbf{R}) that operates by smoothing emission spectra in an asymmetric way, such that it not only identifies signal dips in the spectrum but directly provides a spectral baseline\footnote{In this paper, all instances of the word ``baseline'' refer to the absorption-free spectrum that is used as a spectral baseline to extract clean HISA.} of potential self-absorption features. It works in multiple iterations, such that both narrow and broad absorption components can be recovered. An optimization step has been implemented that is designed to tune the amount of smoothing that is required to recover self-absorption features, irrespective of spectral resolution and line width. To test the performance and applicability of the algorithm, we apply \saber{} to the known sample of GMFs \citep{2014A&A...568A..73R} since they serve as a good laboratory to investigate the presence of CoAt gas. The properties of \ion{H}{i} self-absorption toward two of these molecular filaments have already been investigated in dedicated case studies employing previous HISA extraction methods (GMF20.0-17.9 in \citealp{2020A&A...642A..68S} and GMF38.1-32.4 in \citealp{2020A&A...634A.139W}).

The paper is organized as follows: In Sect.~\ref{sec:methods} we briefly introduce the data used in this analysis and outline the methods of our newly developed \ion{H}{i} self-absorption extraction routine and Gaussian decomposition. In Sect.~\ref{sec:results} we present the kinematic and column density properties derived from the HISA extraction and spectral decomposition. We discuss the kinematic and spatial relationship between the CoAt gas and molecular gas as well as the column density properties in Sect.~\ref{sec:discussion}. We furthermore elaborate on some of the limitations of our HISA extraction method before concluding with our summary in Section~\ref{sec:conclusions}.

\section{Methods and observations}\label{sec:methods}
\subsection{\ion{H}{i}, CO, and continuum observations}\label{sec:methods_observation}
The following analysis of the HISA properties toward molecular clouds is based on the \ion{H}{i} and 1.4\,GHz continuum observations as part of The \ion{H}{i}/OH Recombination line survey of the inner Milky Way \citep[THOR;][]{2016A&A...595A..32B,2020A&A...634A..83W}. The final THOR-\ion{H}{i} and 1.4\,GHz continuum data include observations taken with the Karl G. Jansky Very Large Array (VLA) in C-configuration that were combined with the \ion{H}{i} Very Large Array Galactic Plane Survey \citep[VGPS;][]{2006AJ....132.1158S}, which consists of VLA D-configuration data. To account for missing flux on short $uv$ spacings, the VGPS also includes single-dish observations of \ion{H}{i} and 1.4\,GHz continuum taken with the Green Bank and Effelsberg 100m telescope, respectively. The final \ion{H}{i} emission data, from which the continuum has been subtracted during the data reduction, have an angular and spectral resolution of $\Delta\Theta=40\arcsec$ and $1.5\rm\,km\,s^{-1}$, respectively. The rms noise in emission-free channels is $\sigma_{\mathrm{rms}}\sim4\rm\,K$.

We selected six GMF regions to investigate the presence of CoAt gas. Our selection is based on the findings of \citet{2014A&A...568A..73R} and \citet{2018ApJ...864..153Z}. \citet{2014A&A...568A..73R} identified seven mid-infrared extinction features as giant filaments that exhibit corresponding \element[][13]{CO} emission and velocity coherence over their full length. Of these seven GMFs, six of the fields are covered by the THOR survey. We present an overview of the six fields covering the filament regions in Table~\ref{tab:overview}. The indices of the source names refer to the approximate range in Galactic longitude the giant filaments cover. The selected filament regions are in close proximity to the Galactic midplane and are located in the inner disk of the Milky Way, a site where HISA is more likely to occur. These GMFs serve as a good laboratory to investigate the relationship between the atomic and molecular gas as they are molecular concentrations of lengths ${>}50\rm\,pc$ and likely to be at an early evolutionary stage having formed out of the large-scale diffuse ISM \citep{2018ApJ...864..153Z}. More details about each region can be found in \citet{2014A&A...568A..73R}.
    \begin{table*}[!htbp]
     \caption{Properties of studied filament regions.}
     \renewcommand*{\arraystretch}{1.3}
     \centering
     \begin{tabular}{l c c c c c }
     (1) & (2) & (3) & (4) & (5) & (6) \\
     \hline\hline
     Source name\tablefootmark{(a)} & Glon [\degr.\degr] & Glat [\degr.\degr] & $\varv_{\mathrm{LSR}}$ [$\rm km\,s^{-1}$] & $d_{\mathrm{near}}$\tablefootmark{(b)} [$\rm kpc$] & $D_{\mathrm{GC}}$\tablefootmark{(b)} [$\rm kpc$] \\
     GMF20.0-17.9 & 17.80 -- 20.60 & $-$1.00 -- +0.30 & 37 -- 50 & 3.2 & 5.2 \\
     GMF26.7-25.4 & 25.10 -- 26.90 & $+$0.40 -- +1.20 & 41 -- 51 & 2.9 & 5.7 \\
     GMF38.1-32.4a & 33.30 -- 37.30 & $-$1.00 -- +0.60 & 50 -- 60 & 3.2 & 5.9 \\
     GMF38.1-32.4b & 33.30 -- 37.30 & $-$1.00 -- +0.60 & 43 -- 46 & 2.6 & 6.2 \\
     GMF41.0-41.3 & 40.80 -- 41.50 & $-$0.70 -- +0.50 & 34 -- 42 & 2.2 & 6.7 \\
     GMF54.0-52.0 & 52.30 -- 54.20 & $-$0.50 -- +0.40 & 20 -- 26 & 1.4 & 7.4 \\\hline
     \end{tabular}
     \tablefoot{
     Columns (2) and (3) give the Galactic longitude range and latitude range of the filament regions, respectively. Column (4) gives the line-of-sight velocity range of each GMF as defined in \citet{2014A&A...568A..73R}. Columns (5) and (6) give the kinematic near distance from us and the Galactocentric distance, respectively.\\
     \tablefoottext{a}{as in \citet{2014A&A...568A..73R}.}\\
     \tablefoottext{b}{The distances are not taken from \citet{2014A&A...568A..73R} but have been recalculated using the more recent spiral arm model by \citet{2019ApJ...885..131R}.}
     }
     \label{tab:overview}
    \end{table*}
In Sect.~\ref{sec:HI_emission_optical_depth}, we correct for optical depth effects to compute the atomic hydrogen column densities from \ion{H}{i} emission. The optical depths are taken from the measurements provided by \citet{2020A&A...634A..83W} and have been obtained from VLA C-configuration data only that have an angular resolution of $\sim$15\arcsec{} and effectively filter out large-scale emission, such that \ion{H}{i} absorption against discrete continuum sources can yield a direct measurement of the optical depth of atomic hydrogen.

In order to provide a comprehensive description of the kinematic and spatial relationship between the atomic gas and the molecular gas, we investigate the molecular gas properties toward the GMF regions using two different data sets. The kinematic information is based on the \element[][13]{CO}(1--0) data of the Galactic Ring Survey \citep[GRS;][]{2006ApJS..163..145J}, with an angular and spectral resolution of 46\arcsec{} and $0.21\rm\,km\,s^{-1}$, respectively. \citet{2020A&A...633A..14R} present an overview of a Gaussian decomposition of the entire GRS using the fully automated \textsc{GaussPy+} algorithm \citep{2019A&A...628A..78R}. Since the decomposition results are publicly available, we use these data to investigate the kinematic properties of the clouds.

In Sect.~\ref{sec:CO_column_dens} we compute the \element[][13]{CO} column densities from the \element[][12]{CO}(1--0) and \element[][13]{CO}(1--0) emission line data taken from the Milky Way Imaging Scroll Painting survey \citep[MWISP;][]{2019ApJS..240....9S}. The GRS does not include \element[][12]{CO} observations, that are required to estimate the CO excitation temperatures and ultimately \element[][13]{CO} column densities. We therefore use both the \element[][13]{CO} and \element[][12]{CO} from the MWISP data to derive the column density properties in a consistent way, and to reduce systematic errors arising from observational biases. The MWISP \element[][12]{CO} and \element[][13]{CO} data have an angular resolution of $\sim$55\arcsec{} and an rms noise of 0.5\,K and 0.3\,K at a spectral resolution of $0.16\rm\,km\,s^{-1}$ and $0.17\rm\,km\,s^{-1}$, respectively. The \element[][12]{CO} data have been reprojected onto the same spectral grid as the \element[][13]{CO} data to infer the excitation temperatures on a voxel-by-voxel basis. The rms noise of the \element[][12]{CO} data is then reduced to 0.4\,K.

\subsection{Absorption baseline reconstruction}\label{sec:baseline_reconstruction}
In this section we describe the \saber{} method that we used to obtain self-absorption baselines to recover HISA features. The basic workflow of \saber{} is the following: 1) Generating mock \ion{H}{i} spectra to use as ``training data''\footnote{While the terms ``test data'' and ``training data'' are commonly used in the context of machine learning algorithms, we note that the accuracy of \saber{} is not tested on unseen data but the underlying concepts are the same, such that these concepts can be used to integrate them in a machine learning algorithm.} (described in Sect.~\ref{sec:optimization}), 2) finding optimal smoothing parameters using gradient descent (described in Sect.~\ref{sec:optimization} and Appendix~\ref{sec:gradient_descent_app}), 3) applying baseline extraction with optimal smoothing parameters found in step 2) (described in Sect.~\ref{sec:asymmetric_least_squares}).

\subsubsection{Asymmetric Least Squares Smoothing}\label{sec:asymmetric_least_squares}
The publicly available python-based \saber{} algorithm (\url{https://github.com/astrojoni89/astrosaber}) is an automated baseline extraction routine that is designed to recover baselines of absorption features that are superposed with \ion{H}{i} emission spectra. In the following, the \saber{} algorithm is described in
detail. A description of \saber{} parameters used throughout the paper, including their keywords and default values, can be found in Appendix~\ref{sec:keywords_app}. The \saber{} method utilizes asymmetric least squares smoothing first proposed by \citet{2004Eilers} in the context of Raman spectroscopy. The algorithm progresses iteratively in two cycles to obtain a smoothed baseline, the major (outer) cycle and the minor (inner) cycle executed at each iteration of the major cycle. The basis of the minor cycle is to find a solution that minimizes the penalized least squares function
\begin{equation}
     F(\mathbf{z}) = (\mathbf{y} - \mathbf{z})^\top\mathbf{W}(\mathbf{y} - \mathbf{z})+\lambda\,\mathbf{z}^\top\mathbf{D}^\top\mathbf{D}\mathbf{z} \: ,
     \label{equ:least_squ}
\end{equation}
\noindent where $\mathbf{y}$ is the input signal (e.g., the observed \ion{H}{i} spectrum) and $\mathbf{z}$ is the asymmetrically smoothed baseline to be found. The first and second term on the right-hand side describe the fitness of the data and the smoothness of $\mathbf{z}$ defined by the second order differential matrix $\mathbf{D}$, respectively. The parameter $\lambda$, which is a two-dimensional vector by default (see below), adjusts the weight of the smoothing term. The regularized smoothing allows the detection of less significant absorption features that would otherwise be missed by finite-difference detection methods (see the discussion in Appendix~\ref{sec:second_deriv_app}). In order to correct the baseline with respect to peaks and dips in the spectrum, the asymmetry weighting matrix $\mathbf{W} = \mathrm{diag}(\mathbf{w})$ is introduced. The asymmetry weights are initialized to be $w_i=1$. After a first iteration of the minor cycle with equal weights, the weights for channels containing signal are then assigned as follows:
\begin{equation}
      w_i = \begin{cases} p, & y_i > z_i \\ 1-p, & y_i \leq z_i \end{cases} \: .
      \label{equ:weighting}
\end{equation}
\noindent The asymmetry parameter $p\in[0,1]$ is set to favor either peaks or dips while smoothing the spectra. Given both the parameters $\lambda$ and $p$, a smoothed baseline $\mathbf{z}$ is updated iteratively. Depending on $p$ and the deviation of $\mathbf{z}$ from $\mathbf{y}$ after each iteration, peaks (dips) in the spectrum will be retained while dips (peaks) will be given less weight during the smoothing. Since we only aim to asymmetrically smooth real signals, spectral channels containing only noise will be given equal weights of $0.5$, hence the baseline will be within the noise in emission-free channels. The signal range estimation is described in Sect.~\ref{sec:signalrange}. As can be seen in Eq.~\eqref{equ:least_squ}, there is a degeneracy in the solution of the least squares function introduced by the weighting factors $\mathbf{W}(p)$ and $\lambda$. It is then sensible to keep one of these parameters fixed while finding the best-fit solution for the other parameter in order to optimize the smoothing (see Sect.~\ref{sec:optimization}). In the case of self-absorption features, we therefore chose to fix the asymmetry parameter at $p=0.9$.

After $n_{\mathrm{minor}}$ iterations, the minor cycle converges, such that the iteratively updated baseline $\mathbf{z}$ will not change anymore given the input spectrum $\mathbf{y}$. However, in order to effectively smooth out dips while still retaining real signal peaks in the spectra, the smoothed baseline $\mathbf{z}$ is then passed to the next iteration of the major cycle as an input (i.e. now $\mathbf{y}$) for its minor cycle smoothing.

After evaluating the THOR-\ion{H}{i} data, the minor cycle has shown to already converge after three iterations. Hence, the number of minor cycle iterations has been fixed at $n_{\mathrm{minor}}=3$ in the algorithm. This parameter affects the output of \mbox{\saber{}} only mildly since the final smoothed baseline is mostly dependent on the number of iterations in the major cycle and on the $\lambda$ parameter that tunes the smoothing (see Sect.~\ref{sec:optimization}).

The algorithm stops as soon as a convergence criterion in the major cycle is met, or if the maximum number of iterations $n_{\mathrm{major}}$ is reached. The convergence criterion is met if the change in baseline from one major cycle iteration to the next is below a threshold set by $s_{\mathrm{thresh}}\cdot\sigma_{\mathrm{rms}}$ for at least some number of iterations $n_{\mathrm{converge}}$. The default values set by \saber{} are $n_{\mathrm{major}}=20$, $s_{\mathrm{thresh}}=1$, and $n_{\mathrm{converge}}=3$. There is a slight degeneracy between the actual number of iterations needed to make the baseline converge and a fixed smoothing parameter $\lambda$ used for every smoothing iteration. For $\lambda$ sufficiently high, fewer iterations are needed to smooth out sharp kinks and dips in the spectrum. In the case of the THOR-\ion{H}{i} data where the continuum has been subtracted during data reduction, the maximum number of iterations can be reached for emission spectra that are contaminated by imperfect continuum subtraction toward very strong continuum sources. Inspecting the number of iterations can therefore serve as an additional quality check of the spectra. For high-sensitivity data at a spectral resolution that is much smaller than the HISA line width, the optimal smoothing parameter $\lambda$ might be too large to make the algorithm converge since the change in baseline will be significant after every major cycle iteration. It can then be sensible to decrease the convergence threshold or to reduce the maximum number of iterations to force the algorithm to terminate and thus break down the aforementioned degeneracy.

In order to recover both narrow and broad features and to account for the possibility of an absorption baseline that exceeds the intensity of that in adjacent velocity channels, the \saber{} routine can be set to add a residual $\mathcal{R_+}$, which is the absolute difference between the first and last iteration of the major cycle. An example of this is an isolated emission feature with a Gaussian shape that has an absorption dip at the line center, or the  ``flat-top'' spectrum observed in [\ion{C}{ii}] emission toward the \ion{H}{ii} region RCW120 \citep[][see their Fig.~6]{2022A&A...659A..36K}. To add flexibility to the baseline reconstruction, the very first major cycle iteration can be set to operate with its own individual smoothing parameter $\lambda_1$ while all following iterations use a smoothing parameter $\lambda_2$. A $\lambda_2$ smoothing parameter close to zero is then effectively equal to a spectral smoothing without adding the residual. In Sect.~\ref{sec:optimization} we investigate how to optimize the smoothing parameters using mock-\ion{H}{i} data.

\begin{figure*}
     \centering
     \includegraphics[width=1.0\textwidth]{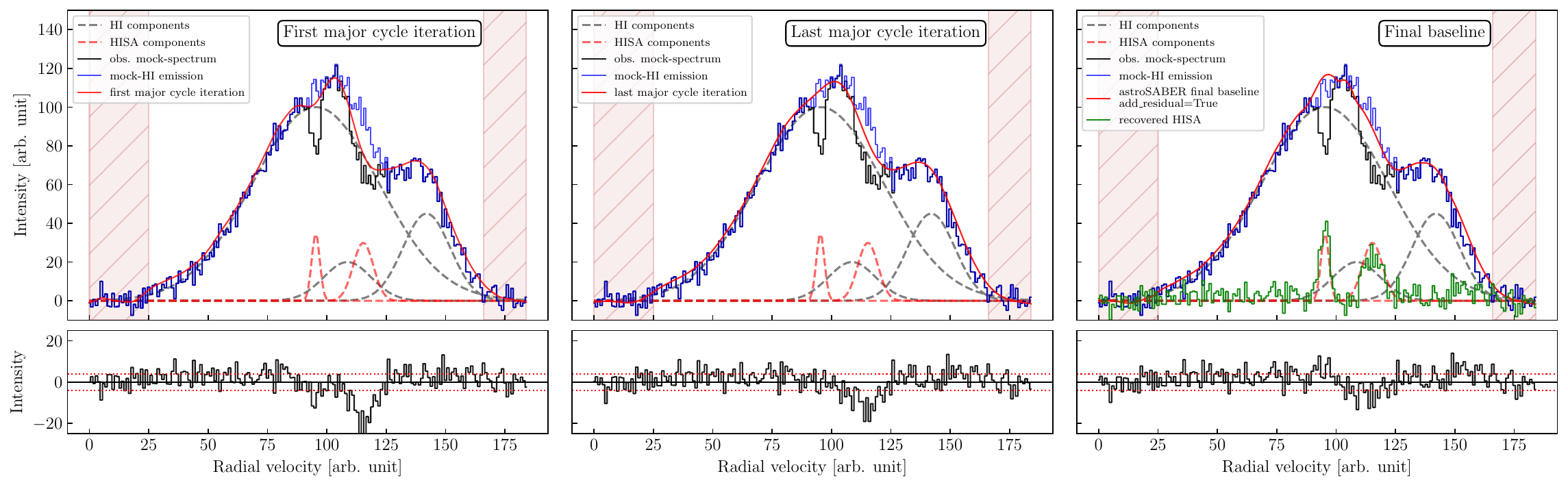}
        \caption[]{Baseline extraction workflow of \saber{}. In each panel, the black mock spectrum represents the observed \ion{H}{i} emission spectrum, which is the sum of the three gray dashed components,  with self-absorption features (two red dashed components) superposed. The blue spectrum shows the ``pure emission'' spectrum that is to be recovered by the \saber{} algorithm. The algorithm is then applied to the observed spectrum using the optimal smoothing parameters $(\lambda_1,\lambda_2)$. Hatched red areas indicate spectral channels that are masked out due to missing signal. \textit{Left panel:} The \saber{} baseline (red) after the first major cycle iteration, that is, after the minor cycle smoothing converged given the input mock spectrum (i.e. after Eq.~\eqref{equ:least_squ} has been solved for $\mathbf{z}$). \textit{Middle panel:} The \saber{} baseline (red) after the last major cycle iteration, that is, after the major cycle smoothing converged and before adding the residual, which is the absolute difference between the first and last major cycle iteration. \textit{Right panel:} The final \saber{} baseline (red) after adding the residual. The baseline so obtained reproduces the pure emission spectrum (blue) well. The resulting HISA features expressed as equivalent emission features are shown in green, and show a good match with the the real HISA absorption features. The smaller subpanels in each column show the residual, which is the difference between the red baseline and the blue emission spectrum, with the horizontal dotted red lines marking values of $\pm\sigma_\mathrm{rms}$.}
        \label{fig:mock_spectrum}
   \end{figure*}

Figure~\ref{fig:mock_spectrum} shows a step-by-step baseline extraction of a mock spectrum to illustrate the major cycle workflow. The mock-\ion{H}{i} contains three emission components where two absorption features of different line widths have been added. Given the observed spectrum (black spectrum in Fig.~\ref{fig:mock_spectrum}), \saber{} is run with optimal smoothing parameters $(\lambda_1,\lambda_2)$ (see Sect.~\ref{sec:optimization}). The left panel in Fig.~\ref{fig:mock_spectrum} shows the baseline after the first major cycle iteration, that is after the minor cycle smoothing converged given the input spectrum (i.e. after Eq.~\eqref{equ:least_squ} has been solved for $\mathbf{z}$). The middle panel then shows the converged baseline after the last major cycle iteration before adding the residual. The right panel presents the final baseline obtained by \saber{} after adding the residual. The baseline so obtained is able to recover the pure emission spectrum well within the uncertainties. We note that if the $\mathcal{R_+}$ setting is turned off, the smoothing parameters obtained during the optimization (Sect.~\ref{sec:optimization}) will be adjusted to have larger values in order to recover the baseline. The differences in baseline between these settings are likely to be small at the velocities of the absorption signals. However, real signal is then also more likely to be smoothed out by the higher smoothing weight.

\begin{figure*}
     \centering
     \includegraphics[width=1.0\textwidth]{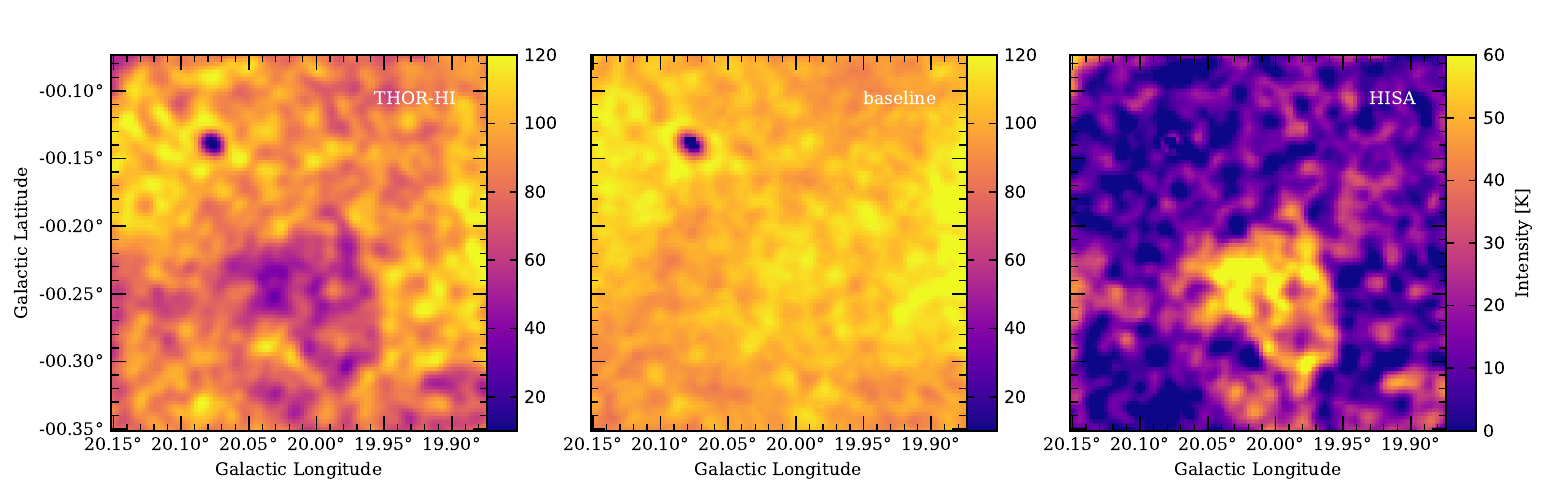}
        \caption[]{Example \ion{H}{i} self-absorption extraction. The \textit{left panel} shows the observed THOR-\ion{H}{i} emission channel map toward a $(100\times 100)\rm\,pixel$ subsection of the giant filament GMF20.0-17.9 at the velocity $44.5\rm\,km\,s^{-1}$. The \textit{middle panel} shows the map of the self-absorption baseline obtained with optimized smoothing parameters. The \textit{right panel} gives the resulting HISA map, which is the difference between the baseline map and the \ion{H}{i} emission map. The HISA feature in the bottom half of the map could be successfully recovered by \saber{}, while the strong continuum source in the top left was masked during a quality check of the spectra.}
        \label{fig:test_maps}
   \end{figure*}

An example of the final output of the extraction step is shown in Fig.~\ref{fig:test_maps}. The figure shows maps of an example region toward a $(100\times 100)\rm\,pixels$ subsection of GMF20.0-17.9 \citep[see][]{2014A&A...568A..73R,2020A&A...642A..68S} that is also made publicly available with the \saber{} code. The maps present the \ion{H}{i} emission data, the  baselines obtained with optimized smoothing parameters, and the resulting \ion{H}{i} self-absorption data, respectively.

\subsection{Gaussian decomposition}\label{sec:gausspyplus}
After \saber{} has been applied to all six giant filament regions with optimized smoothing parameters, in each case the resulting output gives four data cubes containing the reconstructed baseline spectra, the self-absorption features (i.e. the \ion{H}{i} emission spectra subtracted from the baselines), a map of the number of iterations that were required for the baselines to converge, and a map with flags for spectra that did not meet the convergence criteria, either due to missing signal in the spectra or having reached the maximum number of iterations set by the user. Spectra are flagged with ``missing signal'' if there is no significant emission (defined by $s_{\mathrm{signal}}\cdot\sigma_{\mathrm{rms}}$) in more than a specified number of consecutive spectral channels $\Delta\varv_{\mathrm{LSR}}$ (in units of $\rm km\,s^{-1}$), and these spectra are removed from the final data cubes by default, as is done for the strong continuum source in Fig.~\ref{fig:test_maps}. For the THOR-\ion{H}{i} data, we applied the default settings $s_{\mathrm{signal}}=6$ and $\Delta\varv_{\mathrm{LSR}}=15\rm\,km\,s^{-1}$ (see Table~\ref{tab:all_parameters}).

The final self-absorption data cubes obtained from \saber{} contain what we refer to as HISA ``candidates'' since all signal dips have been extracted from the emission spectra. In the following steps, we decomposed the HISA candidate data cubes into their spectral components using the fully automated Gaussian decomposition algorithm \textsc{GaussPy+}\footnote{\url{https://github.com/mriener/gausspyplus}} \citep{2019A&A...628A..78R} and identified ``real'' \ion{H}{i} self-absorption by cross-matching the centroid velocities with the molecular kinematics of the GMF regions given in \citet{2014A&A...568A..73R}.
    
\textsc{GaussPy+} is a multicomponent Gaussian decomposition tool based on the earlier \textsc{GaussPy} algorithm \citep{2015AJ....149..138L}, and it provides additional preparatory steps and quality checks to improve the quality of the spectral decomposition. \textsc{GaussPy} has the fully automated means to decompose spectra using a supervised machine-learning technique. The algorithm automatically determines initial guesses for Gaussian fit components using derivative spectroscopy. To decompose the spectra, the spectra require smoothing to remove noise peaks while retaining real signal. The optimal smoothing parameters are found by employing a machine-learning algorithm that is trained on a few hundred well-fit spectra that are taken from a subsection of each data set. \textsc{GaussPy+} builds upon these results and introduces quality checks of the identified Gaussian components, such as FWHM values, signal-to-noise ratio, significance, and goodness of fit estimation \citep[see Sect.~3.2 in][]{2019A&A...628A..78R}. These criteria are used to decide whether spectra are discarded or refit. Optional quality checks for broad or blended Gaussian components can also be imposed, depending on the specific data set and expected physical cause of the spectral lines.
    
It is essential to reliably estimate the noise in the spectra to obtain good fit results. As we described above, \textsc{GaussPy+} comes with an automated noise estimation routine as a preparatory step for the decomposition that also considers the median absolute deviation (MAD) of negative spectral channels to identify narrow spikes in the spectra that are masked before estimating the rms noise.\footnote{This step in the noise estimation of \textsc{GaussPy+} is not included in \saber{} since we only want to identify signal ranges that are broad enough such that they can be used for generating mock self-absorption spectra.} 
    
For the full decomposition run of each data set, we used the default parameters and standard quality control of \textsc{GaussPy+} if not explicitly stated otherwise. A detailed description of all parameters and in-built quality checks is given in \citet{2019A&A...628A..78R}. For each data cube, we ran the \textsc{GaussPy+} training step with 300 randomly selected spectra from the HISA candidate data to find the optimal parameters for the fitting, as recommended in \citet{2019A&A...628A..78R}. Owing to the absorption properties of the \ion{H}{i} gas, we would naturally expect HISA to probe very cold gas so we opt to refit broad components in the \textsc{GaussPy+} routine. \textsc{GaussPy+} flags a component in a spectrum as broad if its line width is larger than the line width of the second broadest component by a user-defined factor (default: 2). We do not set a specific value as a line width limit. An absolute value that is used as a limiting line width might lead to unphysical fit solutions or artifacts, or can be difficult to determine since the range of expected values is not known.
    
After the initial fitting, we apply the two-phase spatial coherence check implemented by \textsc{GaussPy+} that can optimize the fit by refitting the components based on the fit results of neighboring pixels \citep[see Sect. 3.3 in][]{2019A&A...628A..78R}. Mostly one velocity component was fit by \textsc{GaussPy+} in the given velocity ranges of the filament regions. Only for some small isolated regions and single pixels more than one component was fit to the HISA spectra. As we show in a test environment in Appendix~\ref{sec:kinematics_app}, the centroid velocities recovered by the extraction and subsequent spectral decomposition are robust and have an uncertainty of $\sim$0.35$\rm\,km\,s^{-1}$.

Spectra where the maximum number of iterations $n_{\mathrm{major}}$ is reached during the baseline extraction are flagged but not removed from the \saber{} routine. The affected spectra are usually toward positions where continuum emission contaminates the detection of self-absorption. We removed these spectra manually by masking pixels where there is strong continuum emission $T_{\rm cont}\geq 100\rm K$. Due to the systematic uncertainty in the baselines and to ensure we only report reliable HISA features that are well detected, we additionally masked all pixels of the fit result maps where the corresponding fit amplitude is below $5\sigma_{\mathrm{hisa}}=\sqrt{2}\cdot5\sigma_{\mathrm{rms}}$, with $\sigma_{\mathrm{rms}}$ being the rms noise of the THOR-\ion{H}{i} emission data. The factor $\sqrt{2}$ accounts for the uncertainty in HISA amplitude that is due to the difference between the extracted HISA baseline and the \ion{H}{i} emission.

\section{Results}\label{sec:results}
We show in Table~\ref{tab:overview} an overview of the filament regions analyzed in this paper, which are motivated by the results of \citet{2014A&A...568A..73R}. We will use their designated names (and shortened versions thereof) to refer to these regions throughout this paper.

We detect HISA toward all six filament regions. However, toward GMF26 and GMF41 only a small amount of CoAt gas could be recovered as HISA. The HISA-traced gas toward GMF26 does not appear to trace the distribution of the molecular gas well. Toward GMF20, GMF38a, and GMF38b we recovered a large cold atomic counterpart to the molecular gas within the filaments.
   
\subsection{Kinematics}
In this section, we discuss the kinematic properties of both HISA and their molecular counterpart as traced by \element[][13]{CO} emission. As an example, we show the detected HISA and corresponding \element[][13]{CO} emission map toward GMF20 in terms of their centroid velocities in Fig.~\ref{fig:kin1}. The kinematic maps of the remaining filament regions can be found in Appendix~\ref{sec:kin_app}.
   \begin{figure*}
     \centering
     \includegraphics[width=1.0\textwidth]{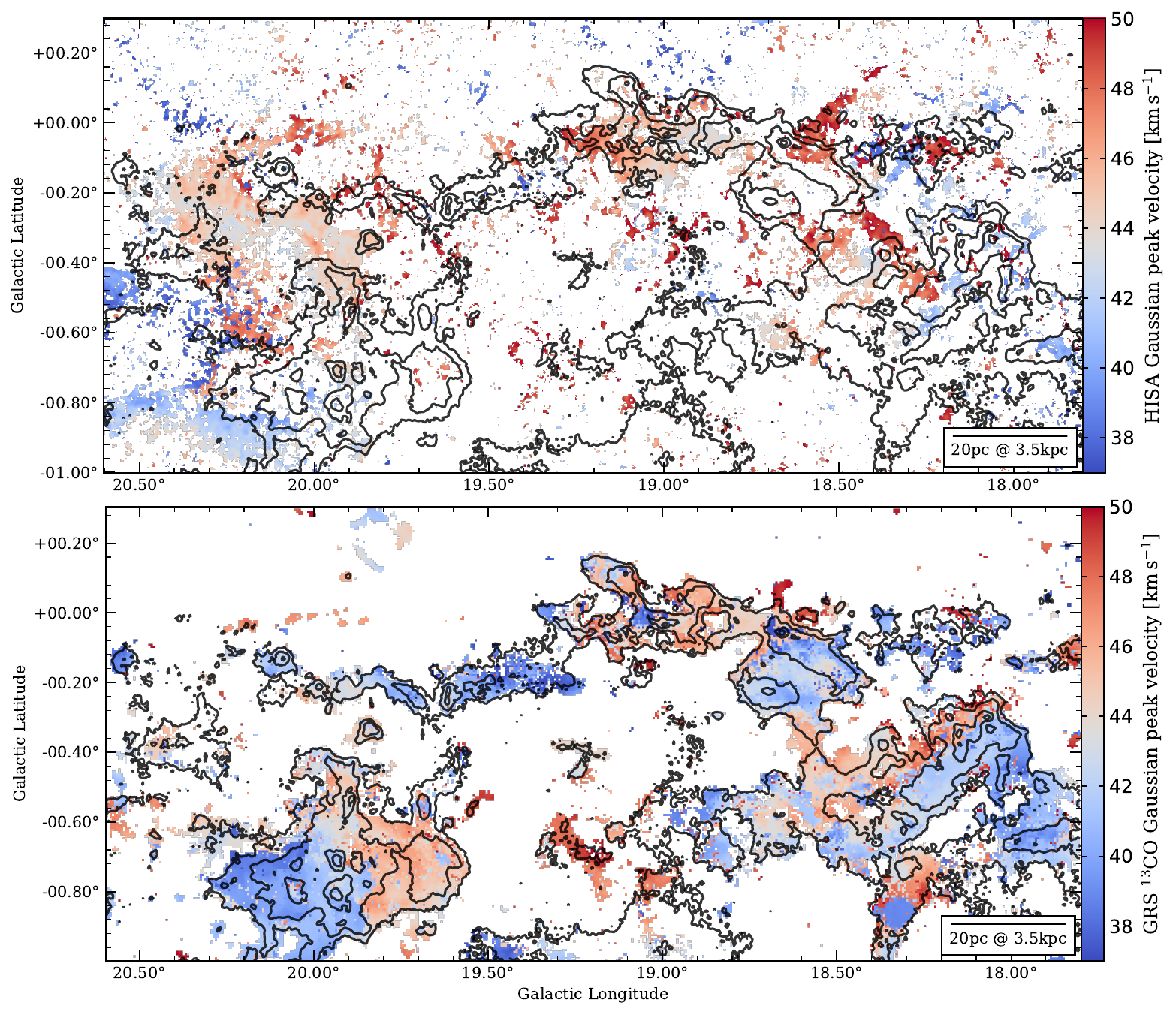}
        \caption[]{Fit peak velocity toward GMF20. These maps show the peak velocities of fit components with amplitudes $\geq 5\sigma_{\mathrm{rms}}$ derived from the \textsc{GaussPy+} decomposition of the spectra. If multiple components are present in a single pixel spectrum within the velocity range of the filament region, the component with the lowest peak velocity is shown. The black contours in both panels show the integrated GRS \element[][13]{CO} emission at the levels 8.0, 16.0, 32.0, and 42.0$\rm\,K\,km\,s^{-1}$. \textit{Top panel:} Fit HISA peak velocity. \textit{Bottom panel:} Fit \element[][13]{CO} peak velocity.}
        \label{fig:kin1}
   \end{figure*}
As we show in Appendix~\ref{sec:kinematics_app}, the centroid velocities and line widths are not heavily affected by our \saber{} routine and have an uncertainty of $0.4\rm\,km\,s^{-1}$ and $1.0\rm\,km\,s^{-1}$ (FWHM), respectively. Since the beam size of the THOR-\ion{H}{i} data is similar to the one of the GRS survey (40\arcsec{} and 46\arcsec{}, respectively), we chose to keep the original resolution (both spatially and spectrally) when comparing the kinematic maps. We tested smoothing the \ion{H}{i} maps to the common beam size of 46\arcsec{}, which had a negligible effect.

\begin{figure*}
     \centering
     \includegraphics[width=1.0\textwidth]{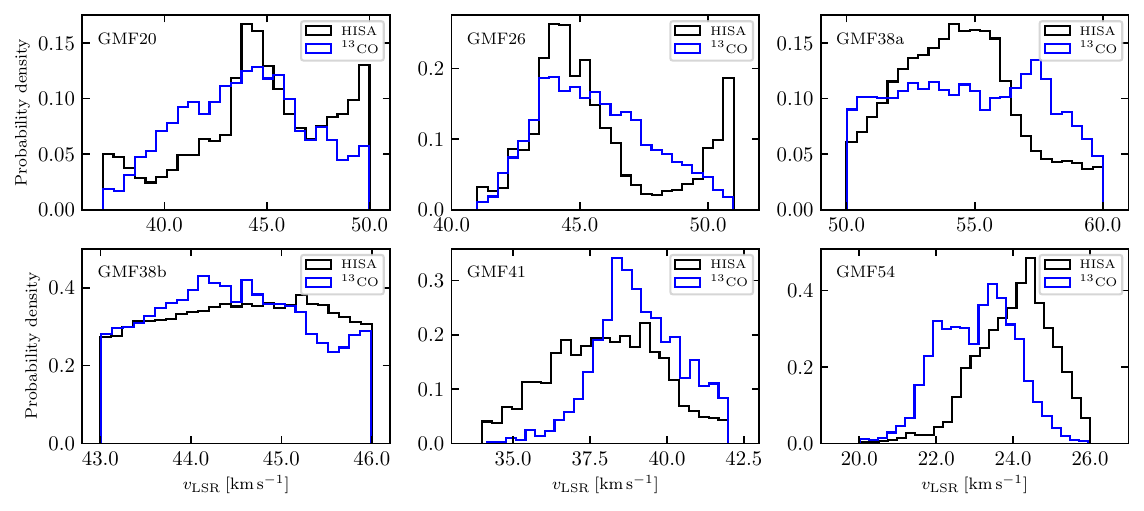}
        \caption[]{Histograms of fit peak velocities. The panels show for each of the six giant filament regions the normalized histogram of peak velocities of HISA and \element[ ][13]{CO} in black and blue, respectively.}
      \label{fig:hist1}
    \end{figure*}

\begin{figure*}
     \centering
     \includegraphics[width=1.0\textwidth]{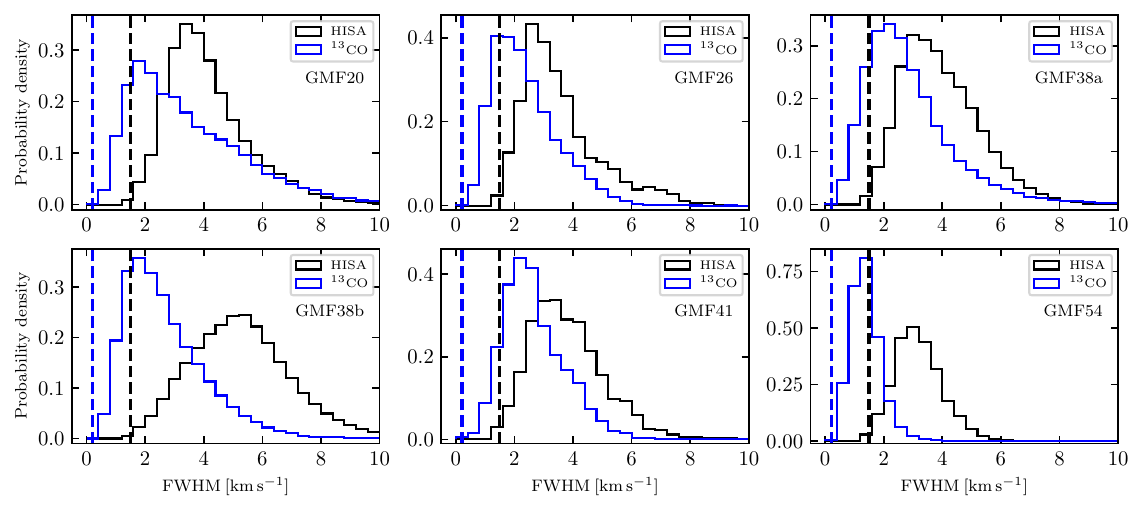}
        \caption[]{Histograms of fit line widths. The panels show for each of the six giant filament regions the normalized histogram of FWHM line widths of HISA and \element[ ][13]{CO} in black and blue, respectively. The black and blue vertical dashed lines mark the spectral resolution limit of the HISA and \element[][13]{CO} data, respectively.}
      \label{fig:hist1line}
    \end{figure*}

For each of the kinematic histograms, we show every fit component along each line of sight within the velocity range of each filament region, thus taking into account multiple components if present. We furthermore only report fit components with an amplitude above the $5\sigma$ noise of the respective data cube ($\sigma\sim5.7\rm\,K$ for HISA, and $\sigma\sim0.3\rm\,K$ for \element[][13]{CO}). The histograms of the centroid velocities of HISA and \element[][13]{CO} show correlation for most of the filament regions (Fig.~\ref{fig:hist1}). The median peak velocity toward GMF20 is $\varv_{\mathrm{LSR}}=44.7\rm\,km\,s^{-1}$ for HISA and $44.0\rm\,km\,s^{-1}$ for \element[][13]{CO}, which is in very good agreement with the results obtained in \citet{2020A&A...642A..68S}. Particularly in the case of HISA, the histogram is mildly affected by components at higher velocities that might not be associated with the giant filament region or that might be troughs between two emission features erroneously picked up by the \saber{} routine. This effect is also evident in the histogram of GMF26. However, the median peak velocities also do agree toward GMF26, with $\varv_{\mathrm{LSR}}=44.9\rm\,km\,s^{-1}$ and $45.4\rm\,km\,s^{-1}$ for HISA and \element[][13]{CO}, respectively. Toward GMF38a the histogram of peak velocities obtained with both \saber{} and the automated spectral line decomposition \textsc{GaussPy+} reproduces the results presented in \citet{2020A&A...634A.139W}, with the median peak velocities agreeing to within $0.5\rm\,km\,s^{-1}$ ($\varv_{\mathrm{LSR}}=54.3\rm\,km\,s^{-1}$ and $54.8\rm\,km\,s^{-1}$ for HISA and \element[][13]{CO}, respectively). The median HISA peak velocity of $\varv_{\mathrm{LSR}}=44.6\rm\,km\,s^{-1}$ toward GMF38b agrees with the \element[][13]{CO} velocity of $\varv_{\mathrm{LSR}}=44.4\rm\,km\,s^{-1}$ within the uncertainty of our HISA extraction method. However, we caution that this agreement might be the result of a selection bias that only takes into account velocities in a rather narrow range. Since the GMF38b filament region is identified in the narrow velocity range between $43.0\rm\,km\,s^{-1}$ and $46.0\rm\,km\,s^{-1}$, it is clear that the selection of velocities will show a smaller deviation between the two tracers. Toward GMF41 and GMF54 there is a more pronounced difference in median peak velocity. Within the GMF41 filament region, the median velocity traced by HISA is $\varv_{\mathrm{LSR}}=38.0\rm\,km\,s^{-1}$ while the median \element[][13]{CO} velocity is $\varv_{\mathrm{LSR}}=39.0\rm\,km\,s^{-1}$. The median peak velocities toward GMF54 are $\varv_{\mathrm{LSR}}=24.2\rm\,km\,s^{-1}$ and $23.1\rm\,km\,s^{-1}$ for HISA and \element[][13]{CO}, respectively.
We show the histograms of line width in Fig.~\ref{fig:hist1line}. The peaks of the line width distributions are well resolved and above the spectral resolution limit, so the spectral resolution does not heavily affect the statistics of the kinematics. We find in general higher observed line widths in HISA than in \element[][13]{CO}. The \element[ ][13]{CO} line widths are 1.3--$3.0\rm\,km\,s^{-1}$ while the HISA line widths are between $3.1\rm\,km\,s^{-1}$ and $5.2\rm\,km\,s^{-1}$. The kinematic properties of the clouds are also summarized in Table~\ref{tab:summary_kinematics}. 
    \begin{table*}[!htbp]
     \caption{Kinematic properties of the giant filament regions.}
     \renewcommand*{\arraystretch}{1.3}
     \centering
     \begin{tabular}{l | c c c | c c c }
     & & HISA & & & \element[][13]{CO} & \\
     (1) & (2) & (3) & (4) & (5) & (6) & (7) \\
     \hline\hline
     Source name & $\langle\varv\rangle$ [$\rm km\,s^{-1}$] & $\langle\Delta\varv\rangle$ [$\rm km\,s^{-1}$] & $\langle\mathcal{M}\rangle$ & $\langle\varv\rangle$ [$\rm km\,s^{-1}$] & $\langle\Delta\varv\rangle$ [$\rm km\,s^{-1}$] & $\langle\mathcal{M}\rangle$ \\
     GMF20.0-17.9 & 44.7 & 3.9 & 4.8 & 44.0 & 3.0 & 9.5 \\
     GMF26.7-25.4 & 44.9 & 3.2 & 3.7 & 45.4 & 2.1 & 7.9 \\
     GMF38.1-32.4a & 54.3 & 3.8 & 4.1 & 54.8 & 2.5 & 9.0 \\
     GMF38.1-32.4b & 44.6 & 5.2 & 6.0 & 44.4 & 2.4 & 7.0 \\
     GMF41.0-41.3 & 38.0 & 3.6 & 3.7 & 39.0 & 2.5 & 8.2 \\
     GMF54.0-52.0 & 24.2 & 3.1 & 3.7 & 23.1 & 1.3 & 4.8 \\\hline
     \end{tabular}
     \tablefoot{
     Columns (2) and (5) give the median peak velocity as traced by HISA and \element[][13]{CO} for all six filament regions, respectively. Similarly, columns (3) and (6) present the median line width as traced by HISA and \element[][13]{CO}, respectively. Columns (4) and (7) give the median sonic Mach number of HISA and \element[][13]{CO}, respectively, which are computed in Sect.~\ref{sec:machnumber} using the sound speed at the temperatures estimated in Sects.~\ref{sec:HISA_coldens} and \ref{sec:CO_column_dens}.}
     \label{tab:summary_kinematics}
    \end{table*}
Assuming a kinetic temperature, we can estimate the expected thermal line width. In local thermodynamic equilibrium (LTE), the thermal line width (FWHM) is given by $\Delta\varv_{\mathrm{th}} = \sqrt{8\,\mathrm{ln}2\,k_{\mathrm{B}}T_{\mathrm{k}}/(\mu m_{\mathrm{H}})}$, where $k_{\mathrm{B}}$, $T_{\mathrm{k}}$, and $\mu$ are the Boltzmann constant, kinetic temperature, and the mean molecular weight of \ion{H}{i} ($\mu_{\mathrm{\ion{H}{i}}}=1.27$) and the CO molecule \citep[$\mu_{\mathrm{CO}}=2.34$;][]{1973asqu.book.....A,2000asqu.book.....C} in terms of the mass of a hydrogen atom $m_{\mathrm{H}}$, respectively. The kinetic temperature can be well approximated by the estimated spin or excitation temperature of HISA and \element[][13]{CO}, given the low temperatures and high densities of the cold gas (see Sects.~\ref{sec:HISA_coldens} and \ref{sec:CO_column_dens}).  If different line broadening effects are uncorrelated, the total observed line width will be
\begin{equation}
     \Delta \varv_{\mathrm{obs}} = \sqrt{\Delta\varv_{\mathrm{th}}^2 + \Delta\varv_{\mathrm{nth}}^2 + \Delta\varv_{\mathrm{res}}^2} \: ,
\end{equation}{}
\noindent where $\Delta\varv_{\mathrm{nth}}$ is the line width due to nonthermal effects and $\Delta\varv_{\mathrm{res}}$ is the line width introduced by the spectral resolution of the data. The thermal line widths are on the order of $\sim$0.5$\rm\,km\,s^{-1}$ for \element[][13]{CO} and $\sim$1.0$\rm\,km\,s^{-1}$ for HISA at the given temperatures. The observed line widths of both HISA and \element[][13]{CO} therefore show that the line widths cannot be explained by thermal broadening alone. Nonthermal effects such as turbulent motions have a significant effect on the observed line widths and are most likely the dominant driver for the broadening of the lines. We investigate the turbulent Mach number of the gas in Sect.~\ref{sec:machnumber}.

\subsection{Column density and mass}\label{sec:coldens_mass}
In this section, we compute the column densities toward each filament region using HISA, \element[][13]{CO}, and \ion{H}{i} emission as a tracer for the CNM, molecular hydrogen, and bulk atomic hydrogen, respectively. For the derivation of the column density maps we integrated each filament region over the velocity range given in Table~\ref{tab:overview}. The column density maps of each tracer can be found in Appendix~\ref{sec:coldens_app}.
\subsubsection{CNM column density traced by HISA}\label{sec:HISA_coldens}
Following the derivation given in \citet{2000ApJ...540..851G}, we compute the optical depth of HISA as
    \begin{equation}
        \tau_{\mathrm{HISA}} = -\mathrm{ln}\left(1-\frac{T_{\mathrm{on}}-T_{\mathrm{off}}}{T_{\mathrm{HISA}} - p_{\mathrm{bg}}T_{\mathrm{off}} - T_{\mathrm{cont}}}\right) \: ,
        \label{equ:T_ON-T_OFF}
    \end{equation}
\noindent with the dimensionless parameter $p_{\mathrm{bg}}\equiv T_{\mathrm{bg}}\,\left(1-e^{-\tau_{\mathrm{bg}}}\right)/T_{\mathrm{off}}$ describing the fraction of background emission in the optically thin limit \citep{1993A&A...276..531F}. Assuming a HISA spin temperature $T_{\mathrm{s}}$ ($=T_{\rm HISA}$), we can then calculate the \ion{H}{i} column density of the cold \ion{H}{i} gas using the general form \citep{2013tra..book.....W} 
    \begin{equation}
        \frac{N_{\mathrm{H}}}{\rm cm^{-2}} = 1.8224\times 10^{18}\,\, \frac{T_{\mathrm{s}}}{\rm K}\,\int\tau\left(T_{\mathrm{s}},\varv\right)\,\left( \frac{\mathrm{d}\varv}{\rm km\,s^{-1}}\right) \: ,
        \label{equ:HI_column_density}
    \end{equation}  
\noindent where $T_{\mathrm{s}}$ is the spin temperature of atomic hydrogen and $\tau\left(T_{\mathrm{s}},\varv\right)$ describes the optical depth. We estimate the column density uncertainty by setting $T_{\mathrm{on}}=T_{\mathrm{off}} - \Delta T$ in Eq.~\eqref{equ:T_ON-T_OFF} as the limit at which we can detect \ion{H}{i} self-absorption, where $\Delta T$ is the rms noise in emission-free channels.

We can estimate the amount of background emission from the radial \ion{H}{i} volume density distribution in the Galactic plane. For Galactocentric radii $7\lesssim D_{\mathrm{GC}}\lesssim 35\rm\,kpc$ \citet{2008A&A...487..951K} report an average mid-plane volume density distribution of $n(R)\sim n_0\,e^{-(D_{\mathrm{GC}}-D_{\odot})/D_n}$ with $n_0=0.9\rm\,cm^{-3}$, $D_{\odot}=8.5\rm\,kpc$, and with the radial scale length $D_n=3.15\rm\,kpc$ (IAU recommendations). We assume a constant volume density distribution in the inner Galaxy of $n(D_{\mathrm{GC}}<7\,\mathrm{kpc})=n(D_{\mathrm{GC}}=7\rm\,kpc)$ as the volume density distribution is flattening off at lower Galactocentric distances \citep[see Fig.~5 in][]{2008A&A...487..951K}. This relation gives the averaged distribution of the northern and southern Galactic plane and could hold systematic differences in some regions \citep{2009ARA&A..47...27K}.

In principle, only atomic gas located at a distance that corresponds to the radial velocity of HISA is relevant to the estimation of the background. Since we usually place any observed HISA at the near kinematic distance, most of the emission must stem from the background and thus move $p_{\mathrm{bg}}$ close to 1 at any given finite spectral resolution element. However, due to the velocity dispersion of \ion{H}{i}, in particular the WNM, atomic hydrogen emission that has radial velocities slightly offset from the HISA velocity can blend into the velocity channels under consideration. Atomic hydrogen emission in the foreground or background that corresponds to radial velocities around HISA can therefore contribute to the observed feature and affect the optical depth computation. We estimate the background fraction using the volume densities in the kinematic near and far distance intervals $\Delta d$ corresponding to radial velocity intervals around the mean velocities of the clouds. The length of the distance interval is estimated from an average velocity dispersion of $\sim$10$\rm\,km\,s^{-1}$ that falls between typical CNM and WNM velocity dispersions found in the Milky Way \citep{2003ApJ...586.1067H,2007A&A...466..555H}. Equal steps in radial velocity $\Delta\varv_{\mathrm{LSR}}$ will map into unequal steps in distance $\Delta D$ that are proportional to the inverse of the velocity gradient along the line of sight
    \begin{equation}
        \Delta D = \left|\frac{\mathrm{d}D}{\mathrm{d}\varv_{\mathrm{LSR}}}\right|\Delta\varv_{\mathrm{LSR}} \: .
        \label{equ:dist_interval}
    \end{equation}  
\noindent Using the rotation curve by \citet{2007ApJ...671..427M} gives distance intervals between $\sim$0.6\,kpc and $\sim$0.9\,kpc for the considered clouds. The gas density is then integrated from $d_{\mathrm{near}}-\Delta d$ to $d_{\mathrm{near}}$ to obtain a foreground fraction of the emission at the velocity of HISA. The background gas fraction is inferred by adding the integrated gas density in the interval $[d_{\mathrm{near}}, d_{\mathrm{near}} + \Delta d]$ to the gas density integrated on the kinematic far side interval $[d_{\mathrm{far}}-\Delta d,d_{\mathrm{far}}+\Delta d]$. The derived background fractions are between 0.75 and 0.77. If we assume a continued exponential rise in volume density toward the inner Galaxy instead of a constant density distribution, the background fraction increases by up to three percentage points. This effect is strongest for sources at lower Galactic longitude. We do note that while considering many factors in the treatment of the background fraction, the uncertainties are substantial due to non-circular and streaming motions superposed with the Galactic rotation, or systematic differences in the density distribution of \ion{H}{i}. This adds a considerable source of uncertainty to the column density derivation. The column density decreases by a factor of $\sim$2 if the background fraction increases from $p_{\mathrm{bg}}=0.7$ to 0.9 \citep[see a detailed discussion of these uncertainties in][]{2020A&A...634A.139W,2020A&A...642A..68S}. These uncertainties will be revisited later in this section. Since we expect most of the emission background to originate in more diffuse \ion{H}{i} gas, we assume a constant $p_{\mathrm{bg}}$ for each filament region (see Table~\ref{tab:tspin_p}).

Depending on the assumed spin temperature $T_{\mathrm{s}}$ and background fraction $p_{\mathrm{bg}}$, there might be no solution to Eq.~\eqref{equ:T_ON-T_OFF} in some velocity channels of the spectra if the spin temperature is too high. Disregarding these channels in the line-of-sight integration, the column density computed in Eq.~\eqref{equ:HI_column_density} underestimates the true column density toward some regions. To resolve this, we derived the maximum spin temperature limit by $T_{\mathrm{max}}=T_{\mathrm{on}}+T_{\mathrm{cont}}-(1-p_{\mathrm{bg}})T_{\mathrm{off}}$ for $\tau\to\infty$ (see Eq.~\ref{equ:T_ON-T_OFF}), and set the 0.1-percentile of the maximum spin temperature to be the spin temperature of the whole cloud, such that outliers in the extracted baseline data or strong continuum emission do not affect the temperature estimate while Eq.~\eqref{equ:T_ON-T_OFF} gives a solution for 99.9\% of pixels in the integrated column density map. The assumed spin temperatures $T_{\mathrm{s}}$ and estimated background fractions $p_{\mathrm{bg}}$ are shown in Table~\ref{tab:tspin_p} for each GMF source. Since a lower assumed spin temperature (at constant background fraction) producing the same observed HISA feature results in a lower column density (see Eqs.~\ref{equ:T_ON-T_OFF} and \ref{equ:HI_column_density}), the total column density will yet again be underestimated. This shortcoming of the HISA column density computation is addressed in the HISA simulations conducted by \citet{2022MNRAS.512.4765S}. However, assuming a constant spin temperature for the entire cloud appears to be the best approach to qualitatively recover the true column density structure of the cloud \citep{2022MNRAS.512.4765S}.

If the spin temperature is varied by 10\,K, the column density, and consequently mass, traced by HISA changes by a factor of $\sim$2. Hence, the largest uncertainty arises from the assumption of a spin temperature and the background fraction that is coupled to the optical depth computation. Even for an assumed spin temperature that comes close to the limit at which the optical depth computation gives an analytic solution (see Eq.~\ref{equ:T_ON-T_OFF}), the column density will still be underestimated due to line-of-sight variations in spin temperature and observational noise. By assuming an optically thick HISA cloud with $\tau\rightarrow\infty$, we are able to determine the spin temperature limit above which the line-of-sight geometry does not allow the computation of the column density. The uncertainty in column density and mass is further amplified by the background fraction $p_{\mathrm{bg}}$ in Eq.~\eqref{equ:T_ON-T_OFF}. If the background fraction is lowered, the column density will increase as the cold \ion{H}{i} cloud would be more efficient in producing the same observed HISA feature given the weaker background. A moderate variation in the background fraction of 10\% at fixed spin temperature results in a $\sim$30\% change in column density and therefore mass. We derive a HISA mass uncertainty by varying the background fraction by 10\% and adjusting the spin temperature accordingly, such that we have again a solution for most pixels in the map. We report these uncertainties in Table~\ref{tab:masses} as well. For a more detailed discussion of these uncertainties we refer to \citet{2020A&A...634A.139W} and \citet{2020A&A...642A..68S}.

The aforementioned statistical uncertainties add to the intrinsic systematic effects of the HISA method. We are generally limited by the background emission that enables the observation of HISA. Furthermore, HISA is only sensitive to gas that is colder than the gas that contributes to the emission background. The CNM is reported to have spin temperatures up to $\sim$300\,K \citep[e.g.,][]{2003ApJ...586.1067H,2009ARA&A..47...27K}, rendering the HISA detection of the CNM in many cases impossible given the observed brightness temperatures. According to the simulations conducted by \citet{2022MNRAS.512.4765S}, the HISA-traced mass underestimates the mass of the CNM that could in principle be observed through HISA by a factor of 3--10. This underestimation is generally attributed to two effects. The proper estimation of the spin temperature that is required to compute the HISA properties is a challenging task because of its variation within a cold \ion{H}{i} cloud. Due to the line-of-sight geometry, an assumed \ion{H}{i} spin temperature that is too low will result in an underestimate of the optical depth and the true column density (see Eqs.~\ref{equ:T_ON-T_OFF} and \ref{equ:HI_column_density}). An \ion{H}{i} spin temperature that is too high will cause the HISA-traced cloud to have no solution to the optical depth at least for some part of the spectrum (Eq.~\ref{equ:T_ON-T_OFF}). This will again underestimate the integrated column density as individual spectral channels are omitted. Varying the spin temperature along the line-of-sight or spatially can lead to an even larger deviation and might recover a column density structure that does not reflect the true distribution qualitatively. The challenges of unknown spin temperature consequently give rise to a large systematic uncertainty in the determination of the column density and mass \citep{2022MNRAS.512.4765S}.
    \begin{table}[!htbp]
        \caption{Assumed spin temperatures and background fractions.}
        \renewcommand*{\arraystretch}{1.3}
        \centering
        \begin{tabular}{l c c}
             \hline\hline
             Source & $T_{\mathrm{s}}$ [K] & background fraction $p_{\mathrm{bg}}$ \\\hline
             GMF20 & 26 & 0.75 \\
             GMF26 & 27 & 0.75 \\
             GMF38a & 32 & 0.75 \\
             GMF38b & 33 & 0.75 \\
             GMF41 & 37 & 0.76 \\
             GMF54 & 24 & 0.77 \\\hline
        \end{tabular}
        \tablefoot{
        The second column gives the spin temperature $T_{\mathrm{s}}$ assumed toward each GMF region. The third column gives the background fraction $p_{\mathrm{bg}}$ that is estimated from the ratio of foreground and background column density along the line of sight (see Sect.~\ref{sec:HISA_coldens}).
        }
        \label{tab:tspin_p}
    \end{table}

\subsubsection{$\rm H_2$ column density}
We computed the \element[][13]{CO} column densities following the standard procedure given in \citet{2013tra..book.....W}. Details of the derivation are given in Appendix~\ref{sec:CO_column_dens}.
In order to convert the \element[][13]{CO} column densities to $\rm H_2$ column densities, we used Galactocentric distance-dependent abundance relations to estimate an $[\rm H_2]/[\element[][13]{CO}]$ conversion factor for each source. \citet{2014A&A...570A..65G} give a \element[][12]{CO}-to-\element[][13]{CO} abundance relation of $[\element[][12]{CO}]/[\element[][13]{CO}]=6.2D_{\mathrm{GC}} + 9.0$, and the $\rm H_2$-to-\element[][12]{CO} abundance given in \citet{2012MNRAS.423.2342F} is $[\mathrm{H_2}]/[\element[][12]{CO}]=[8.5\times10^{-5}\,\mathrm{exp}(1.105\,-0.13D_{\mathrm{GC}})]^{-1}$, where $D_{\mathrm{GC}}$ is the Galactocentric distance in units of $\rm kpc$. We estimate the uncertainty in $\rm H_2$ column density to be at least 50\% due to the large uncertainties in these relations. Furthermore, CO might not always be a good tracer of $\rm H_2$ as "CO-dark $\rm H_2$" could account for a significant fraction of the total $\rm H_2$ \citep{2008ApJ...679..481P,2009ApJ...692...91G,2013A&A...554A.103P,2014MNRAS.441.1628S,2016A&A...593A..42T}, particularly at low column densities and early evolutionary stages as molecular clouds might not have become CO-bright yet \citep{2008ApJ...680..428G,2011A&A...536A..19P}. The $[\rm H_2]/[\element[][13]{CO}]$ conversion factor for each source is given in Table~\ref{tab:T_ex_limits}.
\begin{table}[!htbp]
        \caption{Limits of CO excitation temperatures and optical depths.}
        \renewcommand*{\arraystretch}{1.2}
        \centering
        \begin{tabular}{l c c c c}
             \hline\hline
             Source & $T_{\mathrm{ex,low}}$ [K] & $T_{\mathrm{ex,up}}$ [K] & $\tau_{\mathrm{low}}$ & $X(\mathrm{[H_2]/[\element[][13]{CO}]})$ \\\hline
             GMF20 & 5 & 29 & 0.05 & $3.1\times 10^5$ \\
             GMF26 & 5 & 16 & 0.11 & $3.6\times 10^5$ \\
             GMF38a & 5 & 21 & 0.08 & $3.9\times 10^5$ \\
             GMF38b & 5 & 18 & 0.10 & $4.1\times 10^5$ \\
             GMF41 & 5 & 12 & 0.26 & $4.7\times 10^5$ \\
             GMF54 & 5 & 36 & 0.04 & $5.6\times 10^5$ \\\hline
        \end{tabular}
        \tablefoot{
        The second and third column gives the lower limit and upper limit of the \element[][]{CO} excitation temperature, respectively. The fourth column shows the lower limit of the optical depth estimated from the $5\sigma$ \element[][13]{CO} noise and the highest excitation temperature found toward each source (see also Appendix~\ref{sec:CO_column_dens} for details). The last column gives the \element[][13]{CO}-to-$\rm H_2$ conversion factor that we have used for each source. \\
        }
        \label{tab:T_ex_limits}
    \end{table}{}

\subsubsection{Atomic gas column density seen in \ion{H}{i} emission}\label{sec:HI_emission_optical_depth}
In addition to the cold atomic gas traced by HISA, we investigated the properties of the total atomic hydrogen gas budget (WNM+CNM) by measuring the column density from \ion{H}{i} emission and correcting for optical depth effects and diffuse continuum. As the optically thin assumption might not hold for some regions, we can utilize strong continuum emission sources to directly measure the optical depth. \ion{H}{i} continuum absorption (HICA) is a classical method to derive the properties of the CNM \citep[e.g.,][]{2004ApJ...603..560S,2003ApJ...586.1067H}. This method uses strong continuum sources, such as Galactic \ion{H}{ii} regions or active galactic nuclei (AGNs), to measure the optical depth of \ion{H}{i}. As these sources have brightness temperatures that are larger than typical spin temperatures of cold \ion{H}{i} clouds, we observe the \ion{H}{i} cloud in absorption. The absorption feature is furthermore dominated by the CNM since the absorption is proportional to $T_{\mathrm{s}}^{-1}$ \citep[e.g.,][]{2013tra..book.....W}.

The advantage of this method is the direct measurement of the optical depth. However, the HICA method requires strong continuum emission sources. As most strong continuum sources are discrete point sources, this method results in an incomplete census of optical depth measurements. However, \citet{2020A&A...634A..83W} derived a velocity-resolved optical depth map computed from 228 continuum sources within the THOR survey that are above a $6\sigma$ noise threshold and interpolated the measurements using a nearest-neighbor method. For more details about the optical depth measurement we refer to \citet{2020A&A...634A..83W}. To the first approximation, we can use this optical depth map to correct the \ion{H}{i} column density as confirmed in \citet{2020A&A...642A..68S}, in spite of potential kinematic distance ambiguities and the location of a continuum source along each line of sight that might add or miss optical depth for each line-of-sight velocity, respectively. For each velocity channel, we take the spatial average of the optical depth map measured toward each filament region in order to avoid artifacts introduced by the interpolation.
    
Besides strong continuum sources we observe weak continuum emission throughout the Galactic plane. This component has brightness temperatures between $10$ and $50\rm\,K$. The continuum emission has been subtracted during data reduction as described in Sect.~\ref{sec:methods_observation}. As even weak continuum emission might suppress \ion{H}{i} emission and therefore lead to an underestimate of the column density, we account for the weak emission component when computing the \ion{H}{i} column density \citep[see][Eq.~9]{2015A&A...580A.112B}. We estimate the column density and mass uncertainty by varying the optical depth by 10\%, which roughly corresponds to the $1\sigma$ brightness variation of our weakest continuum sources.

\subsection{Masses}\label{sec:masses}
Based on the column density estimates in the previous sections, we can directly estimate the (cold) atomic and molecular mass toward the filament regions (see Table \ref{tab:masses}). We compute the masses by summing up the mass pixels above a column density threshold corresponding to significant emission or \ion{H}{i} self-absorption. These thresholds are then also used to derive column density PDFs (see Sect.~\ref{sec:coldens_PDF}). The change in mass that comes with varying thresholds is relatively small compared to the uncertainties of the column density derivation itself.

The CNM mass traced by HISA corresponds to 3--9\% of the total atomic gas mass, depending on the region and assumed spin temperature. The HISA mass fraction toward GMF38b is 0.09 and exceeds that found in all other filament regions. We recovered column density regions off the main cloud that is defined as GMF38.1-32.4b (see Fig.~\ref{fig:coldens4}). These regions might not be tightly associated with the molecular gas that defines the GMF. The HISA mass fraction reduces to 3--4\% if we only take into account the gas in close proximity to the main molecular feature of the cloud (gas beyond the lowest contour in Fig.~\ref{fig:coldens4} to within 0.2\degr{} offset), thus being comparable to other filament regions. However, this example also illustrates that we recover cold atomic gas structures that do not have a molecular counterpart. The cold phase of the atomic ISM appears to be much more widespread than the molecular gas in Fig.~\ref{fig:coldens4}. The masses of both GMF20 and GMF38a are similar to the masses found by \citet{2020A&A...642A..68S} and \citet{2020A&A...634A.139W}. Given that we assume a spin temperature of 26\,K for GMF20 (instead of 20\,K and 40\,K in \citealt{2020A&A...642A..68S}), the derived mass falls within the mass range $4.6\times10^3$--$1.3\times10^4\,\rm M_\sun$ obtained in \citet{2020A&A...642A..68S}.

The molecular hydrogen mass is on the order of $10^4$--$10^5\rm\,M_\sun$. The total atomic gas fraction shows large differences among the filament regions. The atomic gas mass is generally comparable to the molecular gas mass. However, for GMF54 the atomic gas seen in \ion{H}{i} emission and HISA accounts to a total that is just one quarter of the total hydrogen mass. With respect to the molecular gas phase, the total atomic gas fraction is found to increase with Galactocentric distance on average \citep[e.g.,][]{2016PASJ...68....5N,2017ApJ...834...57M}. In spite of having the largest Galactocentric distance in our sample, GMF54 appears to have used up much of the atomic gas in which it was embedded to transition into a more complete molecular gas phase.
    \begin{table*}[!htbp]
     \caption{Derived masses of the filament regions.}
     \renewcommand*{\arraystretch}{1.3}
     \centering
     \begin{tabular}{ l c c c c c }
     (1) & (2) & (3) & (4) & (5) & (6)\\
     \hline\hline
     & $M(\mathrm{HISA})$ [$\rm M_\sun$] & $M(\mathrm{\ion{H}{i}})$ [$\rm M_\sun$] & $M(\mathrm{H_2})$ [$\rm M_\sun$] & $f_{\mathrm{HISA}}$ & $f_{\mathrm{atomic}}$ \\
     GMF20.0-17.9 & $7.5\substack{+1.3 \\ -1.9}\times10^3$ & $2.7\substack{+0.2 \\ -0.3}\times10^5$ & $2.3\pm1.2\times10^5$ & 0.03 & 0.55 \\
     GMF26.7-25.4 & $1.1\substack{+0.3 \\ -0.2}\times10^3$ & $3.0\substack{+0.3 \\ -0.3}\times10^4$ & $5.5\pm2.8\times10^4$ & 0.04 & 0.36 \\
     GMF38.1-32.4a & $1.1\substack{+0.2 \\ -0.3}\times10^4$ & $3.6\substack{+0.5 \\ -0.7}\times10^5$ & $3.0\pm1.5\times10^5$ & 0.03 & 0.55 \\
     GMF38.1-32.4b & $6.2\substack{+1.3 \\ -1.6}\times10^3$ & $6.2\substack{+0.7 \\ -0.7}\times10^4$ & $5.5\pm2.8\times10^4$ & 0.09 & 0.55 \\
     GMF41.0-41.3 & $5.3\substack{+0.7 \\ -1.1}\times10^2$ & $1.4\substack{+0.2 \\ -0.3}\times10^4$ & $1.7\pm0.9\times10^4$ & 0.04 & 0.46 \\
     GMF54.0-52.0 & $3.3\substack{+0.7 \\ -1.0}\times10^2$ & $5.2\substack{+1.1 \\ -0.9}\times10^3$ & $1.6\pm0.8\times10^4$ & 0.06 & 0.26 \\
     \hline
     \end{tabular}
     \tablefoot{
     The masses were calculated from the column density maps shown in Appendix~\ref{sec:coldens_app}. Column (2) presents the mass of the cold atomic hydrogen traced by HISA. The uncertainties are statistical errors arising from the uncertainties in background fraction and spin temperature and do not include the systematic uncertainties due to the detection method. Column (3) shows the atomic hydrogen mass inferred from the optical depth and continuum corrected \ion{H}{i} emission. The uncertainties are estimated from variations in the optical depth measurement. Column (4) gives the molecular hydrogen mass as traced by \element[ ][13]{CO} emission along with a conservative 50\% uncertainty owing to the large uncertainties in the CO-$\rm H_2$ conversion. Column (5) gives the mass traced by HISA as a fraction of the total atomic gas mass $M(\mathrm{HISA}) + M(\mathrm{\ion{H}{i}})$. Column (6) is the fraction of the total atomic gas mass traced by HISA and \ion{H}{i} emission with respect to the total gas mass $M(\mathrm{HISA}) + M(\mathrm{\ion{H}{i}}) + M(\mathrm{H_2})$.}
     \label{tab:masses}
    \end{table*}

\section{Discussion}\label{sec:discussion}
\subsection{Column density PDF}\label{sec:coldens_PDF}
We employ the probability density function (PDF) of the column density to investigate the physical processes acting within the filament regions. The shape of column or volume density PDFs are commonly used as a means to describe the underlying physical mechanisms of a cloud \citep[e.g.,][]{2013ApJ...763...51F,2014prpl.conf...77P,2014Sci...344..183K,2015A&A...575A..79S, 2022A&A...666A.165S}. Turbulence is considered to be the dominant driver of a cloud's structure if its PDF shows a log-normal shape. Furthermore, the width of a log-normal PDF is linked to the Mach number as it changes with the magnitude of the turbulence driving the cloud's structure \citep[e.g.,][]{1997MNRAS.288..145P,1998PhRvE..58.4501P,2002ApJ...576..870P,2007ApJ...665..416K,2008PhST..132a4025F,2012ApJ...761..149K,2012MNRAS.423.2680M,2014Sci...344..183K,2022MNRAS.517.5003B}, while noting that the turbulence driving scale and CNM-WNM mass ratio also affect the width of the PDF \citep{2017ApJ...843...92B}.

Molecular clouds that are subject to the increasing effect of self-gravity develop high-density regions, producing a power-law tail in their PDF \citep[e.g.,][]{2000ApJ...535..869K,2014ApJ...781...91G,2017ApJ...834L...1B}. Many star-forming molecular clouds have been confirmed to show such power-law tails \citep{2009A&A...508L..35K,2013ApJ...766L..17S,2016A&A...587A..74S,2022A&A...666A.165S}. Even before the effects of gravity become dominant, gravitationally unbound clumps can exhibit power-law tails due to pressure confinement from the surrounding medium \citep{2011A&A...530A..64K}.

\begin{figure*}
     \centering
     \includegraphics[width=0.9\textwidth]{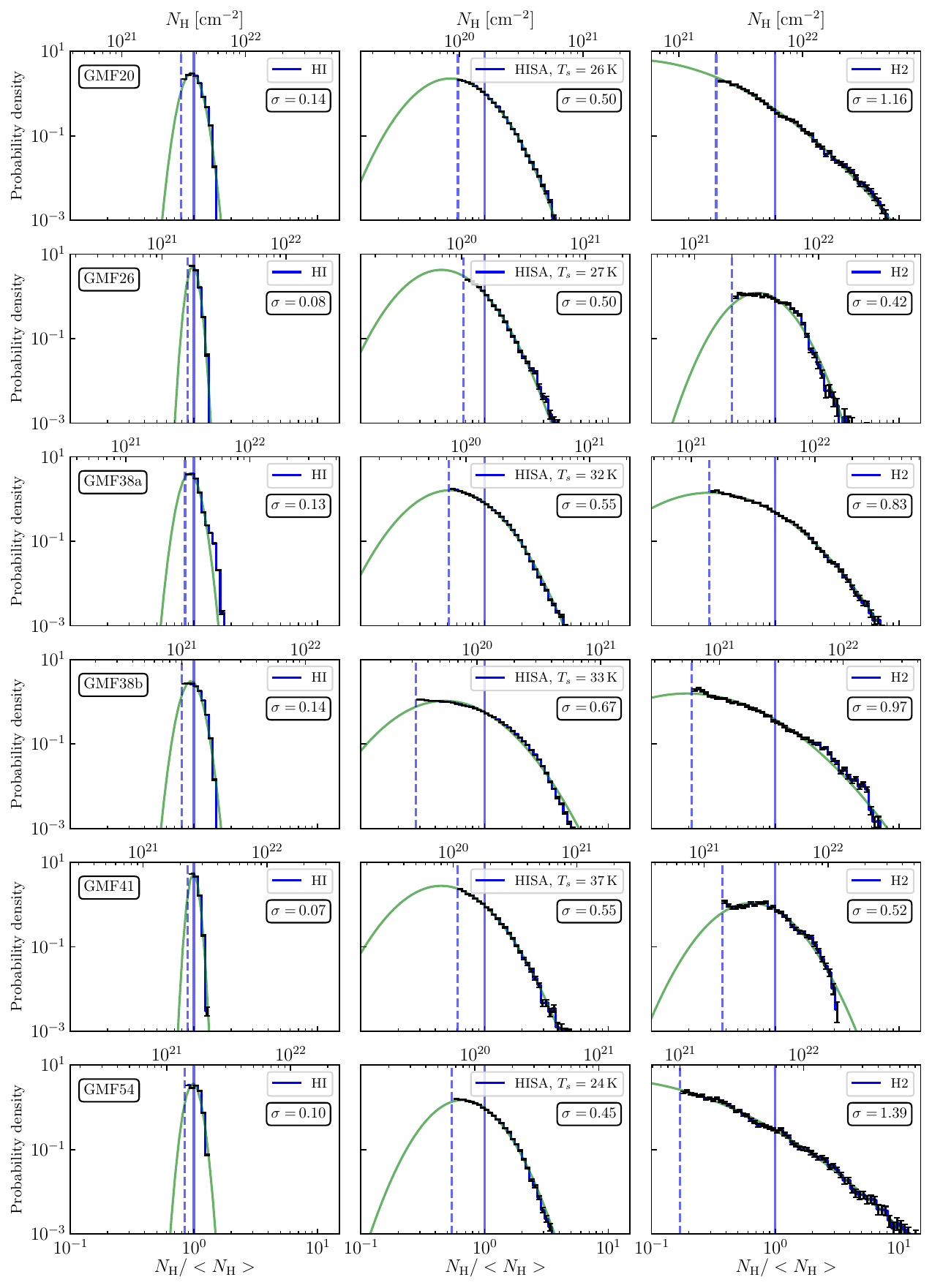}
        \caption[]{Normalized column density PDFs of \ion{H}{i}, HISA, and $\rm H_2$ toward the giant filament regions. Each row shows the \ion{H}{i}, HISA, and $\rm H_2$ N-PDF toward one GMF region. \textit{Left panels:} N-PDFs traced by \ion{H}{i} emission. The distributions are derived from the \ion{H}{i} column densities that have been corrected for optical depth and continuum emission. \textit{Middle panels:} The N-PDFs of the gas traced by HISA. \textit{Right panels:} $\rm H_2$ N-PDFs traced by \element[ ][13]{CO} in units of hydrogen atoms per $\rm cm^{-2}$. The green curves indicate a log-normal fit to the distributions. The blue vertical dashed and solid lines mark the column density threshold and mean column density, respectively.}
        \label{fig:N-PDFs}
   \end{figure*}

We show in Fig.~\ref{fig:N-PDFs} the column density PDFs (N-PDFs) of all filament regions as traced by \ion{H}{i} emission, HISA, and \element[][13]{CO}. We take into account only column densities above the noise threshold of each tracer and find that the widths of each N-PDF do not change significantly when considering higher thresholds. The column density thresholds are $\sim$2$\times 10^{21}\rm\,cm^{-2}$ for \ion{H}{i} emission, $\sim$8$\times 10^{19}\rm\,cm^{-2}$ for HISA, and $\sim$1$\times 10^{21}\rm\,cm^{-2}$ for molecular hydrogen. We fit all column density PDFs with a log-normal function and report their widths in Fig.~\ref{fig:N-PDFs}. Since we use a consistent way in deriving the PDFs, systematic differences between the distributions should be small, such that they can be well compared in relative terms. All N-PDFs are well described by a log-normal function.

Toward all filaments, the HISA-traced cold atomic gas shows a column density distribution that is broader than the narrow distribution of the diffuse atomic gas (left panels of Fig.~\ref{fig:N-PDFs}). The mean column densities of molecular hydrogen are at least an order of magnitude higher than the column densities traced by HISA. We note the narrow distributions in the molecular gas phase toward GMF26 and GMF41 that are comparable to the HISA distributions. Toward the other filament regions, the molecular gas N-PDF has a larger width than the HISA N-PDF, highlighting the spatially more concentrated distribution of the molecular gas. The relatively narrow distributions of GMF26 and GMF41 might be related to the low excitation temperatures we find toward these clouds. This might be an indication of an early evolutionary stage where gravity has not yet become dynamically important. This is further supported by the low number of YSOs identified toward GMF41 \citep[see][]{2019A&A...622A..52Z}.
   
The narrow log-normal shaped N-PDFs are commonly observed in the diffuse \ion{H}{i} emission toward well-known molecular clouds \citep{2015ApJ...811L..28B,2016ApJ...829..102I,2017MNRAS.472.1685R,2022A&A...666A.165S}. The HISA N-PDFs that trace the CNM show broader distributions, indicative of the clumpy structure and higher degree of turbulence. Considering the column density PDFs, HISA appears to trace the cold atomic gas phase that connects the diffuse state of the atomic ISM with the transition of a cloud becoming molecular.

\subsection{Mach number distribution}\label{sec:machnumber}
In the following, we derive the turbulent Mach number distributions using the constant HISA spin temperatures given in Table~\ref{tab:tspin_p} and the excitation temperatures of \element[][13]{CO} derived in Sect.~\ref{sec:CO_column_dens}. Given the low temperature regime, we approximate the kinetic gas temperatures of the CNM and the molecular gas with the spin temperature and mean excitation temperature, respectively. We then estimate the three-dimensional scale-dependent Mach number of the filaments assuming isotropic turbulence with $\mathcal{M}=\sqrt{3}\,\sigma_{\mathrm{turb}}/c_s$, where $\sigma_{\mathrm{turb}}$ and $c_s$ are the turbulent one-dimensional velocity dispersion and sound speed, respectively.
The turbulent line width is calculated by subtracting the thermal line width contribution from the observed line width as
   \begin{equation}
       \sigma_{\rm turb} = \sqrt{\sigma_{\rm obs}^2 - \sigma_{\rm th}^2 - \sigma_{\rm res}^2} \: ,
   \end{equation}
\noindent where $\sigma_{\rm obs}$, and $\sigma_{\rm th}$ are the observed, and thermal velocity dispersion, respectively. For completeness, we also account for the broadening introduced by the spectral resolution $\sigma_{\rm res}$. Since the thermal line width and sound speed scale as $T_{\mathrm{k}}^{1/2}$, the variation with spin temperature is moderate and does not change the Mach number significantly. \citet{2022MNRAS.512.4765S} showed that the Mach number estimate inferred through HISA is robust and can be determined with an accuracy within a factor of $\sim$2.

\begin{figure*}
     \centering
     \includegraphics[width=1.0\textwidth]{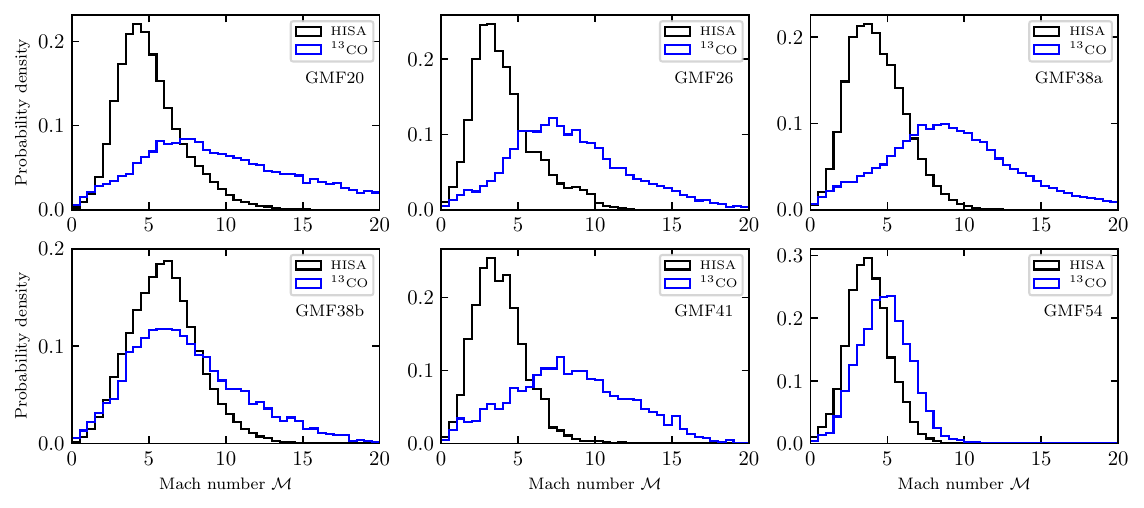}
        \caption[]{Histograms of Mach numbers. The panels show for each of the six filament regions the normalized histograms of the Mach numbers of HISA and \element[ ][13]{CO} in black and blue, respectively.}
      \label{fig:hist1mach}
    \end{figure*}

Five of the six filament regions show very similar Mach number distributions (Fig.~\ref{fig:hist1mach}). The Mach number distributions traced by HISA are generally much narrower than those traced by \element[][13]{CO} emission, and peak round $\mathcal{M}\sim3-6$, with few values as high as $\sim$10. Our findings are in very good agreement with recent HISA observations \citep{2015ApJ...811L..28B,2019ApJ...880..141N,2020A&A...634A.139W,2020A&A...642A..68S} and the simulations conducted by \citet{2022MNRAS.512.4765S}.

With the exception of GMF54, the molecular gas is highly supersonic, and has median Mach numbers between $\mathcal{M}\sim7-10$. The molecular gas toward GMF54 is moderately supersonic and has a median Mach number around $\sim$5. The total observed line widths are generally small with a few $\sim$km$\rm\,s^{-1}$ (see Fig.~\ref{fig:kin12}) and we do find the highest excitation temperatures up to $\sim$35\,K in GMF54. As the HISA Mach numbers are also smallest toward GMF54, we consider this an imprint of a different physical mechanism dominating the dynamics of the cloud. In combination with the high excitation temperatures, low atomic mass fraction, and the most pronounced power-law tail in its column density distribution that we find in our sample, GMF54 appears to be at a much more advanced stage in its evolution, at which gravity seems to be the dominant driver of the cloud's dynamics. \citet{2019A&A...622A..52Z} also find a star formation rate surface density that is among the highest in their sample of giant molecular filaments.
\subsection{Spatial correlation between atomic and molecular gas}\label{sec:HOG}
According to the classical idealized photodissociation region (PDR) picture, we would expect cold atomic gas to be spatially associated with its molecular counterpart \citep[e.g.,][]{1988ApJ...334..771V,1991ApJ...366..464A}. We therefore employed the Histogram of Oriented Gradients tool\footnote{\url{https://github.com/solerjuan/astroHOG}} \citep[HOG;][]{2019A&A...622A.166S} to investigate the spatial correlation between HISA and \element[][13]{CO} emission. The HOG method is based on machine vision to examine the spatial correlation between two spectral line tracers across their spectral domain. A detailed description of the HOG is given in \citet{2019A&A...622A.166S}.

The underlying principle is the computation of intensity gradients in each velocity channel map of the respective line tracer. The relative angles between the intensity gradients of the line tracers (here HISA and \element[][13]{CO}) are then computed for each pair of velocity channel maps. To statistically evaluate the significance of spatial correlation in terms of relative orientation between intensity gradients, the HOG uses the projected Rayleigh statistic $V$ as a metric, which is a test to determine if the distribution is non-uniform and centered around $0\degr$. It is tuned such that the sign of $V$ is indicative of the angle distribution having a peak around $\theta=0\degr$ ($V>0$) or $\theta=90\degr$ ($V<0$) \citep{2018MNRAS.474.1018J}. The absolute value of $V$ indicates the significance of that preferred orientation in the angle distribution. The projected Rayleigh statistic is therefore
\begin{equation}
    V = \frac{\sum_{ij}^{m,n} w_{ij} \mathrm{cos}(2\theta_{ij})}{\sqrt{\sum_{ij}^{m,n}w_{ij}/2}} \: ,
\end{equation}
\noindent where the indices $i$ and $j$ run over the pixel locations in the two
spatial dimensions for a given velocity channel and $w_{ij}$ is the statistical weight of each angle $\theta_{ij}$. We account for the spatial correlation between pixels introduced by the telescope beam and set the statistical weights to $w_{ij}=(\delta x/\Delta)^2$, where $\delta x$ is the pixel size and $\Delta$ is the diameter of the derivative kernel that we used to calculate the gradients. We set the derivative kernel to $\Delta=92\arcsec$, which is twice the beam size of the GRS.

We smoothed the extracted HISA cubes to a common beam size of 46\arcsec{} and reprojected them onto the same spatial grid as the \element[][13]{CO} data to run the HOG. Furthermore, we restricted the radial velocity range to $\varv_{\mathrm{LSR,low}}-25\rm\,km\,s^{-1}$ and $\varv_{\mathrm{LSR,up}}+25\rm\,km\,s^{-1}$ to save computational cost, where $\varv_{\mathrm{LSR,low}}$ and $\varv_{\mathrm{LSR,up}}$ are the lower and upper velocity limits given in Table~\ref{tab:overview}, respectively. The extension of the velocity range $\pm 25\rm\,km\,s^{-1}$ provides a baseline measure of $V$ (assuming there are signal-free channels over this velocity range). The projected Rayleigh statistic $V$ should be $\sim$0 for these channels.

We use Monte Carlo sampling to propagate the errors introduced by the uncertainties in the flux measurement in each velocity channel \citep[see e.g.][]{2020A&A...642A.163S}. For each velocity channel map, we generated 10 random realizations per tracer with the same mean intensity and observational noise. Using this sampling, the uncertainty of the correlation can be determined by the variance of the correlation of different Monte Carlo realizations. Since we expect a contribution from non-gaussian noise introduced by the observation, we report only $\geq 5\sigma$ confidence levels.
    \begin{figure*}
     \centering
     \includegraphics[width=1.0\textwidth]{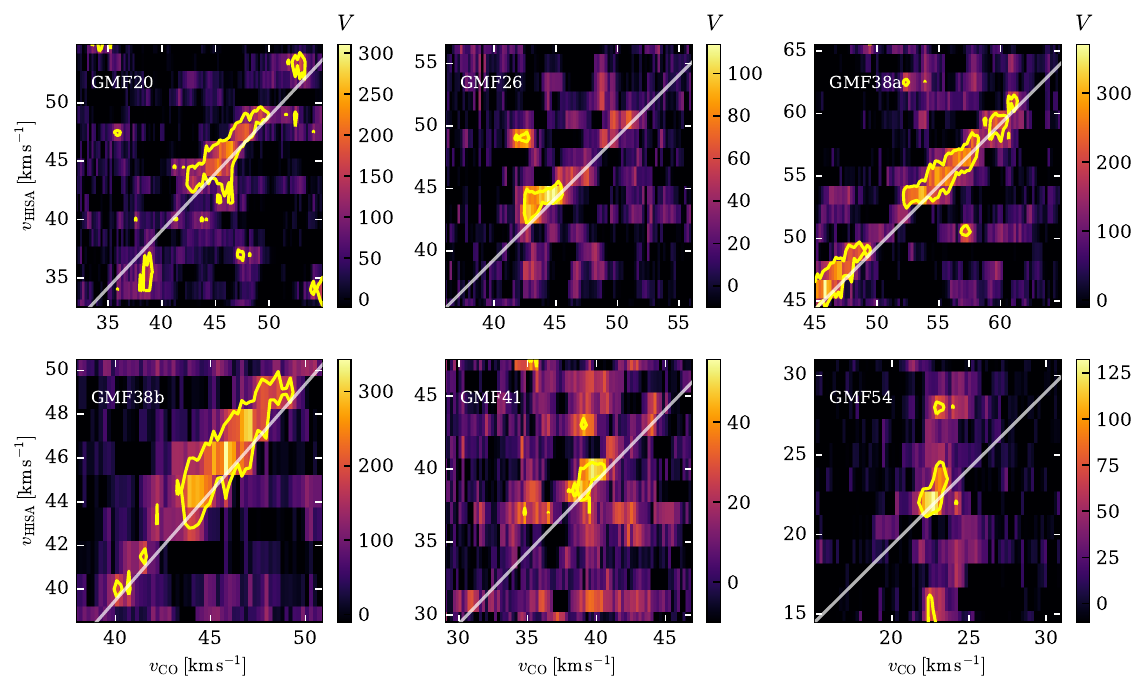}
        \caption[]{Correlation in the distribution of HISA and \element[][13]{CO} emission toward the six GMF regions as quantified by the projected Rayleigh statistic ($V$) in the HOG method \citep{2019A&A...622A.166S}. The panels present the computed spatial correlation between HISA and \element[][13]{CO} across velocities in terms of the projected Rayleigh statistic $V$ for each filament region. The values of $V$ are indicated by the color bar to the right of each panel. The white line in each panel shows the 1-to-1 correlation. The yellow contours show the $5\sigma$ threshold estimated from the Monte Carlo sampling. Large values of $V$ indicate a high spatial correlation. Values of $V$ close to zero indicate a random orientation of the HISA structures with respect to \element[][13]{CO} emission.}
        \label{fig:HOG_vplane}
    \end{figure*}
We show in Fig.~\ref{fig:HOG_vplane} the computed spatial correlation in terms of the projected Rayleigh statistic $V$ for each filament region. We observe a strong spatial correlation between HISA and \element[][13]{CO} toward GMF20, GMF38a, and GMF38b across multiple velocity channels. Even toward GMF26, GMF41, and GMF54 we detect significant spatial correlation in few velocity channels, despite little HISA detection. Although the spatial correlation between HISA and \element[][13]{CO} appears to be poor in Fig.~\ref{fig:kin1}, we note that the intensity gradients of both tracers are in fact aligned where significant signal is overlapping, even if the intensity peaks do not exactly match. Small deviations in velocity are reflected by the width of the 2D distribution across the 1-to-1 correlation. We were able to reproduce the result obtained by \citet{2020A&A...642A..68S} toward GMF20, showing a significant spatial correlation within the entire filament region. Despite detecting significant spatial correlation between HISA and \element[][13]{CO} toward all filament regions in our sample, we note that there might be little to no correlation when investigating subsections of filament regions, as observed in \citet{2020A&A...642A..68S} where the western part of the filament showed a strong agreement between the spatial distribution of HISA and \element[][13]{CO} while the eastern part entirely lacks correlation.
    
We conclude that the CNM traced by HISA generally appears to be associated with molecular gas in the giant filament regions on a large spatial scale. However, toward specific sub-regions within each filament systematic differences in spatial correlation can be evident that could be indicative of local events of star forming activity \citep[see e.g.][]{2020A&A...642A.163S,2021A&A...651L...4S}.

\section{Conclusions}\label{sec:conclusions}
We have investigated the properties of the cold atomic gas and molecular gas toward a sample of six giant molecular filament regions. We traced the cold atomic gas phase by \ion{H}{i} self-absorption and obtained these features using the newly developed baseline extraction algorithm \saber{}. The kinematic properties of both the cold atomic gas and molecular gas were obtained using the spectral decomposition tool \textsc{GaussPy+} \citep{2019A&A...628A..78R}. The main results are summarized as follows:
\begin{enumerate}
    \item We detect HISA toward all giant filament regions. The mass traced by HISA accounts to a few percent of the total atomic hydrogen mass traced by \ion{H}{i} self-absorption and emission. The total atomic mass is in most cases comparable to the molecular mass. Deviations from these mass fractions can be linked to different evolutionary stages of the clouds.
    \item On a global average, the median centroid velocities of identified HISA and \element[][13]{CO} appear to be similar, even though the agreement might be partially imposed by the restricted velocity range under consideration. A future large-scale HISA survey will facilitate an unbiased comparison of the global kinematics of HISA with molecular gas tracers. The well-resolved observed line widths of HISA are systematically larger than those of \element[][13]{CO}. The CoAt gas traced by HISA is found to be moderately supersonic with Mach numbers of a $\sim$few, while the molecular gas within the majority of the filaments is driven by highly supersonic dynamics.
    \item The derived column densities of the CoAt gas traced by HISA are on the order of $\sim$10$^{20}\rm\,cm^{-2}$ and the column density distributions of the CoAt gas can be well described by a log-normal. The HISA-traced column density distributions are broader than the N-PDFs of the diffuse atomic gas traced by \ion{H}{i} emission, indicating a spatially more concentrated cold gas distribution. The molecular gas has comparable or larger N-PDF widths than its cold atomic counterpart.
    \item The recovered HISA features show a spatial correlation with the molecular gas toward many regions within the filaments. The Histogram of Oriented Gradients analysis \citep{2019A&A...622A.166S} confirms a significant spatial correlation between HISA and \element[][13]{CO} toward all filament regions at similar velocities.
\end{enumerate}
Probing the cold atomic gas by means of \ion{H}{i} self-absorption toward molecular clouds is a powerful tool to investigate the dynamical and physical interplay between the atomic and molecular gas during cloud formation. While molecular clouds are ideal targets to investigate the properties of HISA, we are looking to extend our findings and identify HISA without the bias of corresponding molecular line emission. We will investigate the global distribution of HISA in the inner Galactic plane in an upcoming paper.
  
\begin{acknowledgements}
      J.S., H.B., and R.S.K. acknowledge support from the Deutsche Forschungsgemeinschaft (DFG, German Research Foundation) -- Project-ID 138713538 -- SFB 881 (``The Milky Way System'', subproject A01, B01, B02, B08). This research was carried out in part at the Jet Propulsion Laboratory, California Institute of Technology, under a contract with the National Aeronautics and Space Administration (80NM0018D0004). R.S.K. and J.D.S. thank for support from the European Research Council via the ERC Synergy Grant ``ECOGAL'' (project ID 855130). R.S.K. thanks support from the Heidelberg Cluster of Excellence (EXC 2181 - 390900948) ``STRUCTURES'', funded by the German Excellence Strategy, and from the German Ministry for Economic Affairs and Climate Action in project ``MAINN'' (funding ID 50OO2206). R.S.K. also is grateful for computing resources provided by the Ministry of Science, Research and the Arts (MWK) of the State of Baden-W\"{u}rttemberg through bwHPC and DFG through grant INST 35/1134-1 FUGG and for data storage at SDS@hd through grant INST 35/1314-1 FUGG. M.R.R. is a Jansky Fellow of the National Radio Astronomy Observatory. This research made use of the data from the Milky Way Imaging Scroll Painting (MWISP) project, which is a multi-line survey in 12CO/13CO/C18O along the northern galactic plane with PMO-13.7m telescope. We are grateful to all the members of the MWISP working group, particularly the staff members at PMO-13.7m telescope, for their long-term support. MWISP was sponsored by National Key R\&D Program of China with grant 2017YFA0402701 and CAS Key Research Program of Frontier Sciences with grant QYZDJ-SSW-SLH047.\\
      We thank the anonymous referee for the detailed comments and appreciate the
effort that has clearly gone into reviewing our work. We also thank M.B.~Scheuck for his valuable contributions to the code development.
\end{acknowledgements}

%
%

\bibliographystyle{aa_url} 
\bibliography{references} 

\begin{appendix}
\section{\saber{} optimization and parameters}
\subsection{Test data and smoothing optimization}\label{sec:optimization}
We implemented a gradient descent method \citep[see e.g.,][]{2016arXiv160904747R} that uses generated mock data to find the optimal smoothing parameters $\lambda_1$ and $\lambda_2$. The $\vec\lambda=(\lambda_1,\lambda_2)$ parameter generally depends on the spectral resolution, noise level, line width of the absorption features, and on variations in the emission background. In a preparatory step \saber{} generates ``pure emission'' and ``observed'' mock data that are based on the actual observational data. The pure emission data represent emission line spectra that do not contain any absorption features and those are used as the test data that are to be recovered by the \saber{} algorithm. The observed data contain spectra where randomly generated (but known) absorption features were added to the pure emission data, and those are used as the input data for \saber{}. The mock data are generated by randomly selecting a defined number of spectra $N_{\mathrm{train}}$ taken from the real observational data on which the baseline extraction is to be applied later. The algorithm then uses asymmetric least squares smoothing with predefined and fixed parameters $(\lambda_1,\lambda_2)=(2.0,2.0)$ (equivalent to one-phase smoothing with $\lambda_1=2.0$), without adding a residual, to smooth the spectra in the training set. The algorithm then adds a user-defined noise level to the spectra, thus creating spectra that will be used as pure emission data to be recovered. The reason for the smoothing in this preparatory step is to remove any dips that are present in the real data, such that the test data are free of absorption and that generated absorption features can be added anywhere in the spectrum. A moderate setting of the parameters $p$ and $\vec\lambda$ in the preparation step does not heavily affect the optimization as these parameters are only used to generate data that show similarity to the overall structure of the real spectra as we show in Appendix~\ref{sec:testdata_params_app}. We have taken samples of 200 spectra for each filament region to use as test data to find the optimal smoothing parameters.
    \begin{table}[!htbp]
        \caption{Optimal smoothing parameters.}
        \renewcommand*{\arraystretch}{1.3}
        \centering
        \begin{tabular}{l c c}
             \hline\hline
             Source & $\lambda_1$ & $\lambda_2$ \\\hline
             GMF20.0-17.9 & 3.10 & 0.56 \\
             GMF26.7-25.4 & 3.40 & 0.46 \\
             GMF38.1-32.4a & 2.76 & 0.50 \\
             GMF38.1-32.4b & 2.76\tablefootmark{(a)} & 0.50\tablefootmark{(a)} \\
             GMF41.0-41.3 & 3.45 & 0.43 \\
             GMF54.0-52.0 & 3.70 & 0.39 \\\hline
        \end{tabular}
        \tablefoot{
        The second column and third column give the best-fit $\lambda_1$ and $\lambda_2$ smoothing parameters obtained during the optimization step of \saber{}, respectively. Minor differences in these optimal parameters are expected due to different noise and fluctuations in the emission spectra.\\
        \tablefoottext{a}{Since the optimal smoothing parameters are obtained from the same data cube as GMF38.1-32.4a, the $\vec\lambda$ values are the same.}
        \label{tab:lam_opt}}
    \end{table}
The randomly generated absorption spectra are created using Gaussian functions whose 1) amplitude, 2) mean, and 3) standard deviation parameters are drawn from normal distributions with the following mean and standard deviation ($\mu$ and $\sigma$) by default (see Table~\ref{tab:all_parameters}): 1) the amplitude values follow a normal distribution with $\mu_{\mathrm{amp}}=7\sigma_{\mathrm{rms}}$, and $\sigma_{\mathrm{amp}}=1\sigma_{\mathrm{rms}}$, where $\sigma_{\mathrm{rms}}$ is the noise of the observational data, 2) the mean velocity values follow a normal distribution where the mean $\mu_{\mathrm{mean}}$ is set by the central velocity at which there is signal in each spectrum, its standard deviation is set accordingly, such that $3\sigma_{\mathrm{mean}}$ is at the edge of the signal range of the spectrum, 3) the magnitude of the $\vec\lambda$ parameters that is required for smoothing is crucially dependent on the width of the absorption features, so the standard deviation values of the absorption features drawn from a normal distribution have to be defined by the user. In the case of the THOR-\ion{H}{i} data, \citet{2020A&A...634A.139W} and \citet{2020A&A...642A..68S} report HISA FWHM values of $\sim$4\,$\rm km\,s^{-1}$. We have therefore set the mean and standard deviation of the line width distribution to $\mu_{\mathrm{lw}}=4\rm\,km\,s^{-1}$ and $\sigma_{\mathrm{lw}}=1\rm\,km\,s^{-1}$, respectively. The number of self-absorption components that are added to each spectrum are drawn from a normal distribution with $\mu_{\mathrm{n}}=2.0$ and $\sigma_{\mathrm{n}}=0.5$ by default, where all samples below $0.5$ are set to $1.0$ to add at least one self-absorption feature to each spectrum. In Appendix~\ref{sec:testdata_params_app} we discuss our fiducial parameter set.

Once the mock spectra have been generated, a gradient descent algorithm is run to find the optimal smoothing parameters $\lambda_1$ and $\lambda_2$. The gradient descent is designed to minimize the residual between the actual test data (pure emission mock spectra) and the baselines obtained by the \saber{} smoothing routine. Since we do not expect large variations in the emission spectra and absorption baselines within single HISA regions, we aim to find single $\lambda_1$ and $\lambda_2$ values that we then apply to the whole filament region in each case. We make use of the statistics of the training data and select the median of the reduced chi square values as the cost function for the optimization. The median value is robust against individual outliers in the training data and represents on average the best solution for the entire training data set. The reduced chi square is only evaluated in channels where artificial absorption features have been added. More details about the gradient descent method applied in this paper are given in Appendix~\ref{sec:gradient_descent_app}.

With increasingly complex emission spectra containing multiple broad and narrow emission peaks as well as absorption features, adding a residual to a moderately smoothed spectrum has shown to give the best results for all the mock data that we have tested. In the subsequent analysis, we have therefore used the two-phase smoothing with two $\vec\lambda=(\lambda_1,\lambda_2)$ parameters and added a residual that is the difference between the very first major cycle iteration (using $\lambda_1$) and the last major cycle iteration (using $\lambda_2$). We note, however, that a simpler one-phase smoothing without adding a residual might give generally good results, depending on the signal-to-noise and complexity in the spectrum. The final smoothing parameters obtained in the optimization step of \saber{} that were used for the final baseline reconstruction are listed in Table~\ref{tab:lam_opt}. The inferred smoothing parameters are similar toward all filament regions and compare well to each other. We expect small differences between the samples because of the training data sets containing different emission spectra and noise. These minor differences in the smoothing parameters only have a limited impact on the extraction results (see also Fig.~\ref{fig:parameter_space}). In particular, the accuracy of the fit results does not heavily depend on $\lambda_1$. Figure~\ref{fig:parameter_space} shows similar accuracy in the fits for a range between 3--6.
\subsection{Mock data parameter testing}\label{sec:testdata_params_app}
We tested the final output of the optimization step using different parameter settings for generating the test and training data sets. As mentioned above, we apply a prior smoothing to test spectra in order to remove any pronounced absorption dips, such that absorption features can be added anywhere in the spectrum while generating the training data. The ideal test data should only contain pure emission features free of self-absorption. If an absorption feature were present in the test data, as might be the case if the real observations were used as test data, and an additional absorption feature were added at the same location, the optimization would falsely result in a too small smoothing parameter $\vec\lambda$ to recover the absorption. This might only affect a fraction of test data spectra as we randomly sample self-absorption features but these spectra would not yield reliable results during optimization.
In a first test, we have therefore applied varying asymmetric least squares smoothing weights $\vec\lambda$ to the observational spectra and added noise to generate different test data sets. We used a one-phase smoothing ($\vec\lambda=\lambda_1$), without adding a residual, and smoothing values $\lambda_1=1.0,\,2.0,\,4.0$. In addition, we used a sample of real observations as test data to compare the optimization with the smoothed test data sets. In each case, we selected 100 spectra (same spectra for all tests) for the optimization. We then added self-absorption features and ran the optimization for all training data sets and report the optimal smoothing parameters in Table~\ref{tab:lam_opt_prior_smoothing}. The optimal parameters of the various test and training data sets, that were generated using varying degrees of prior smoothing, show only marginal variations. Only the time for reaching a stable convergence might increase with less prior smoothing due to contamination of the pure emission test data, which can lead to fluctuations in the cost function (see Sect.~\ref{sec:gradient_descent_app}). The final baseline emission shows small differences, that are well within the noise of the observations. We conclude that a moderate prior smoothing of the data does not affect the final result of the optimization but can help achieve a stable convergence toward the optimal parameters more rapidly. We therefore chose to apply a prior smoothing weight of $(\lambda_1,\lambda_2)=(2.0,2.0)$ (equivalent to one-phase smoothing) to generate the test and training data. However, \saber{} users can adjust these predefined smoothing parameters or opt to use real observations as test data instead.
\begin{table}[!htbp]
        \caption{Optimal smoothing parameters for test data with varying prior smoothing.}
        \renewcommand*{\arraystretch}{1.3}
        \centering
        \begin{tabular}{l c c}
             \hline\hline
             Prior smoothing weight & $\lambda_1$ & $\lambda_2$ \\\hline
             None & 3.47 & 0.11 \\
             1.0 & 3.37 & 0.41 \\
             2.0 & 3.84 & 0.45 \\
             4.0 & 3.17 & 0.58 \\\hline
        \end{tabular}
        \tablefoot{The first column gives the prior smoothing weight that was applied to the observations to generate pure emission test data. In the first row, no smoothing has been applied and real observations have been used as test data. The second and third column give the optimal smoothing parameters $\lambda_1$ and $\lambda_2$ that best recover the test data. Minor differences are expected due to noise fluctuations.}
        \label{tab:lam_opt_prior_smoothing}
    \end{table}
    
We have also varied the mean values of the Gaussian distributions, from which the parameters of self-absorption features are sampled, to investigate how this affects the final output of the \saber{} optimization. In each test, we vary one Gaussian distribution from which self-absorption parameters are drawn (i.e. either amplitude, line width, or number of self-absorption features), while fixing the remaining distributions to our fiducial values ($\mu_{\mathrm{amp}}=7\sigma_{\mathrm{rms}}$, and $\sigma_{\mathrm{amp}}=1\sigma_{\mathrm{rms}}$; $\mu_{\mathrm{lw}}=4\rm\,km\,s^{-1}$ and $\sigma_{\mathrm{lw}}=1\rm\,km\,s^{-1}$; $\mu_{\mathrm{n}}=2.0$ and $\sigma_{\mathrm{n}}=0.5$; see Sect.~\ref{sec:optimization}). Table~\ref{tab:lam_opt_gauss_params} shows the final output of the optimization with varying self-absorption input parameters. The optimal smoothing weights $(\lambda_1, \lambda_2)$ slightly increase with increasing absorption depth. When using a finite and fixed asymmetry weight $p$, larger smoothing weights are required to effectively smooth out strong absorption features and recover their baselines. We chose to set a mean amplitude of self-absorption features to $7\sigma_{\mathrm{rms}}$ in order to ensure a good balance between recovering significant (i.e. $\gtrsim 5\sigma_{\mathrm{rms}}$) self-absorption and retaining real emission signals. Stronger self-absorption features will then still be identified and extracted with \saber{}, with their amplitudes being slightly underestimated.

As expected, the optimal smoothing weights increase with increasing line widths of self-absorption features. The output of \saber{} is most sensitive to the input line width of the absorption components, and therefore has to be provided by the user. As described in Sect.~\ref{sec:optimization}, we set our fiducial value to $4\rm\,km\,s^{-1}$ that is in agreement with the reported line widths in \citet{2020A&A...634A.139W} and \citet{2020A&A...642A..68S}.

The output of \saber{} does not significantly change with the number of self-absorption components that are added to each test spectrum. The number of components effectively changes the number of samples that are tested and optimized in the training data. A larger number of samples allows a statistically more robust conclusion of optimal parameters, but comes at the cost of increased time to reach a stable convergence. We have therefore chosen to add an average of two components to each of the 200 test spectra, such that approximately 400 self-absorption features per training data set are evaluated for the optimization. The self-absorption parameter distributions can also be modified by the user (see \textit{optimization parameters} listed in Table~\ref{tab:all_parameters}).
\begin{table*}[!htbp]
     \caption{Optimal smoothing parameters for test data with varying self-absorption parameters.}
     \setlength{\tabcolsep}{10pt}
     \renewcommand*{\arraystretch}{1.4}
     \centering
     \begin{tabular}{l c c c c c c c c c}
     & \multicolumn{3}{c}{(1)} & \multicolumn{3}{c}{(2)} & \multicolumn{3}{c}{(3)} \\
     \hline\hline
      & \multicolumn{3}{|c}{$\mu_{\mathrm{amp}}$ [$\sigma_{\mathrm{rms}}$]} & \multicolumn{3}{|c}{$\mu_{\mathrm{lw}}$ [$\rm km\,s^{-1}$]} & \multicolumn{3}{|c|}{$\mu_{\mathrm{n}}$} \\\hline
     & 5.0 & $7.0^{(*)}$ & 9.0 & 2.0 & $4.0^{(*)}$ & 6.0 & $2.0^{(*)}$ & 3.0 & 4.0\\\hline
     $\lambda_1$ & 3.08 & 3.84 & 4.09 & 2.96 & 3.84 & 4.35 & 3.84 & 3.49 & 3.47 \\
     $\lambda_2$ & 0.29 & 0.45 & 0.60 & 0.20 & 0.45 & 1.26 & 0.45 & 0.80 & 1.15 \\\hline
     \end{tabular}
     \tablefoot{The first test (1) describes training data sets with varying mean amplitude $\mu_{\mathrm{amp}}$ of self-absorption features. The second test (2) shows the optimal smoothing parameters for test data with varying mean line width $\mu_{\mathrm{lw}}$. The third test (3) gives the optimal smoothing with varying numbers of self-absorption components $\mu_{\mathrm{n}}$ that are added to each test spectrum. The remaining parameter distributions of each test are always set to the fiducial values $\mu_{\mathrm{amp}}=7.0\,\sigma_{\mathrm{rms}}$, $\mu_{\mathrm{lw}}=4.0\rm\,km\,s^{-1}$, and $\mu_{\mathrm{n}}=2.0$.\par
     $^{(*)}$ Fiducial values.}
     \label{tab:lam_opt_gauss_params}
    \end{table*}

\subsection{Momentum-driven gradient descent}\label{sec:gradient_descent_app}
The smoothing parameter $\vec\lambda$ (which is in our case a two-component vector by default) is tuned to maximize the fitness of the self-absorption baselines using a batch gradient descent with momentum \citep{2016arXiv160904747R}. We define the median reduced chi square $\langle\chi_{\mathrm{red}}^2\rangle$ as the cost function $\mathcal{C}$ that we wish to minimize in order to achieve the highest goodness of fit result:
    \begin{equation}
       \mathcal{C}(\vec\lambda) =  \langle\chi_{\mathrm{red}}^2\rangle = \left\langle \frac{\sum_{i=1}^N\frac{(y_i-z_i(\vec\lambda))^2}{\sigma_{\mathrm{rms}}^2}}{N-k} \right\rangle \: ,
    \end{equation}
\noindent with $y_i$ and $z_i(\vec\lambda)$ denoting the data and baseline value at channel position $i$,
respectively, $N$ is the sample size (in this case the number of spectral channels containing self-absorption features), $k$ denotes the degrees of freedom, which is in our case $k=1$ for one-phase smoothing or $k=2$ for two-phase smoothing, and $\sigma_{\mathrm{rms}}$ is the rms noise of the data.

\begin{figure}
      \centering
        \resizebox{\hsize}{!}{\includegraphics{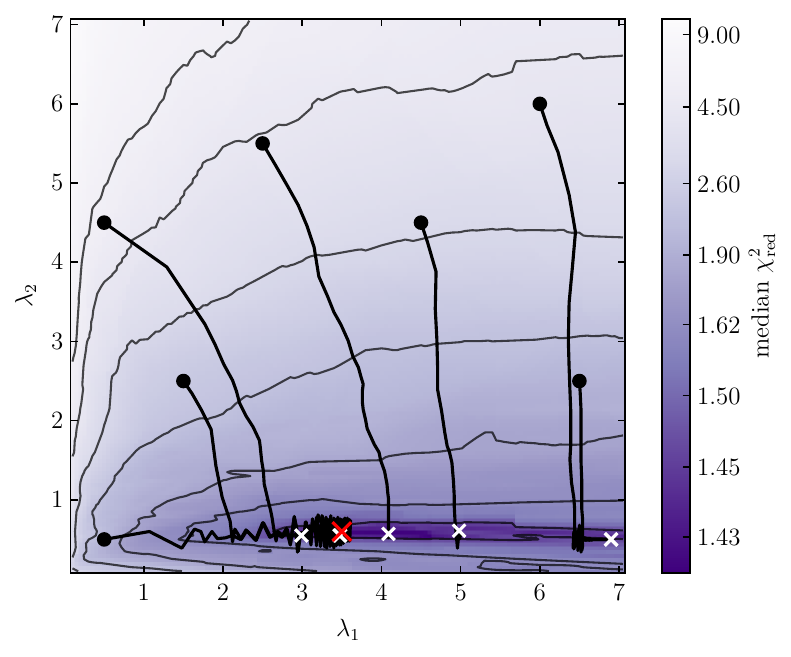}}
      \caption[]{Smoothing parameter optimization using gradient descent. The map shows a sampled representation of the underlying $\vec\lambda$ parameter space in terms of the median value of the reduced chi square results. Initial values, tracks, and convergence locations of the $(\lambda_1,\lambda_2)$ parameters during the optimization are represented by black circles, black lines, and white crosses, respectively. The red cross marks the global minimum in the sampled parameter space. Initial locations that start off too far from the global best solution $(\lambda_1=3.5,\lambda_2=0.6)$ might converge to local minima with less accurate fit results.}
      \label{fig:parameter_space}
   \end{figure}

In a classical gradient descent, updates to the smoothing weight $\vec\lambda$ are made by moving in the direction of greatest decrease in the cost function, that is $\Delta\vec\lambda=-\ell\,\nabla\mathcal{C}(\vec\lambda)$, where the learning rate $\ell$ controls the step size. Since the cost function is usually highly non-convex, we implemented a gradient descent with added momentum to overcome local minima that might be due to noise or fluctuations in the spectra. Therefore, at the $n$-th iteration, the change in $\vec\lambda$ is given by
    \begin{equation}
       \Delta\vec\lambda^{(n)} = -\ell\,\nabla\mathcal{C}(\vec\lambda) + \phi\,\Delta\vec\lambda^{(n-1)}  \: ,
       \label{equ:lambda_step}
    \end{equation}
\noindent where the momentum $\phi$ controls the degree to which the previous step influences the current one. The gradient $\nabla\mathcal{C}(\vec\lambda)$ in Eq.~\eqref{equ:lambda_step} is defined as
    \begin{align}
        \nabla\mathcal{C}(\vec\lambda) &= \begin{pmatrix}\frac{\mathcal{C}(\lambda_1+\epsilon,\lambda_2) - \mathcal{C}(\lambda_1-\epsilon,\lambda_2)}{2\epsilon}\\[\jot]
        \frac{\mathcal{C}(\lambda_1,\lambda_2+\epsilon) - \mathcal{C}(\lambda_1,\lambda_2-\epsilon)}{2\epsilon}
        \end{pmatrix} \: ,
    \end{align}
\noindent where we set the finite-difference step $\epsilon=0.1$. Figure~\ref{fig:parameter_space} shows example tracks of $\vec\lambda=(\lambda_1,\lambda_2)$ when using the gradient descent with different initial values for $\lambda_1$ and $\lambda_2$ during the two-phase optimization on THOR-\ion{H}{i} data. We find that small-scale local optima are ignored effectively during the search for large-scale optima.

\subsection{Noise and signal range estimation}\label{sec:signalrange}
The signal ranges of the spectra are determined by borrowing parts of the noise estimation routine included in the \textsc{GaussPy+} tool described in Sect.~\ref{sec:gausspyplus}. For a detailed description, we refer the reader to Sect.~3.1.1 in \citet{2019A&A...628A..78R}. The underlying assumptions to determine signal or noise ranges in the spectra are as follows: 1) the noise distribution is Gaussian, 2) the spectral channels are uncorrelated, 3) the noise has a mean around zero. The assumption is that a spectrum containing white noise has on average an equal number of negative and positive channels. We can then estimate the probability of a given positive or negative feature observed in consecutive spectral channels to be caused by white noise (as opposed to be due to real signal) using a Markov chain. The routine to determine signal ranges then selects all features in spectral channels that have a probability to be caused by white noise of less than a user-defined threshold. We set this probability limit value to $P_{\mathrm{limit}}=1\%$. In the case of the THOR-\ion{H}{i} data with 185 channels per spectrum, we find that all features with more than 15 consecutive positive channels have a probability to be caused by noise of less than $P_{\mathrm{limit}}=1\%$. To set the mean velocity in the velocity distribution of the absorption features, we additionally clipped the determined signal ranges by five channels on either side to ensure that the signal has sufficient intensity from which absorption features can be subtracted.

\subsection{Symbols, \saber{} keywords, and default values}\label{sec:keywords_app}
Depending on the scientific application and data set, different parameters might be necessary to achieve satisfactory results from the \saber{} extraction. We have designed \saber{} such that most parameters can be easily modified by the user in order to allow a broad applicability of the algorithm. Table~\ref{tab:all_parameters} gives an overview of the parameter settings of \saber{}, listing their corresponding default values and symbols used throughout the text. In order to get first extraction results, only a small number of parameters (listed as \textit{essential parameters}) need to be provided by the user. If the extraction results are deemed not satisfactory after adjusting these parameters, more \textit{advanced settings} might be modified. While the optimization step should yield good results in most cases, the \textit{optimization parameters} listed in Table~\ref{tab:all_parameters} also allow to customize the parameters used to generate the training data.

\begin{table*}[!htbp]
     \caption{\saber{} keywords mentioned throughout the text.}
     \renewcommand*{\arraystretch}{1.3}
     \centering
     \begin{tabular}{ l l c c}\\
     \hline\hline
     Symbol & Description & \saber{} keyword & Default\\\hline
     \multicolumn{4}{c}{\textit{Essential parameters}}\\\hline
     $\Phi$ & Smoothing mode of the extraction (Sect.~\ref{sec:asymmetric_least_squares}) & \code{phase} & \code{`two'}\\
     $\lambda_1$ & \makecell[tl]{If \code{phase=`two'} (\code{=`one'}), smoothing parameter of the\\first major cycle iteration (all major cycle iterations)\\(Sect.~\ref{sec:asymmetric_least_squares})} & \code{lam1} & \code{None}\\
     $p_1$ & \makecell[tl]{If \code{phase=`two'} (\code{=`one'}), asymmetry weight of the\\first major cycle iteration (all major cycle iterations)\\(Sect.~\ref{sec:asymmetric_least_squares})} & \code{p1} & 0.9\\
     $\lambda_2$ & \makecell[tl]{Smoothing parameter of the remaining major cycle iterations\\(Sect.~\ref{sec:asymmetric_least_squares})} & \code{lam2} & \code{None}\\
     $p_2$ & \makecell[tl]{Asymmetry weight of the remaining major cycle iterations\\(Sect.~\ref{sec:asymmetric_least_squares})} & \code{p2} & 0.9\\
     $\sigma_{\mathrm{rms}}$ & Observational noise of the data (Sect.~\ref{sec:asymmetric_least_squares}) & \code{noise} & \code{None}\\\hline
     \multicolumn{4}{c}{\textit{Advanced settings}}\\\hline
     $\mathcal{R_+}$ & \makecell[tl]{Option to add residual\\(= absolute difference between first and last\\major cycle iteration; Sect.~\ref{sec:asymmetric_least_squares})} & \code{add\_residual} & \code{True}\\
     $n_{\mathrm{major}}$ & Maximum number of major cycle iterations (Sect.~\ref{sec:asymmetric_least_squares}) & \code{niters} & 20\\
     $s_{\mathrm{thresh}}$ & \makecell[tl]{Multiplying factor of the noise setting the\\ convergence threshold (Sect.~\ref{sec:asymmetric_least_squares})} & \code{sig} & 1.0\\
     $n_{\mathrm{converge}}$ & \makecell[tl]{Number of iterations of the major cycle\\to determine convergence (Sect.~\ref{sec:asymmetric_least_squares})} & \code{iterations\_for\_convergence} & 3\\
     $s_{\mathrm{signal}}$ & \makecell[tl]{Significance of emission to be considered in the extraction\\(Sect.~\ref{sec:signalrange})} & \code{check\_signal\_sigma} & 6.0\\
     $\Delta \varv_{\mathrm{LSR}}$ & \makecell[tl]{Velocity range of the spectrum containing significant\\emission (set by  $\sigma_{\mathrm{signal}}$) to be considered in the\\baseline extraction (in units of $\rm km\,s^{-1}$)\\(Sect.~\ref{sec:signalrange})} & \code{velo\_range} & 15.0\\
     $P_{\mathrm{limit}}$ & \makecell[tl]{The probability threshold of the Markov chain\\to estimate signal ranges in the spectra\\(Sect.~\ref{sec:signalrange})} & \code{p\_limit} & 0.01\\\hline
     \multicolumn{4}{c}{\textit{Optimization parameters}}\\\hline
     $\mathcal{S_{\mathrm{prior}}}$ & \makecell[tl]{Option to apply prior asymmetric least squares smoothing\\to observations to generate test data (Sect.~\ref{sec:testdata_params_app})} & \code{smooth\_testdata} & \code{True} \\
     $N_{\mathrm{train}}$ & \makecell[tl]{Number of training spectra used for the optimization\\(Sect.~\ref{sec:optimization})} & \code{training\_set\_size} & 100\\
     $\mu_{\mathrm{amp}}$ & \makecell[tl]{Mean of normal distribution to draw\\amplitude values from (in units of $\sigma_{\mathrm{rms}}$; Sect.~\ref{sec:optimization})} & \code{mean\_amp\_snr} & 7.0\\
     $\sigma_{\mathrm{amp}}$ & \makecell[tl]{Standard deviation of normal distribution to draw\\amplitude values from (in units of $\sigma_{\mathrm{rms}}$; Sect.~\ref{sec:optimization})} & \code{std\_amp\_snr} & 1.0\\
     $\mu_{\mathrm{mean}}$ & \makecell[tl]{Mean of normal distribution to draw\\mean velocities from (Sect.~\ref{sec:optimization})} & \code{-} & \makecell[cl]{set by\\signal range}\\
     $\sigma_{\mathrm{mean}}$ & \makecell[tl]{Standard deviation of normal distribution to draw\\mean velocities from (Sect.~\ref{sec:optimization})} & \code{-} & \makecell[cl]{set by\\signal range}\\
     $\mu_{\mathrm{lw}}$ & \makecell[tl]{Mean of normal distribution to draw line width\\(FWHM) values from (in units of $\rm km\,s^{-1}$; Sect.~\ref{sec:optimization})} & \code{mean\_linewidth} & \code{None}\\
     $\sigma_{\mathrm{lw}}$ & \makecell[tl]{Standard deviation of normal distribution to draw line width\\(FWHM) values from (in units of $\rm km\,s^{-1}$; Sect.~\ref{sec:optimization})} & \code{std\_linewidth} & \code{None}\\
     $\mu_{\mathrm{n}}$ & \makecell[tl]{Mean of normal distribution to draw\\number of components from (Sect.~\ref{sec:optimization})} & \code{mean\_ncomponent} & 2.0\\
     $\sigma_{\mathrm{n}}$ & \makecell[tl]{Standard deviation of normal distribution to draw\\number of components from (Sect.~\ref{sec:optimization})} & \code{std\_ncomponent} & 0.5\\\hline
     \end{tabular}
     \tablefoot{A full documentation of all parameters is given at \url{https://github.com/astrojoni89/astrosaber}.}
     \label{tab:all_parameters}
    \end{table*}

\subsection{The \saber{} method and physical implications}
When dealing with finite spectral resolution, one of the shortcomings of classical approaches using finite-difference derivatives is the strong dependence on sensitivity and line width. Noise fluctuations are greatly amplified in second (or higher order) derivatives of a spectrum. Only HINSA with line widths $\lesssim$1$\rm\,km\,s^{-1}$ might be identified using this approach. It is then often assumed that there is a tight physical correlation in temperature between the cold \ion{H}{i} gas traced by self-absorption and the molecular gas within a cloud. This correlation is then used to constrain the baselines of the self-absorption features \citep[see][]{2008ApJ...689..276K}, which is a reasonable approximation given the projected spatial correlation and small line widths often observed toward the central regions of molecular clouds. However, the tight correlation observed through HINSA is likely to trace only the cold ($\sim$10\,K) \ion{H}{i} gas that is well mixed with the molecular gas in well-shielded regions \citep{2003ApJ...585..823L,2005ApJ...622..938G}, where the UV photo-dissociation rate of $\rm H_2$ might become comparable to the cosmic ray dissociation in the central region of a cloud. By construction of the detection method, the CNM traced by HINSA likely results in atomic gas not being detected far beyond the inner regions of a molecular cloud \citep{2007ApJ...654..273G}. However, once it is shown the HINSA-traced gas is coincident with \element[][13]{CO} emission, the uncertainty in kinetic temperature should be considerably less than with our method.

With the newly developed algorithm \saber{} we identify \ion{H}{i} self-absorption in an unbiased way, independent of the occurrence of molecular gas. The \saber{} algorithm can therefore complement the detection of CNM in the outer layers of molecular clouds or even the detection of CoAt gas that has no CO-bright molecular counterpart, which is likely to have larger line widths and would otherwise be missed by a second derivative approach, as we show in Appendix~\ref{sec:second_deriv_app}.

In the following, we discuss some of the limitations and ways that might boost the performance of the \saber{} routine. Since the observed spectra also contain noise where there is signal, the baseline smoothing slightly overestimates the baselines within signal ranges as it also weights the noise asymmetrically that is superposed with the emission. One way to take the noise within signal ranges into account is to adjust the weightings in Eq.~\eqref{equ:weighting} according to the mean and standard deviation of the positive and negative difference values between the spectrum and the baseline after each iteration \citep[see e.g.,][]{C4AN01061B,2022PASA...39...50L}. However, we are only interested in ranges where we expect self-absorption to be present. As we tune the smoothing parameter such that significant dips in the spectra will be smoothed out, any variation of the obtained baselines within emission ranges without absorption should be limited to the noise. Any features in those ranges will therefore not be identified as significant absorption anyway.

As we show in Appendix~\ref{sec:kinematics_app}, the centroid velocities recovered by \saber{} and Gaussian fitting show little deviation from the input velocities within our test environment. The distribution of centroid velocity differences has a mean and standard deviation of $-0.01\rm\,km\,s^{-1}$ and $0.35\rm\,km\,s^{-1}$, respectively. Based on the findings by \citet{2020A&A...634A.139W} and \citet{2020A&A...642A..68S}, the input line widths of $\sim$4$\rm\,km\,s^{-1}$ (FWHM) could be recovered with a standard deviation of $\sim$1$\rm\,km\,s^{-1}$. The larger scatter in line widths is likely due to employing a constant smoothing parameter for both narrow and broad absorption components. The difference in amplitude shows the largest scatter around the mean as a single smoothing parameter is used for the entire region.

Since we set a constant $\vec\lambda$ value for all spectra in each field, we account for significant broad absorption features by performing multiple iterations to obtain their baselines. However, depending on the number of iterations, the final baseline might not reflect the original spectrum within emission ranges accurately as in each iteration an updated baseline is used as an input for the next major cycle iteration. One way to address this is to not have multiple major cycle iterations but instead adjust the smoothing parameter channel by channel as broader absorption features would require more iterations than narrow ones at constant $\vec\lambda$.

With a single iteration, broader absorption features require a larger smoothing parameter $\vec\lambda$ if the asymmetry weighting for negative differences (i.e. absorption dips) is constant but nonzero. This baseline ``drag'' because of nonzero weighting could be corrected for if we introduced another coefficient vector $\vec\alpha$ that adjusts the smoothing parameter in Eq.~\eqref{equ:least_squ} for each channel in the spectrum, with its components being defined as
\begin{equation}
    \alpha_i = \frac{\mathrm{abs}(y_i-z_i)}{\mathrm{max}(\mathrm{abs}(\mathbf{y}-\mathbf{z}))} \: ,
\end{equation}
\noindent where the numerator is the absolute difference of the spectrum and baseline at channel $i$, and the denominator is the maximum of the absolute differences in the spectrum \citep[see e.g.,][]{doi:10.1080/00387010.2020.1730908}. Equation~\eqref{equ:least_squ} would then change to
\begin{equation}
     F(\mathbf{z}) = (\mathbf{y} - \mathbf{z})^\top\mathbf{W}(\mathbf{y} - \mathbf{z})+(\lambda\,\vec\alpha)\,\mathbf{z}^\top\mathbf{D}^\top\mathbf{D}\mathbf{z} \: .
     \label{equ:least_squ_mod}
\end{equation}
\noindent This could be a way to tune the smoothing parameter to an optimum without the need of having to perform multiple iterations. Since the weight curve and smoothing coefficients would also be fixed in that case, the smoothing parameter $\vec\lambda$ would still be the only parameter to be optimized.

In summary, the results could be improved by utilizing parameterized smoothing and asymmetry weights. Ultimately, these training and test data could then be used to feed a machine learning algorithm that sets an optimized smoothing parameter for each spectrum. However, as we have achieved good results with \saber{} in its current state, that already outperforms our traditional approach of using polynomial fits to specific ranges of the emission spectra, we leave the optimization of performance and efficiency to future investigations.

\section{Classical second derivative approach}\label{sec:second_deriv_app}
Another way to identify HISA features uses the second derivative of the observed \ion{H}{i} spectrum as described in \citet{2008ApJ...689..276K}. Pronounced self-absorption features would therefore become readily apparent as signatures in the second derivative representation of the spectrum. In the following, we discuss the limitations of this method and how \saber{} overcomes the issues imposed by finite spectral resolution and noise.

\begin{figure}
      \centering
        \resizebox{\hsize}{!}{\includegraphics{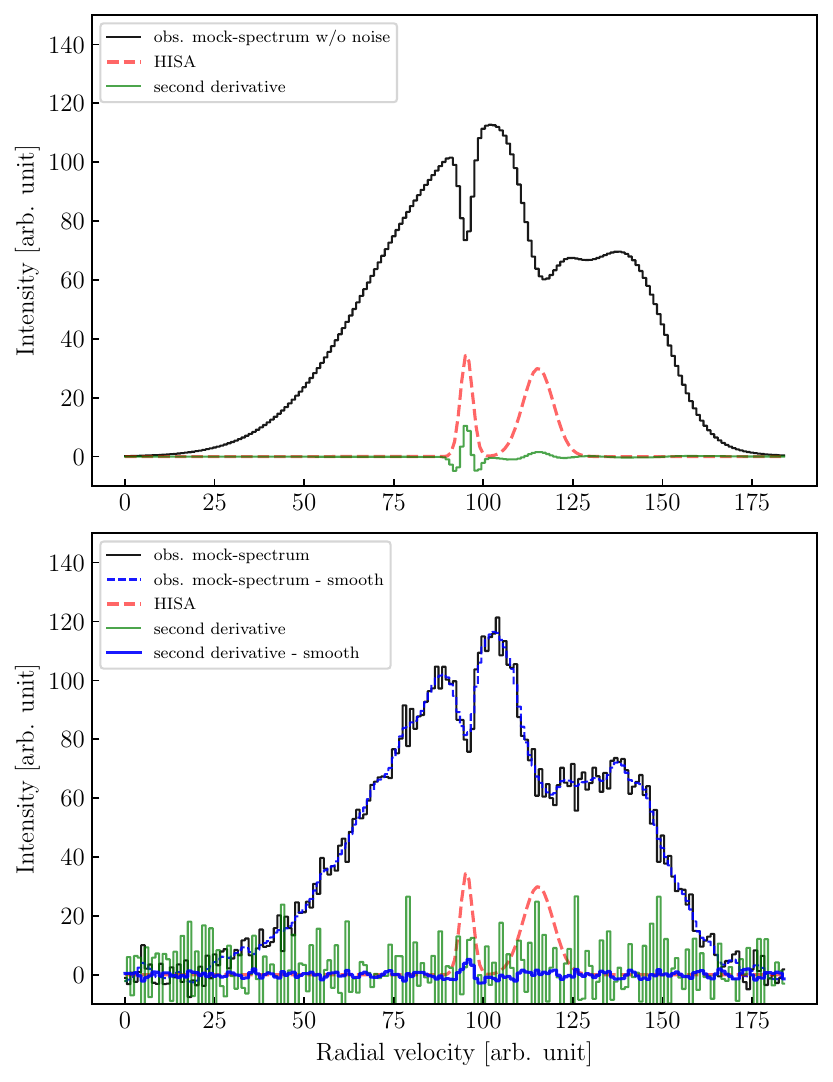}}
      \caption[]{Second derivative representation as a means to identify self-absorption. \textit{Top panel:} The black mock spectrum represents the \ion{H}{i} emission spectrum, with two self-absorption features superposed (red dashed components) and without any observational noise. The green spectrum shows the second derivative of the black mock spectrum, obtained from the finite differences between spectral channels. \textit{Bottom panel:} The black mock spectrum represents the \ion{H}{i} emission spectrum, with two self-absorption features superposed (red dashed components) and with added noise that is comparable to the noise of the THOR-\ion{H}{i} observations (same spectrum as in Fig.~\ref{fig:mock_spectrum}). The green spectrum shows the second derivative of the black mock spectrum, obtained from the finite differences between spectral channels. The dashed blue spectrum represents a regularized least squares solution to the \ion{H}{i} spectrum, which minimizes the second derivative. The corresponding second derivative is shown in blue, which is now less affected by noise fluctuations.}
      \label{fig:second_derivative}
   \end{figure}

Calculating the second (or higher) derivatives using finite-difference techniques might not always give reliable results as noise in the observational data will be greatly amplified. This is illustrated in Fig.~\ref{fig:second_derivative}. While the top panel shows a mock-\ion{H}{i} spectrum including two self-absorption components that does not contain noise, the bottom panel presents the same spectrum with added noise that is comparable to the THOR data (same spectrum as in Fig.~\ref{fig:mock_spectrum}). The green spectrum in each panel shows the finite-difference second derivative of the spectrum. For the noise-less data, the narrow HISA component emerges as a signature in the second derivative. Although less pronounced, even the broader absorption feature can be identified through an enhancement in its second derivative. On the other hand, given the observed spectrum that contains noise (lower panel in Fig.~\ref{fig:second_derivative}) the HISA components do not become visible as noise fluctuations dominate the second derivative of the spectrum. To overcome this, regularized differentiation can be used to mitigate the effect of noise fluctuations. It is a method of regularization of ill-posed problems that commonly occur in models with large numbers of parameters or inverse-solving during optimization. For example, this so-called Tikhonov regularization \citep{Tikhonov:1963} may be used to enforce smoothness of a given vector, giving preference to solutions that minimize the second derivative.

In a similar way, \saber{} uses this type of regularization when it introduces a penalty term to the (asymmetric) least squares function that minimizes the second derivative (see Eq.~\ref{equ:least_squ}). This is demonstrated in a simplified way in the lower panel of Fig.~\ref{fig:second_derivative}. The dashed blue spectrum shows a (in this case symmetric) least squares solution to the mock-\ion{H}{i} spectrum with a regularization term as in Eq.~\eqref{equ:least_squ}. The second derivative of the smooth representation of the spectrum now responds to the narrow absorption feature (blue spectrum) and shows a peak. However, the broader feature cannot be identified in the second derivative. 

For conceptual purposes, if we assume that a self-absorption feature is Gaussian $g(\varv_{\mathrm{LSR}})$, the second derivative of the feature will be
\begin{equation}
    \frac{\mathrm{d}^2g(\varv_{\mathrm{LSR}})}{\mathrm{d}\varv^2_{\mathrm{LSR}}} = \left(-\frac{1}{\sigma^2} + \frac{\varv^2_{\mathrm{LSR}}}{\sigma^4}\right)\,g(\varv_{\mathrm{LSR}})\:,
\end{equation}
\noindent where $\sigma$ is the standard deviation of the Gaussian, thus showing a strong dependence on line width. Narrow self-absorption features can then be identified through their second derivative more easily. In conclusion, the second derivative alone only works reliably well for high sensitivity, sufficient spectral resolution, and HISA line widths that are much smaller than the average emission component.
Furthermore, even if the spectral ranges of HINSA \citep{2003ApJ...585..823L,2005ApJ...622..938G,2007ApJ...654..273G} were determined with second derivatives, the baselines would still need to be inferred using, for example, polynomial fits or making physical assumptions of the HINSA properties \citep[see][]{2008ApJ...689..276K}.

By introducing an asymmetry weighting and an optimized regularization term, that simultaneously mitigates the undesirable effect of noise fluctuations, \saber{} is able to recover baselines while identifying absorption dips without the necessity of assuming a fitting function or a tight physical correlation between the cold \ion{H}{i} gas and the molecular gas. 

\section{Robustness of kinematics}\label{sec:kinematics_app}
To test how well the kinematics of the recovered absorption features match the input data, we ran \saber{} on an example data cube taken from a subsection of GMF20.0-17.9 \citep[see][]{2014A&A...568A..73R,2020A&A...642A..68S}. This example cube is also made available along with \saber{} source code. We created mock data as described in Sect.~\ref{sec:optimization} containing 100 test spectra where known self-absorption have been added. We then ran \saber{} to extract the self-absorption baselines and spectra after finding the optimal smoothing parameters. To obtain the kinematic properties of the extracted self-absorption features, we fit several Gaussian components to the self-absorption spectra, depending on the number of components that were added. In total, 207 self-absorption components have been added while generating the mock spectra.

\begin{figure*}
     \centering
     \includegraphics[width=1.0\textwidth]{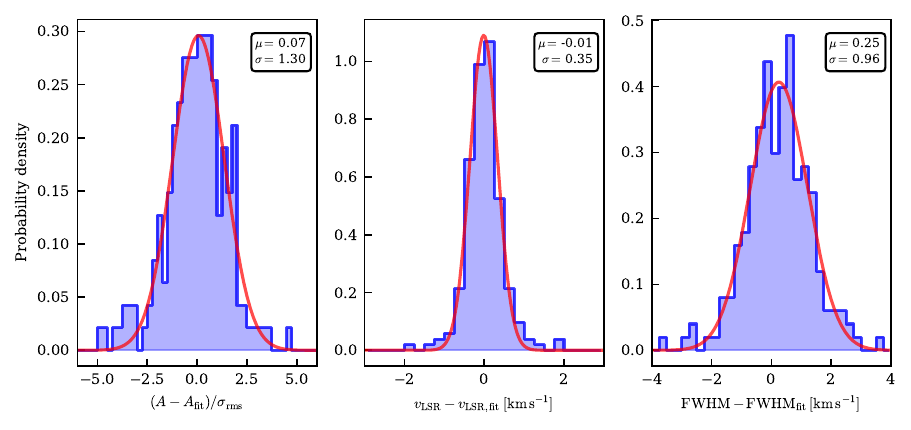}
        \caption[]{Histograms of residuals between input features and their respective fit results. \textit{Left panel:} The distribution shows the residuals between the amplitudes that were used to generate self-absorption features and the fit results (in units of the observational noise) after running \saber{}. \textit{Middle panel:} The distribution shows the residuals between the input velocities of self-absorption features and the recovered fit velocities. \textit{Right panel:} Similarly, the distribution in the right panel shows the residuals of the line widths. The red curve in each panel shows a Gaussian fit to the distribution.}
        \label{fig:robustness_kin}
    \end{figure*}

In Fig.~\ref{fig:robustness_kin} we present histograms showing the residuals between the true amplitudes, centroid velocities, line widths (FWHM) and their respective fit results. All distributions show a mean around zero. The line widths show a weak systematic trend as the mean of the residuals is $0.25\rm\,km\,s^{-1}$, implying that the line width fits slightly underestimate the true line width. As expected, the amplitude distribution shows the largest dispersion as we use only one set of smoothing parameters for the entire region. The recovered centroid velocities are very robust as the histogram shows a mean around zero and a standard deviation of $0.35\rm\,km\,s^{-1}$.

\section{$\rm H_2$ column density traced by \element[ ][13]{CO} emission}\label{sec:CO_column_dens}
Assuming that \element[][12]{CO} is becoming optically thick toward these GMFs, we estimated the column density of molecular hydrogen from \element[][13]{CO} emission. In the optically thin limit, the \element[ ][13]{CO} column density is computed by \citep{2013tra..book.....W} 
    \begin{equation}
     N(\element[ ][13]{CO}) = 3.0\times 10^{14}\,\frac{\int T_{\mathrm{B}}(\varv)\,\mathrm{d}\varv}{1-\mathrm{exp}(-5.3/T_{\mathrm{ex}})} \: ,
     \label{equ:N_CO}
    \end{equation} 
\noindent where $N(\element[ ][13]{CO})$ is the column density of \element[][13]{CO} molecules in $\rm cm^{-2}$, $\mathrm{d}\varv$ is in units of $\rm km\,s^{-1}$, $T_{\mathrm{B}}$ and $T_{\mathrm{ex}}$ are the brightness temperature and excitation temperature of the \element[ ][13]{CO} line in units of Kelvin, respectively. Under the assumption that the excitation temperature $T_{\mathrm{ex}}$ of \element[][12]{CO} and \element[][13]{CO} are the same in LTE, we computed the \element[][13]{CO} excitation temperature from \element[][12]{CO} line emission. Both the \element[][12]{CO} and \element[][13]{CO} data are taken from the high-resolution survey MWISP \citep{2019ApJS..240....9S} to compute the excitation temperatures and column densities, as described in Sect.~\ref{sec:methods_observation}. The excitation temperature is computed as \citep{2013tra..book.....W}
    \begin{equation}
    T_{\mathrm{ex}} = 5.5\cdot\left[\textrm{ln}\left(1+\frac{5.5}{T_{\mathrm{B}}^{12}+0.82}\right)\right]^{-1} \: ,
    \end{equation}  
\noindent where $T_{\mathrm{B}}^{12}$ is the brightness temperature of the \element[][12]{CO} line in units of Kelvin. To calculate the excitation temperature for each voxel, we reprojected the \element[][12]{CO} data cubes onto the same spectral grid as the \element[][13]{CO} data.
    
We set a lower limit to the excitation temperatures for regions where the \element[ ][12]{CO} brightness temperatures reach the $5\sigma$ noise level. We can then derive the optical depth of the \element[][13]{CO} line from the excitation and brightness temperature, using \citep[see e.g.,][]{2013tra..book.....W,2016A&A...587A..74S} 
    \begin{equation}
    \tau = -\mathrm{ln}\left[  1- \frac{T_{\mathrm{B}}}{5.3}\cdot\left(\left[\mathrm{exp}\left(\frac{5.3}{T_{\mathrm{ex}}}\right)-1\right]^{-1}-0.16\right)^{-1}\right] \: .
    \end{equation} 
\noindent We then estimated a lower limit of the optical depth for \element[ ][13]{CO} brightness temperatures at the $5\sigma$ noise level and the highest excitation temperatures we find toward each GMF region. The lower and upper limits to the excitation temperatures as well as the lower limits of the optical depth are listed in Table~\ref{tab:T_ex_limits} for each source. To account for high optical depth effects, we employ a correction factor by replacing the integral in Eq.~\eqref{equ:N_CO} with \citep{1982ApJ...262..590F,1999ApJ...517..209G}  
    \begin{equation}
    \int T_{\mathrm{B}}(\varv)\,\mathrm{d}\varv \rightarrow \frac{\tau}{1-e^{-\tau}}\,\int T_{\mathrm{B}}(\varv)\,\mathrm{d}\varv \: .
    \end{equation}   
\noindent This correction factor is accurate to 15\% for $\tau<2$. 

\section{Kinematics maps}\label{sec:kin_app}
The kinematic properties are presented in this section. The following maps show the fit peak velocities and line widths (FWHM) obtained with \textsc{GaussPy+} for both HISA and \element[][13]{CO} emission toward all remaining filament regions. If multiple components are identified within the velocity range of the filament, we only show the component with the lowest peak velocity.
   \begin{figure*}
     \centering
     \includegraphics[width=1.0\textwidth]{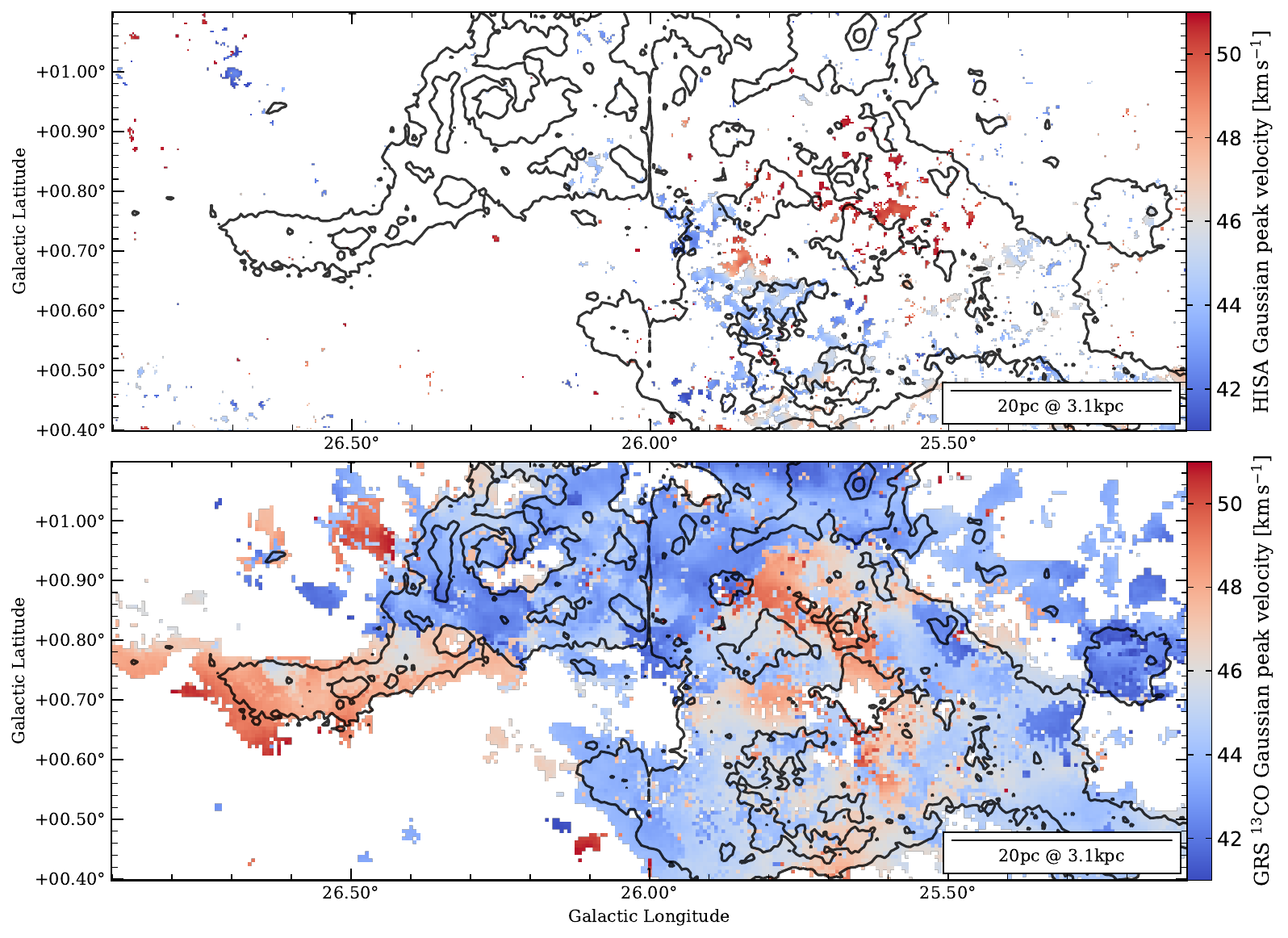}
        \caption[]{Fit peak velocity toward GMF26. These maps show the peak velocities of fit components derived from the \textsc{GaussPy+} decomposition of the spectra. If multiple components are present in a single pixel spectrum within the velocity range of the filament region, the component with the lowest peak velocity is shown. The black contours in both panels show the integrated GRS \element[][13]{CO} emission at the levels 6.0, 12.0, 24.0, and 34.0$\rm\,K\,km\,s^{-1}$. The contour feature at longitude $\ell=26\degr$ is an artifact in the observational data. \textit{Top panel:} Fit HISA peak velocity. \textit{Bottom panel:} Fit \element[][13]{CO} peak velocity.}
        \label{fig:kin2}
   \end{figure*}
   
   \begin{figure*}
     \centering
     \includegraphics[width=1.0\textwidth]{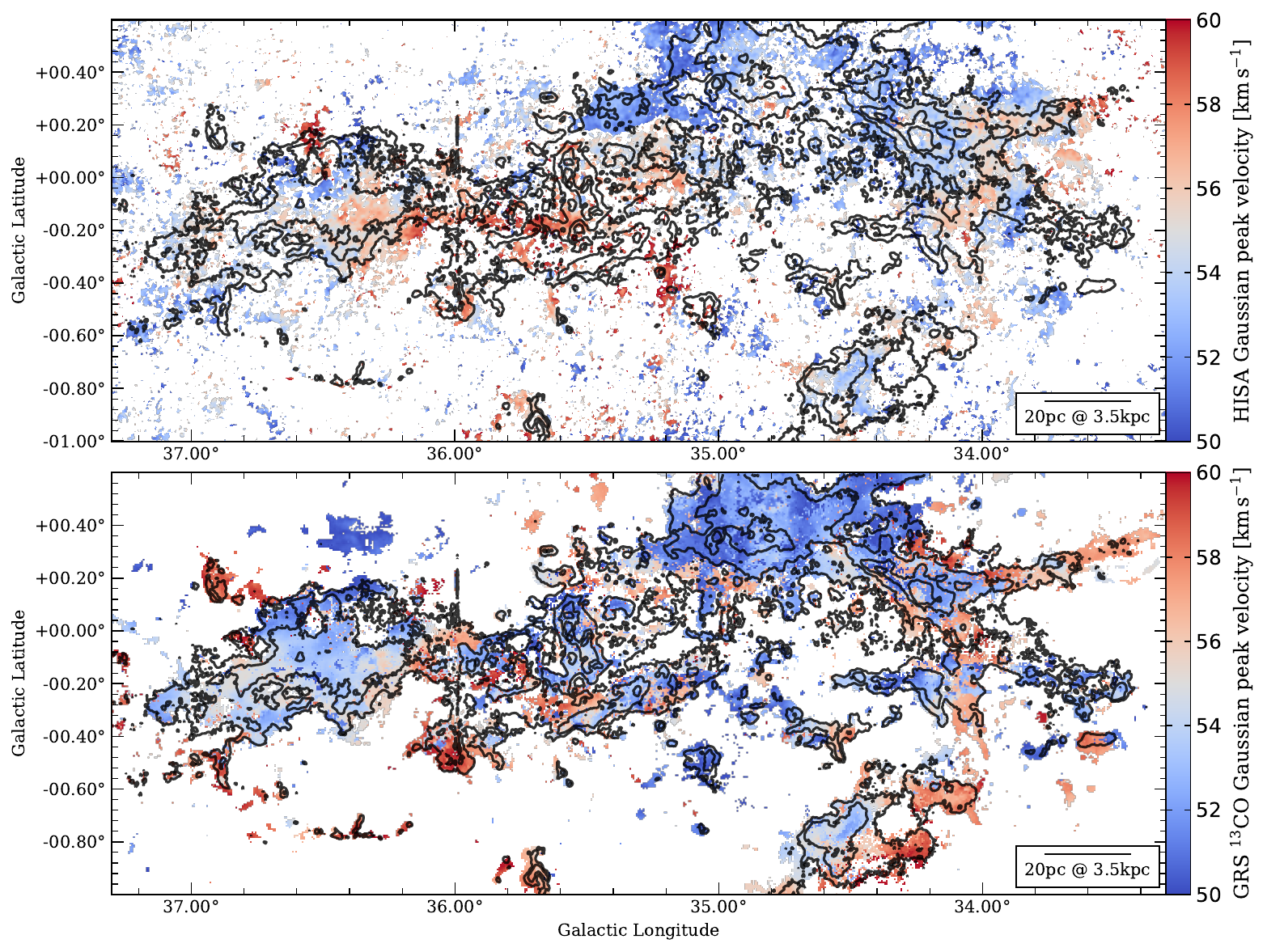}
        \caption[]{Fit peak velocity toward GMF38a. These maps show the peak velocities of fit components derived from the \textsc{GaussPy+} decomposition of the spectra. If multiple components are present in a single pixel spectrum within the velocity range of the filament region, the component with the lowest peak velocity is shown. The black contours in both panels show the integrated GRS \element[][13]{CO} emission at the levels 5.0, 10.0, 20.0, and 30.0$\rm\,K\,km\,s^{-1}$. The contour feature at longitude $\ell=36\degr$ is an artifact in the observational data. \textit{Top panel:} Fit HISA peak velocity. \textit{Bottom panel:} Fit \element[][13]{CO} peak velocity.}
        \label{fig:kin3}
   \end{figure*}
   
   \begin{figure*}
     \centering
     \includegraphics[width=1.0\textwidth]{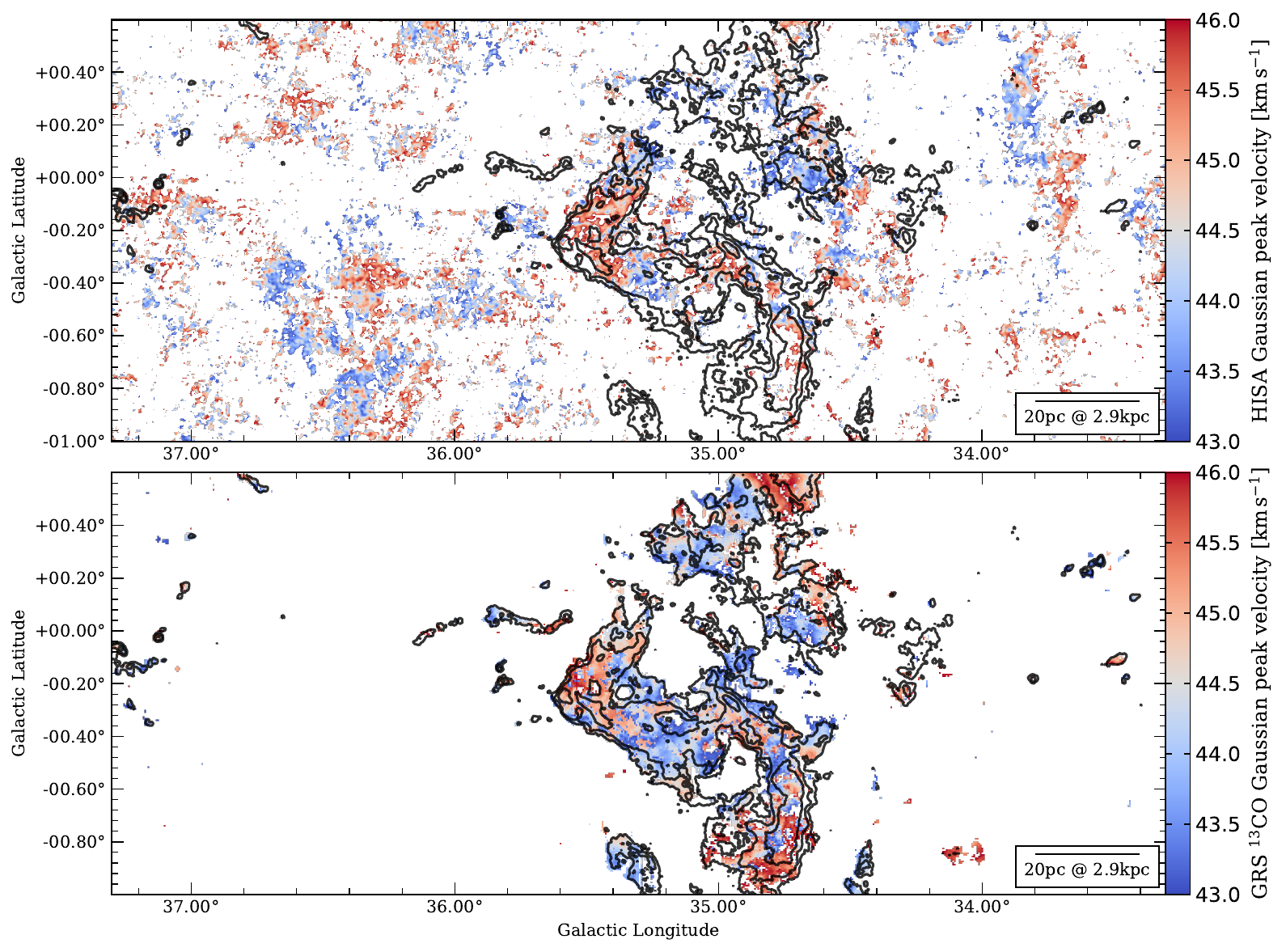}
        \caption[]{Fit peak velocity toward GMF38b. These maps show the peak velocities of fit components derived from the \textsc{GaussPy+} decomposition of the spectra. If multiple components are present in a single pixel spectrum within the velocity range of the filament region, the component with the lowest peak velocity is shown. The black contours in both panels show the integrated GRS \element[][13]{CO} emission at the levels 2.5, 5.0, 10.0, and 20.0$\rm\,K\,km\,s^{-1}$. \textit{Top panel:} Fit HISA peak velocity. \textit{Bottom panel:} Fit \element[][13]{CO} peak velocity.}
        \label{fig:kin4}
   \end{figure*}
   
   \begin{figure*}
     \centering
     \includegraphics[width=1.0\textwidth]{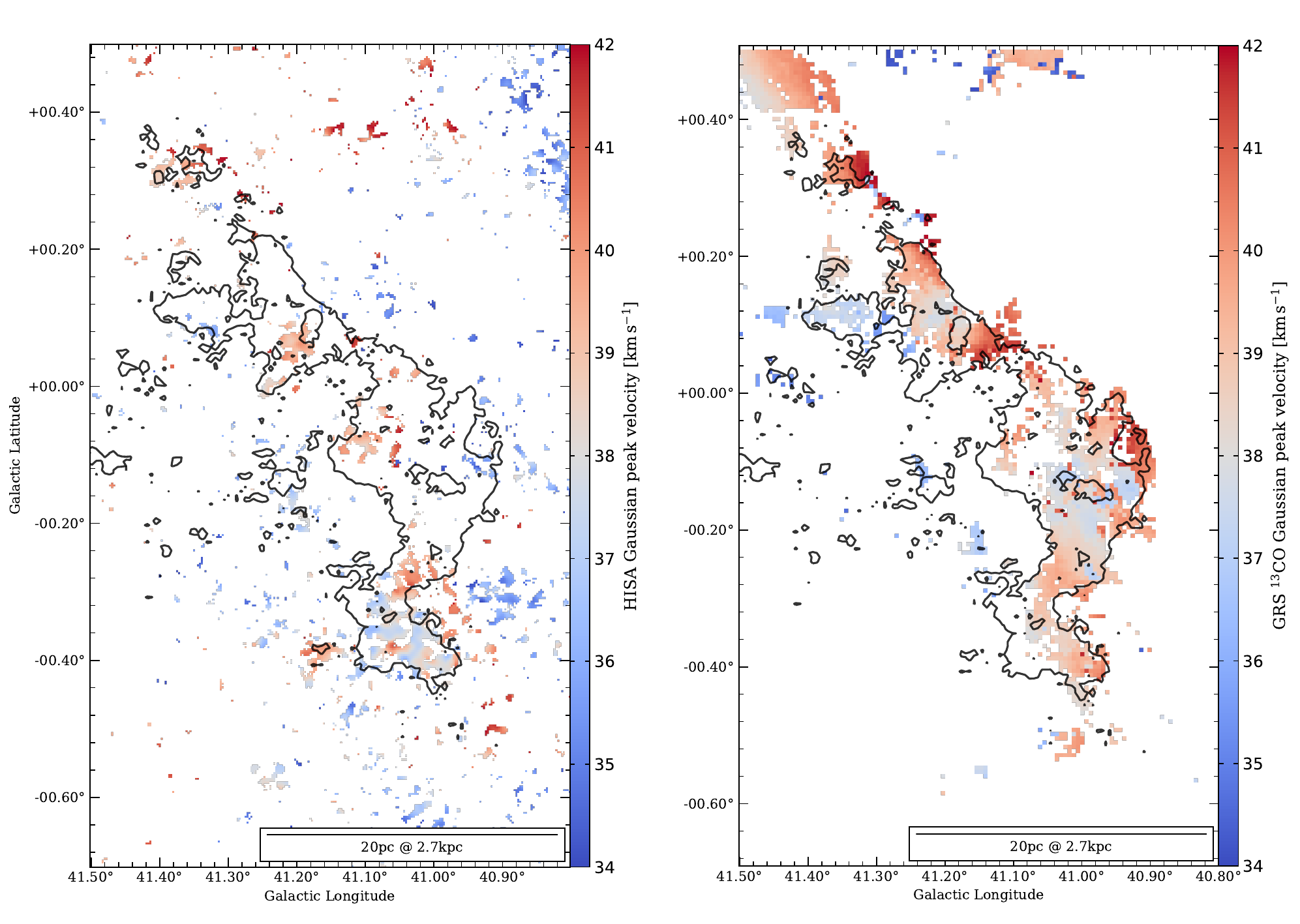}
        \caption[]{Fit peak velocity toward GMF41. These maps show the peak velocities of fit components derived from the \textsc{GaussPy+} decomposition of the spectra. If multiple components are present in a single pixel spectrum within the velocity range of the filament region, the component with the lowest peak velocity is shown. The black contours in both panels show the integrated GRS \element[][13]{CO} emission at the levels 6.0, 12.0, 24.0, and 34.0$\rm\,K\,km\,s^{-1}$. \textit{Top panel:} Fit HISA peak velocity. \textit{Bottom panel:} Fit \element[][13]{CO} peak velocity.}
        \label{fig:kin5}
   \end{figure*}
   
   \begin{figure*}
     \centering
     \includegraphics[width=1.0\textwidth]{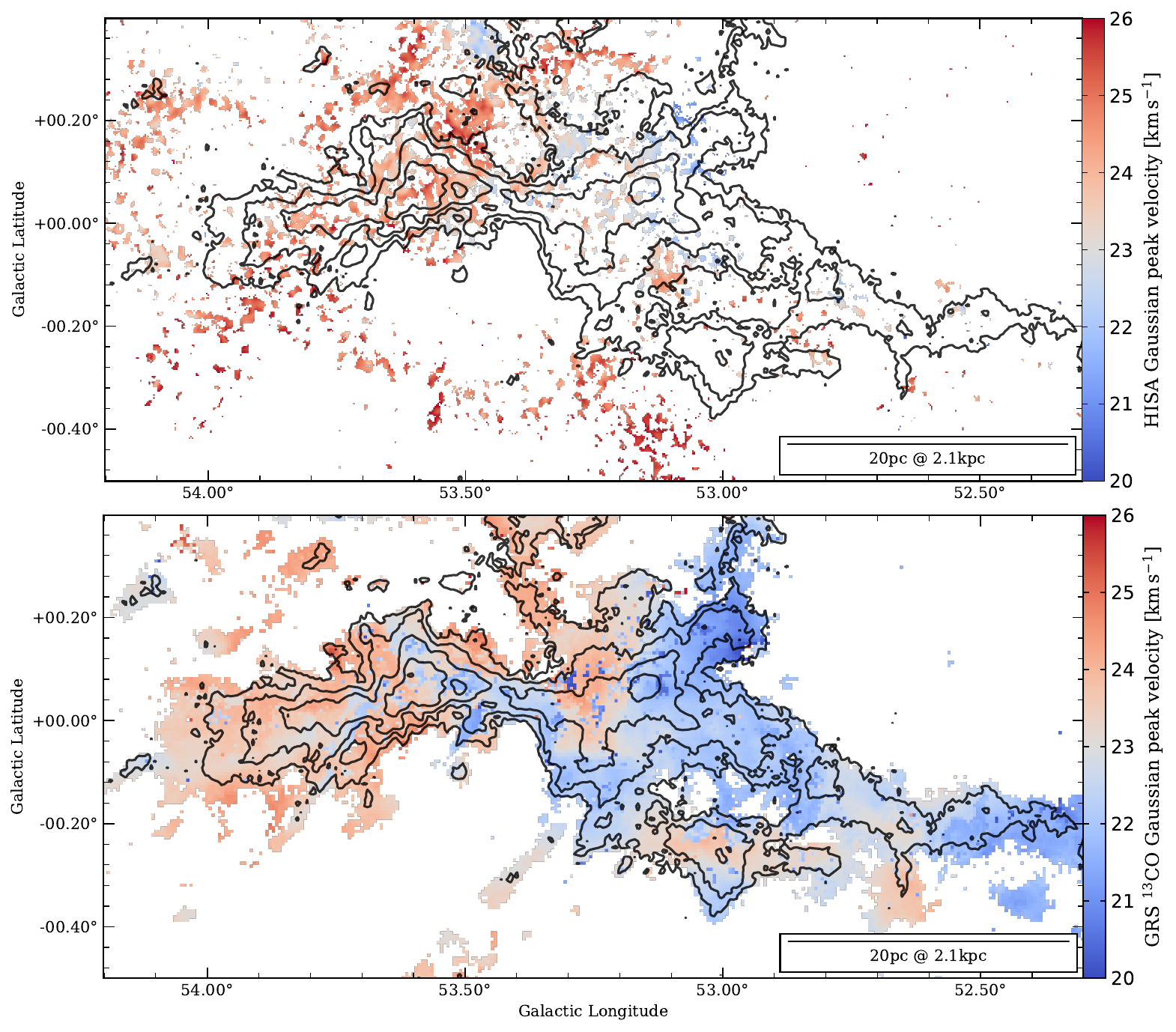}
        \caption[]{Fit peak velocity toward GMF54. These maps show the peak velocities of fit components derived from the \textsc{GaussPy+} decomposition of the spectra. If multiple components are present in a single pixel spectrum within the velocity range of the filament region, the component with the lowest peak velocity is shown. The black contours in both panels show the integrated GRS \element[][13]{CO} emission at the levels 2.5, 5.0, 10.0, and 20.0$\rm\,K\,km\,s^{-1}$. \textit{Top panel:} Fit HISA peak velocity. \textit{Bottom panel:} Fit \element[][13]{CO} peak velocity.}
        \label{fig:kin6}
   \end{figure*}


   \begin{figure*}
     \centering
     \includegraphics[width=1.0\textwidth]{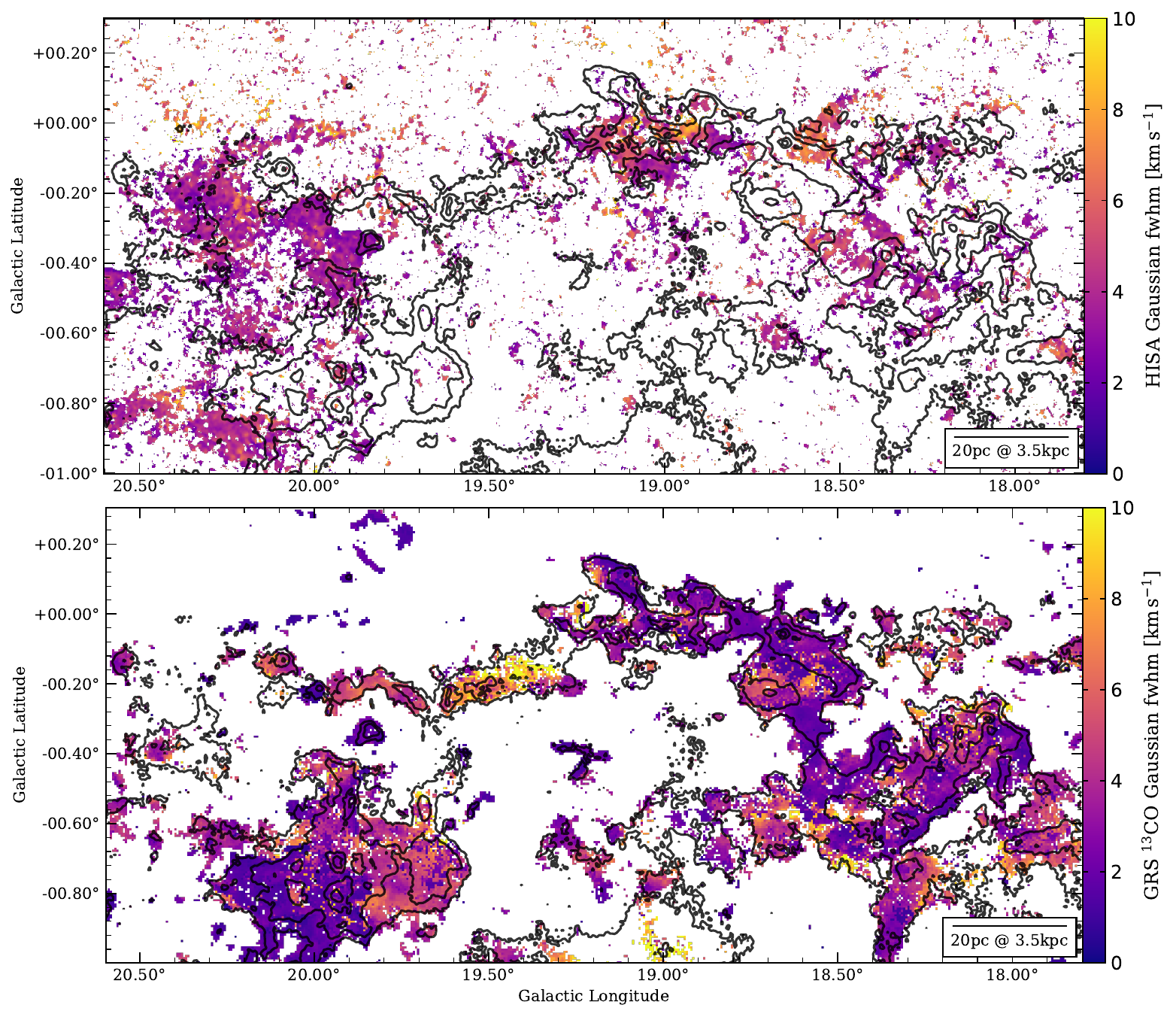}
        \caption[]{Fit line width (FWHM) toward GMF20. These maps show the line widths of fit components derived from the \textsc{GaussPy+} decomposition of the spectra. If multiple components are present in a single pixel spectrum within the velocity range of the filament region, the component with the lowest peak velocity is shown. The black contours in both panels show the integrated GRS \element[][13]{CO} emission at the levels 8.0, 16.0, 32.0, and 42.0$\rm\,K\,km\,s^{-1}$. \textit{Top panel:} Fit HISA line width. \textit{Bottom panel:} Fit \element[][13]{CO} line width.}
        \label{fig:kin7}
   \end{figure*}
   
   \begin{figure*}
     \centering
     \includegraphics[width=1.0\textwidth]{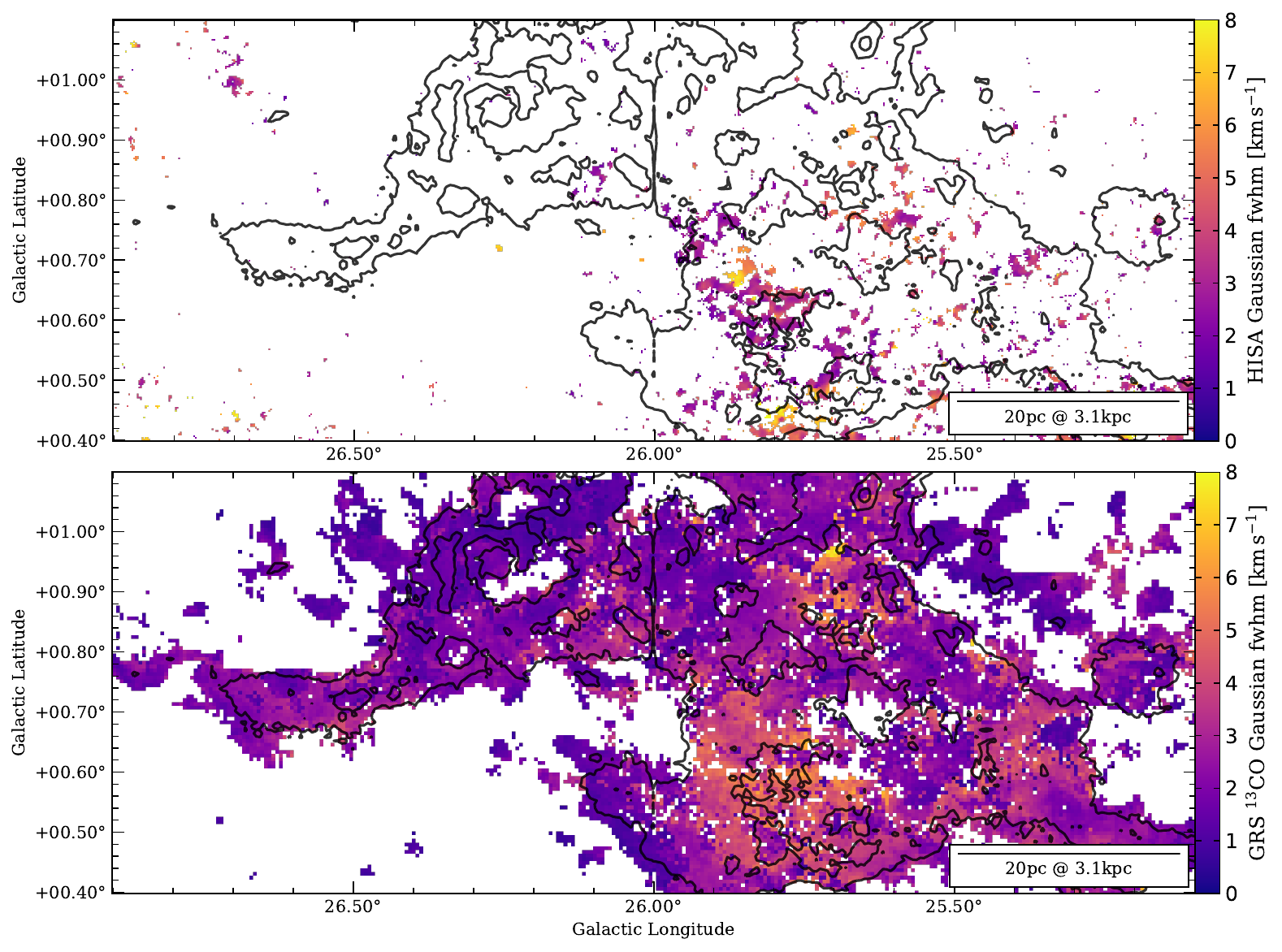}
        \caption[]{Fit line width (FWHM) toward GMF26. These maps show the line widths of fit components derived from the \textsc{GaussPy+} decomposition of the spectra. If multiple components are present in a single pixel spectrum within the velocity range of the filament region, the component with the lowest peak velocity is shown. The black contours in both panels show the integrated GRS \element[][13]{CO} emission at the levels 6.0, 12.0, 24.0, and 34.0$\rm\,K\,km\,s^{-1}$. The contour feature at longitude $\ell=26\degr$ is an artifact in the observational data. \textit{Top panel:} Fit HISA line width. \textit{Bottom panel:} Fit \element[][13]{CO} line width.}
        \label{fig:kin8}
   \end{figure*}
   
   \begin{figure*}
     \centering
     \includegraphics[width=1.0\textwidth]{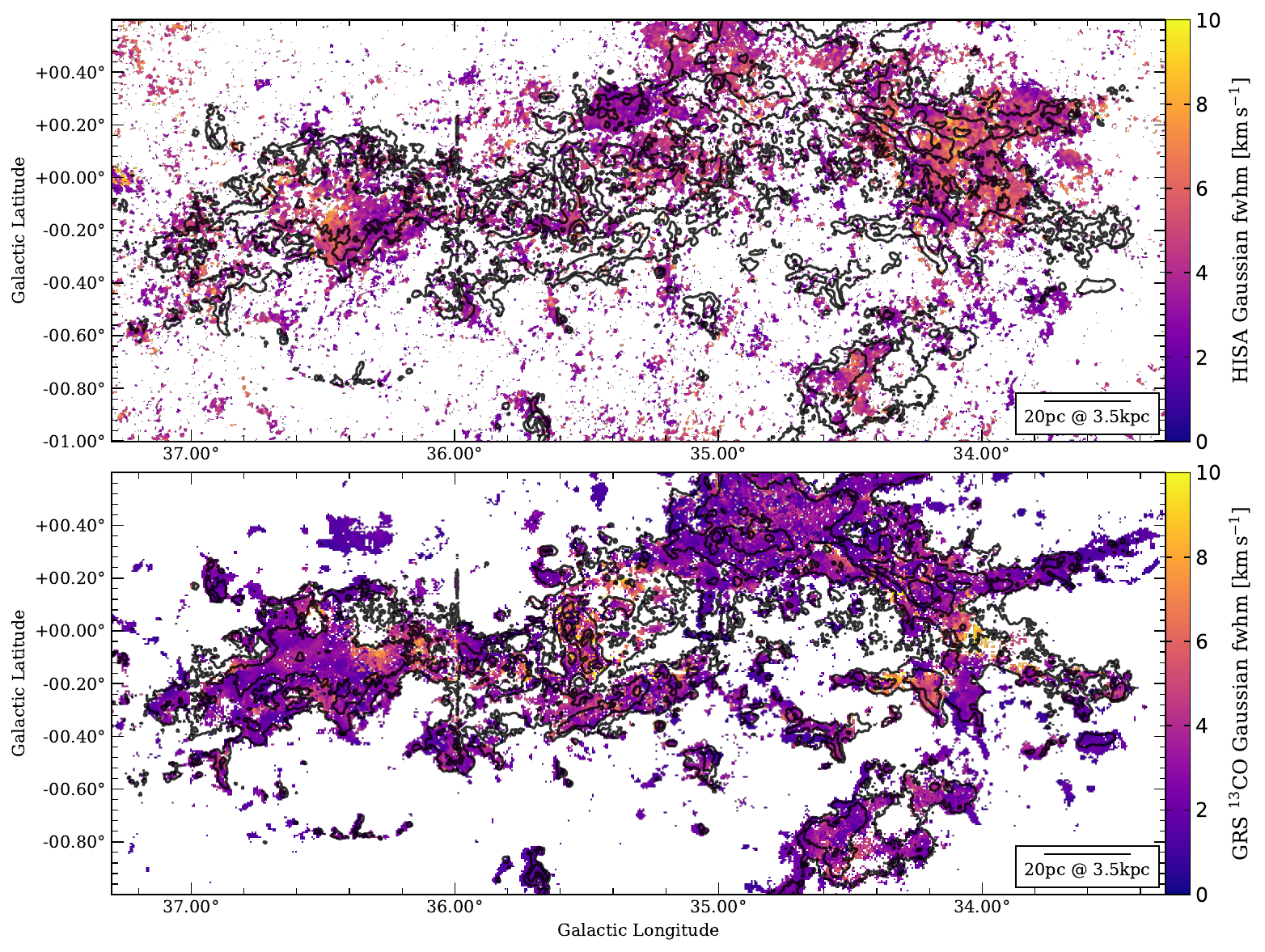}
        \caption[]{Fit line width (FWHM) toward GMF38a. These maps show the line widths of fit components derived from the \textsc{GaussPy+} decomposition of the spectra. If multiple components are present in a single pixel spectrum within the velocity range of the filament region, the component with the lowest peak velocity is shown. The black contours in both panels show the integrated GRS \element[][13]{CO} emission at the levels 5.0, 10.0, 20.0, and 30.0$\rm\,K\,km\,s^{-1}$. The contour feature at longitude $\ell=36\degr$ is an artifact in the observational data. \textit{Top panel:} Fit HISA line width. \textit{Bottom panel:} Fit \element[][13]{CO} line width.}
        \label{fig:kin9}
   \end{figure*}
   
   \begin{figure*}
     \centering
     \includegraphics[width=1.0\textwidth]{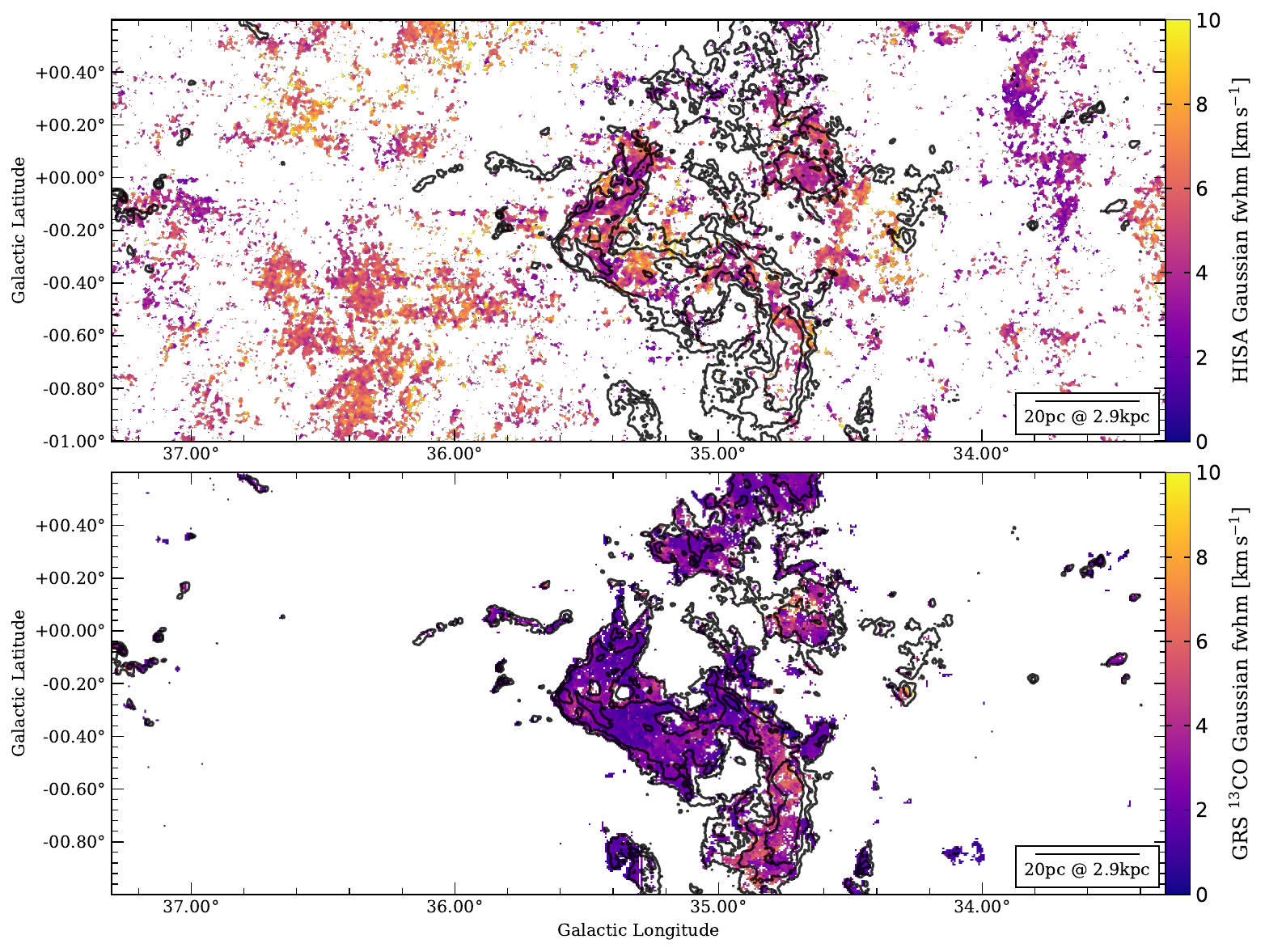}
        \caption[]{Fit line width (FWHM) toward GMF38b. These maps show the line widths of fit components derived from the \textsc{GaussPy+} decomposition of the spectra. If multiple components are present in a single pixel spectrum within the velocity range of the filament region, the component with the lowest peak velocity is shown. The black contours in both panels show the integrated GRS \element[][13]{CO} emission at the levels 2.5, 5.0, 10.0, and 20.0$\rm\,K\,km\,s^{-1}$. \textit{Top panel:} Fit HISA line width. \textit{Bottom panel:} Fit \element[][13]{CO} line width.}
        \label{fig:kin10}
   \end{figure*}
   
   \begin{figure*}
     \centering
     \includegraphics[width=1.0\textwidth]{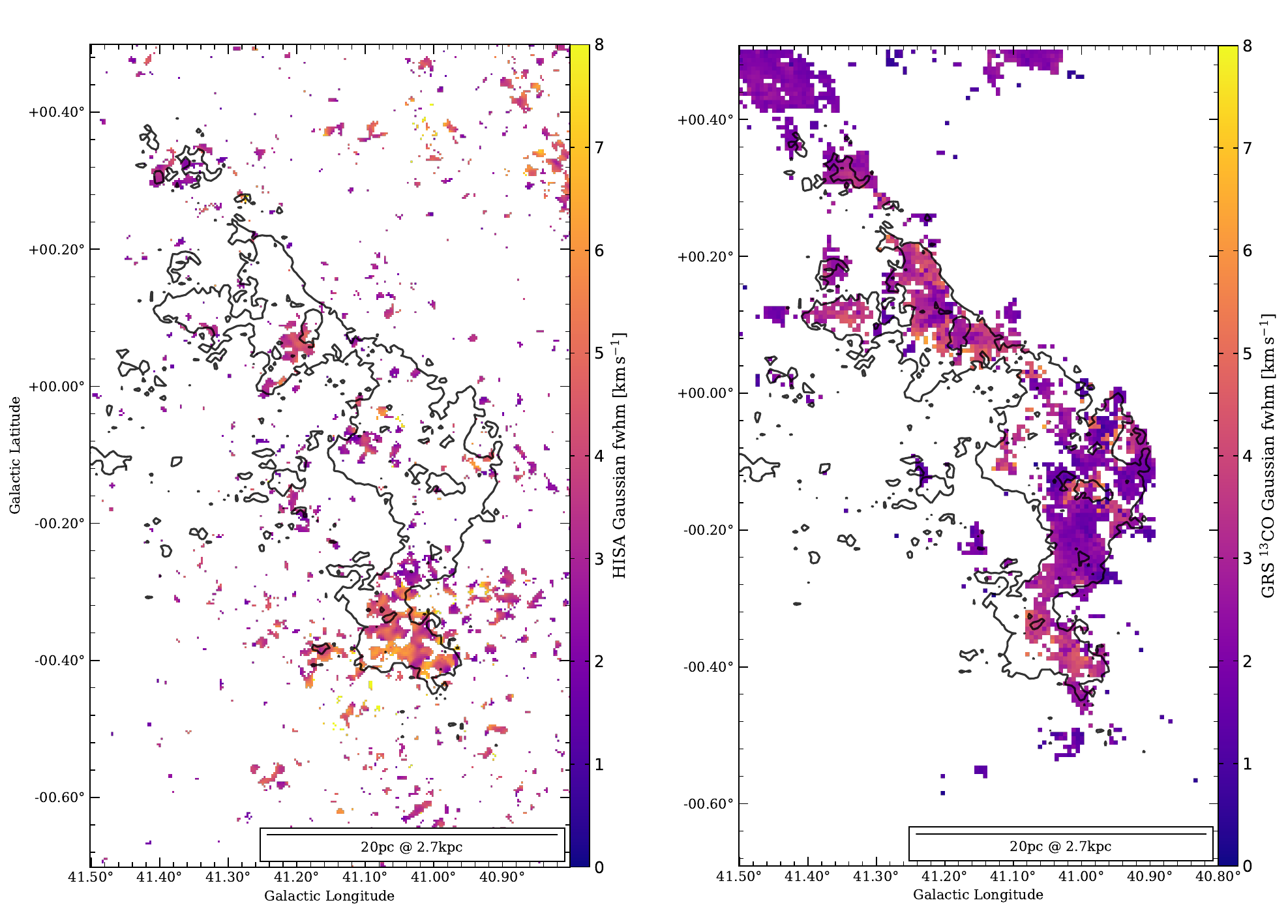}
        \caption[]{Fit line width (FWHM) toward GMF41. These maps show the line widths of fit components derived from the \textsc{GaussPy+} decomposition of the spectra. If multiple components are present in a single pixel spectrum within the velocity range of the filament region, the component with the lowest peak velocity is shown. The black contours in both panels show the integrated GRS \element[][13]{CO} emission at the levels 6.0, 12.0, 24.0, and 34.0$\rm\,K\,km\,s^{-1}$. \textit{Top panel:} Fit HISA line width. \textit{Bottom panel:} Fit \element[][13]{CO} line width.}
        \label{fig:kin11}
   \end{figure*}
   
   \begin{figure*}
     \centering
     \includegraphics[width=1.0\textwidth]{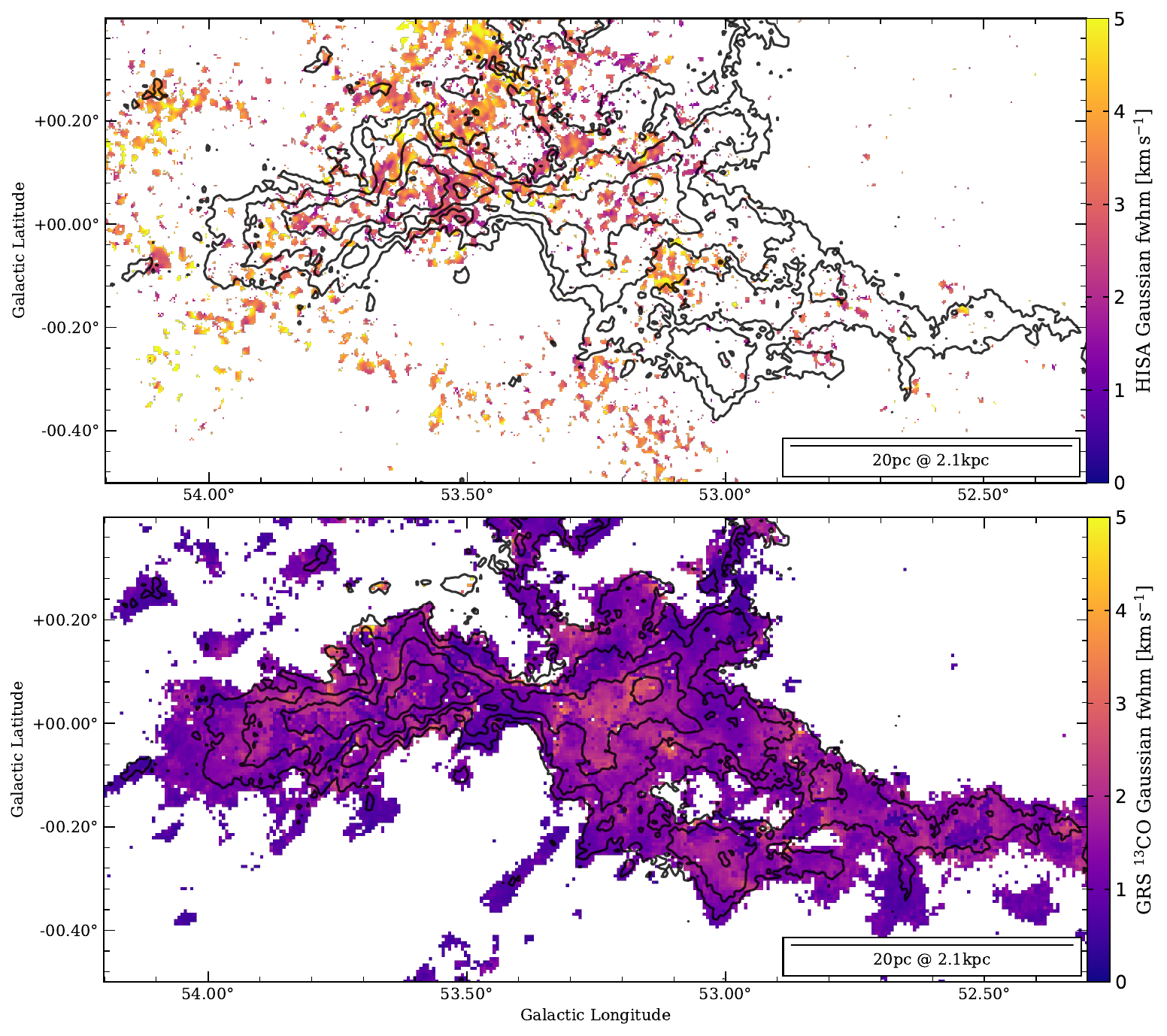}
        \caption[]{Fit line width (FWHM) toward GMF54. These maps show the line widths of fit components derived from the \textsc{GaussPy+} decomposition of the spectra. If multiple components are present in a single pixel spectrum within the velocity range of the filament region, the component with the lowest peak velocity is shown. The black contours in both panels show the integrated GRS \element[][13]{CO} emission at the levels 2.5, 5.0, 10.0, and 20.0$\rm\,K\,km\,s^{-1}$. \textit{Top panel:} Fit HISA line width. \textit{Bottom panel:} Fit \element[][13]{CO} line width.}
        \label{fig:kin12}
   \end{figure*}

\section{Column density maps}\label{sec:coldens_app}
The column density maps are presented in this section. The following maps show the column density maps for both HISA and $\rm H_2$ as traced by \element[][13]{CO} emission integrated over the velocity range of the respective filament region. Details about the column density derivation of each tracer can be found in Sect.~\ref{sec:coldens_mass}.
   \begin{figure*}
     \centering
     \includegraphics[width=1.0\textwidth]{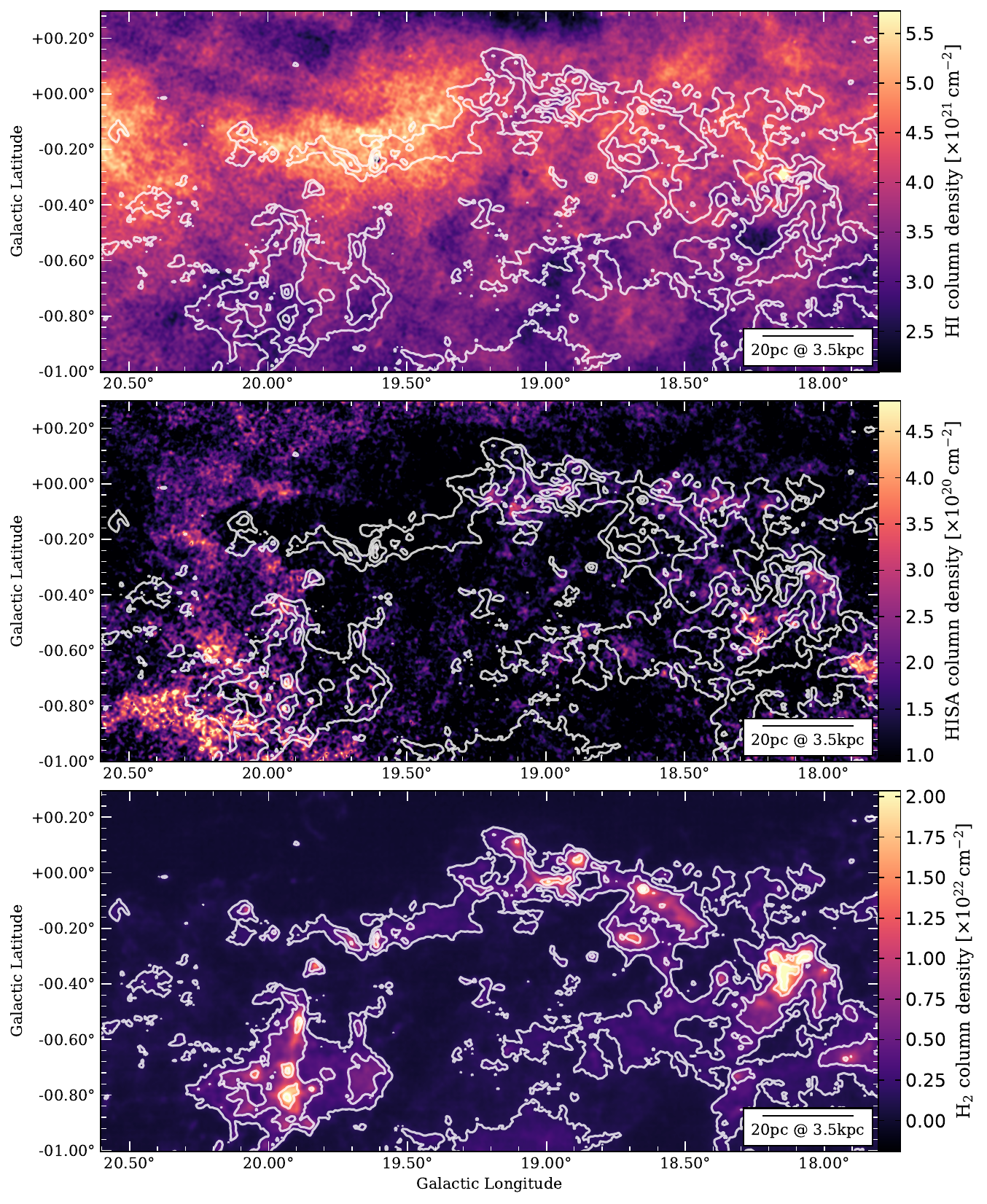}
        \caption[]{Column density toward GMF20. These maps show the column densities of atomic hydrogen traced by \ion{H}{i} emission, the cold hydrogen gas traced by HISA, and molecular hydrogen traced by \element[][13]{CO} emission, respectively. The column densities are integrated over the velocity range of the filament region given in Table~\ref{tab:overview}. The white contours in both panels show the integrated MWISP \element[][13]{CO} emission at the levels 8.0, 16.0, 32.0, and 42.0$\rm\,K\,km\,s^{-1}$. \textit{Top panel:} \ion{H}{i} column density traced by \ion{H}{i} emission. \textit{Middle panel:} HISA column density. \textit{Bottom panel:} $\rm H_2$ column density traced by \element[][13]{CO}.}
        \label{fig:coldens1}
   \end{figure*}
   
   \begin{figure*}
     \centering
     \includegraphics[width=1.0\textwidth]{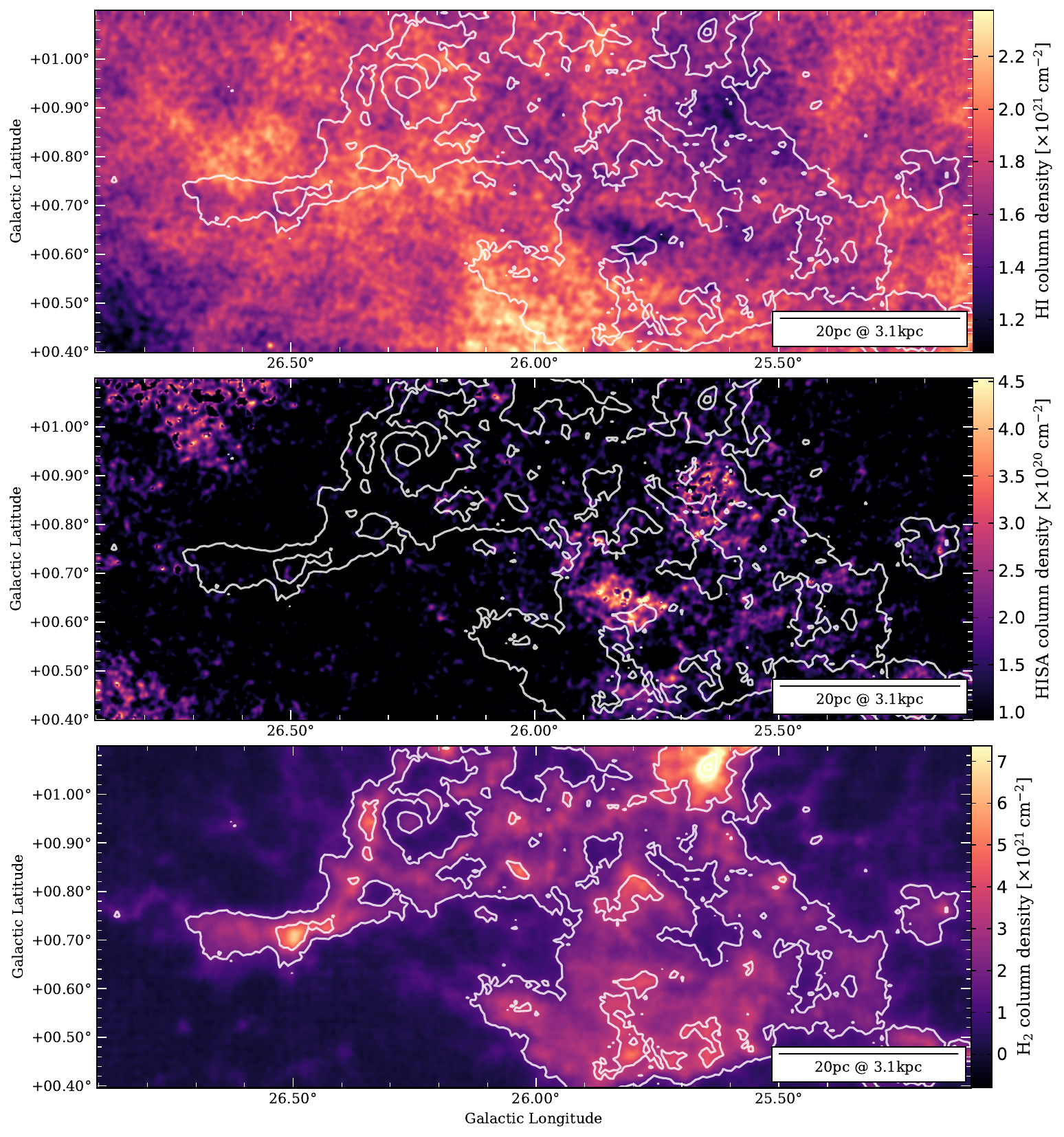}
        \caption[]{Column density toward GMF26. These maps show the column densities of atomic hydrogen traced by \ion{H}{i} emission, the cold hydrogen gas traced by HISA, and molecular hydrogen traced by \element[][13]{CO} emission, respectively. The column densities are integrated over the velocity range of the filament region given in Table~\ref{tab:overview}. The white contours in both panels show the integrated MWISP \element[][13]{CO} emission at the levels 6.0, 12.0, 24.0, and 34.0$\rm\,K\,km\,s^{-1}$. \textit{Top panel:} \ion{H}{i} column density traced by \ion{H}{i} emission. \textit{Middle panel:} HISA column density. \textit{Bottom panel:} $\rm H_2$ column density traced by \element[][13]{CO}.}
        \label{fig:coldens2}
   \end{figure*}
   
   \begin{figure*}
     \centering
     \includegraphics[width=1.0\textwidth]{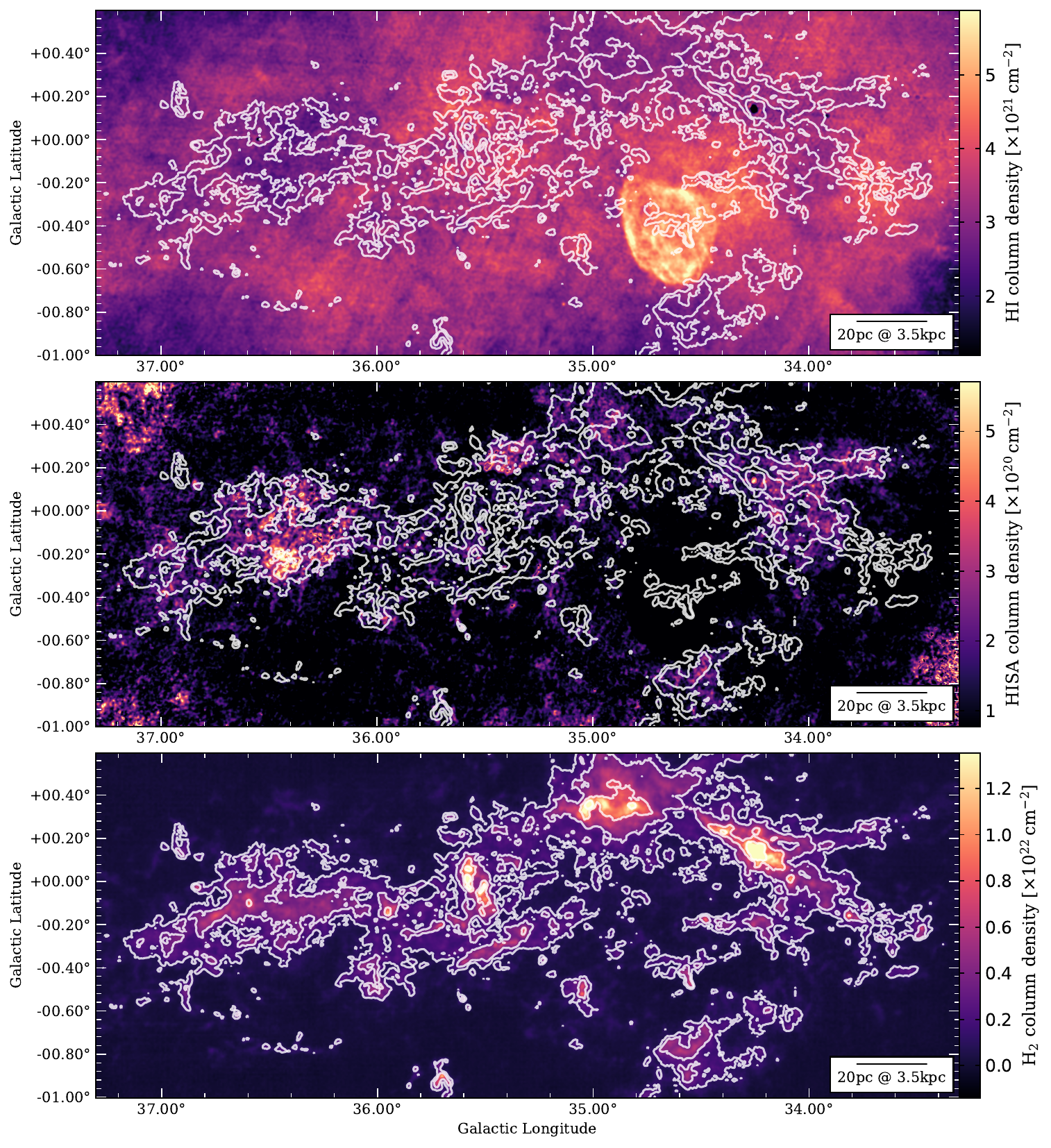}
        \caption[]{Column density toward GMF38a. These maps show the column densities of atomic hydrogen traced by \ion{H}{i} emission, the cold hydrogen gas traced by HISA, and molecular hydrogen traced by \element[][13]{CO} emission, respectively. The column densities are integrated over the velocity range of the filament region given in Table~\ref{tab:overview}. The white contours in both panels show the integrated MWISP \element[][13]{CO} emission at the levels 5.0, 10.0, 20.0, and 30.0$\rm\,K\,km\,s^{-1}$. \textit{Top panel:} \ion{H}{i} column density traced by \ion{H}{i} emission. \textit{Middle panel:} HISA column density. \textit{Bottom panel:} $\rm H_2$ column density traced by \element[][13]{CO}.}
        \label{fig:coldens3}
   \end{figure*}
   
   \begin{figure*}
     \centering
     \includegraphics[width=1.0\textwidth]{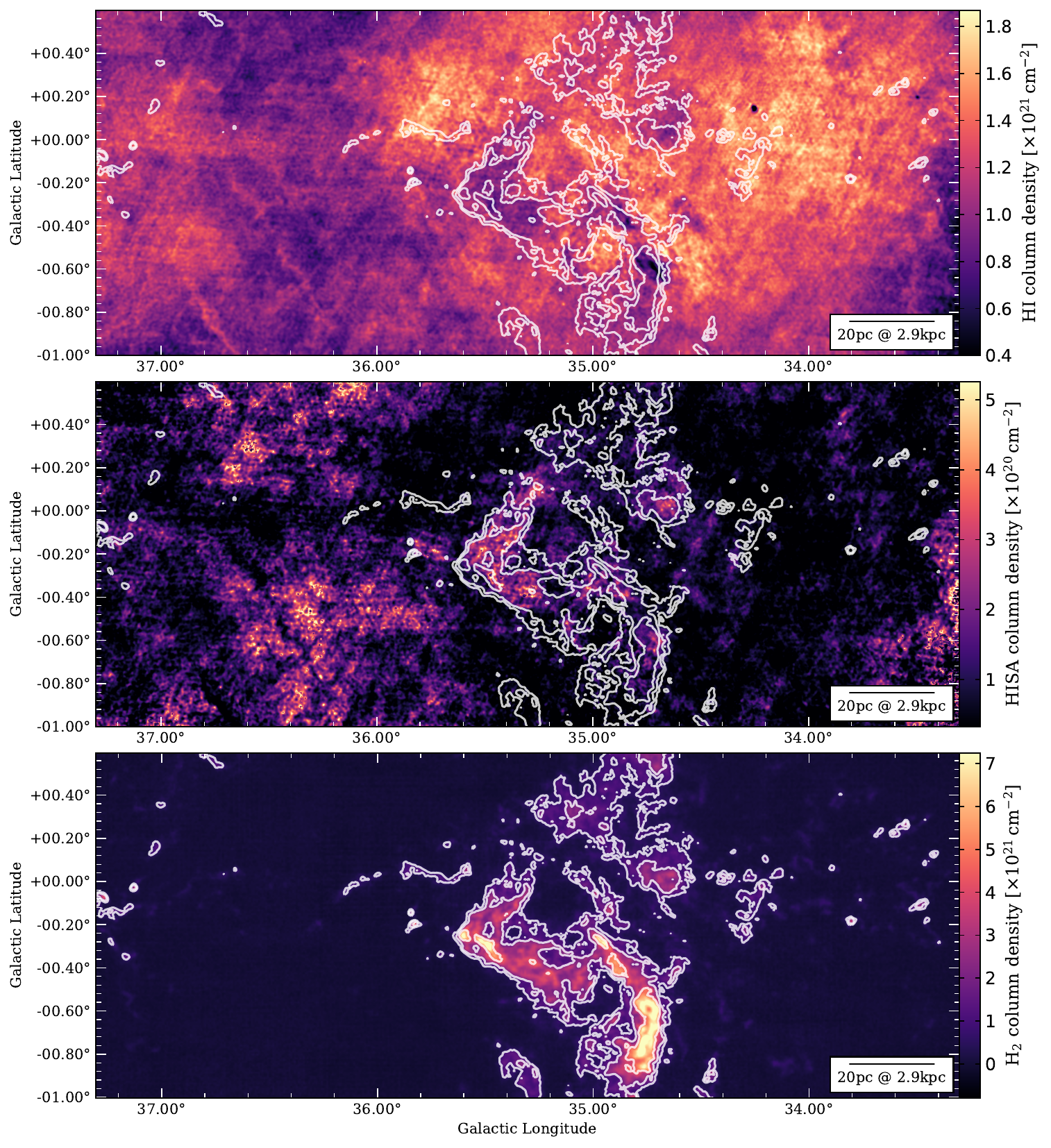}
        \caption[]{Column density toward GMF38b. These maps show the column densities of atomic hydrogen traced by \ion{H}{i} emission, the cold hydrogen gas traced by HISA, and molecular hydrogen traced by \element[][13]{CO} emission, respectively. The column densities are integrated over the velocity range of the filament region given in Table~\ref{tab:overview}. The white contours in both panels show the integrated MWISP \element[][13]{CO} emission at the levels 2.5, 5.0, 10.0, and 20.0$\rm\,K\,km\,s^{-1}$. \textit{Top panel:} \ion{H}{i} column density traced by \ion{H}{i} emission. \textit{Middle panel:} HISA column density. \textit{Bottom panel:} $\rm H_2$ column density traced by \element[][13]{CO}.}
        \label{fig:coldens4}
   \end{figure*}
   
   \begin{figure*}
     \centering
     \includegraphics[width=1.0\textwidth]{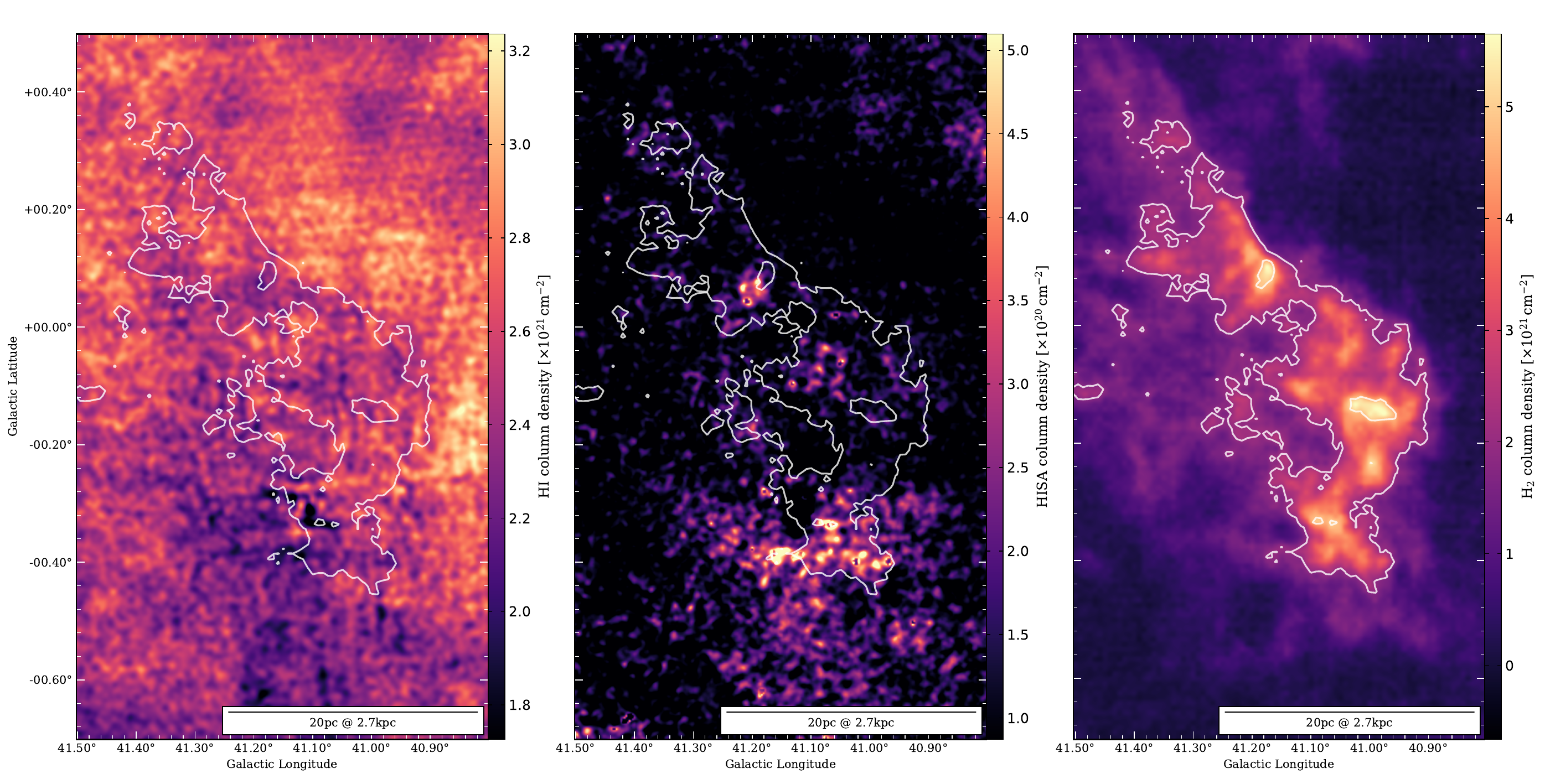}
        \caption[]{Column density toward GMF41. These maps show the column densities of atomic hydrogen traced by \ion{H}{i} emission, the cold hydrogen gas traced by HISA, and molecular hydrogen traced by \element[][13]{CO} emission, respectively. The column densities are integrated over the velocity range of the filament region given in Table~\ref{tab:overview}. The white contours in both panels show the integrated MWISP \element[][13]{CO} emission at the levels 6.0, 12.0, 24.0, and 34.0$\rm\,K\,km\,s^{-1}$. \textit{Left panel:} \ion{H}{i} column density traced by \ion{H}{i} emission. \textit{Middle panel:} HISA column density. \textit{Right panel:} $\rm H_2$ column density traced by \element[][13]{CO}.}
        \label{fig:coldens5}
   \end{figure*}
   
   \begin{figure*}
     \centering
     \includegraphics[width=1.0\textwidth]{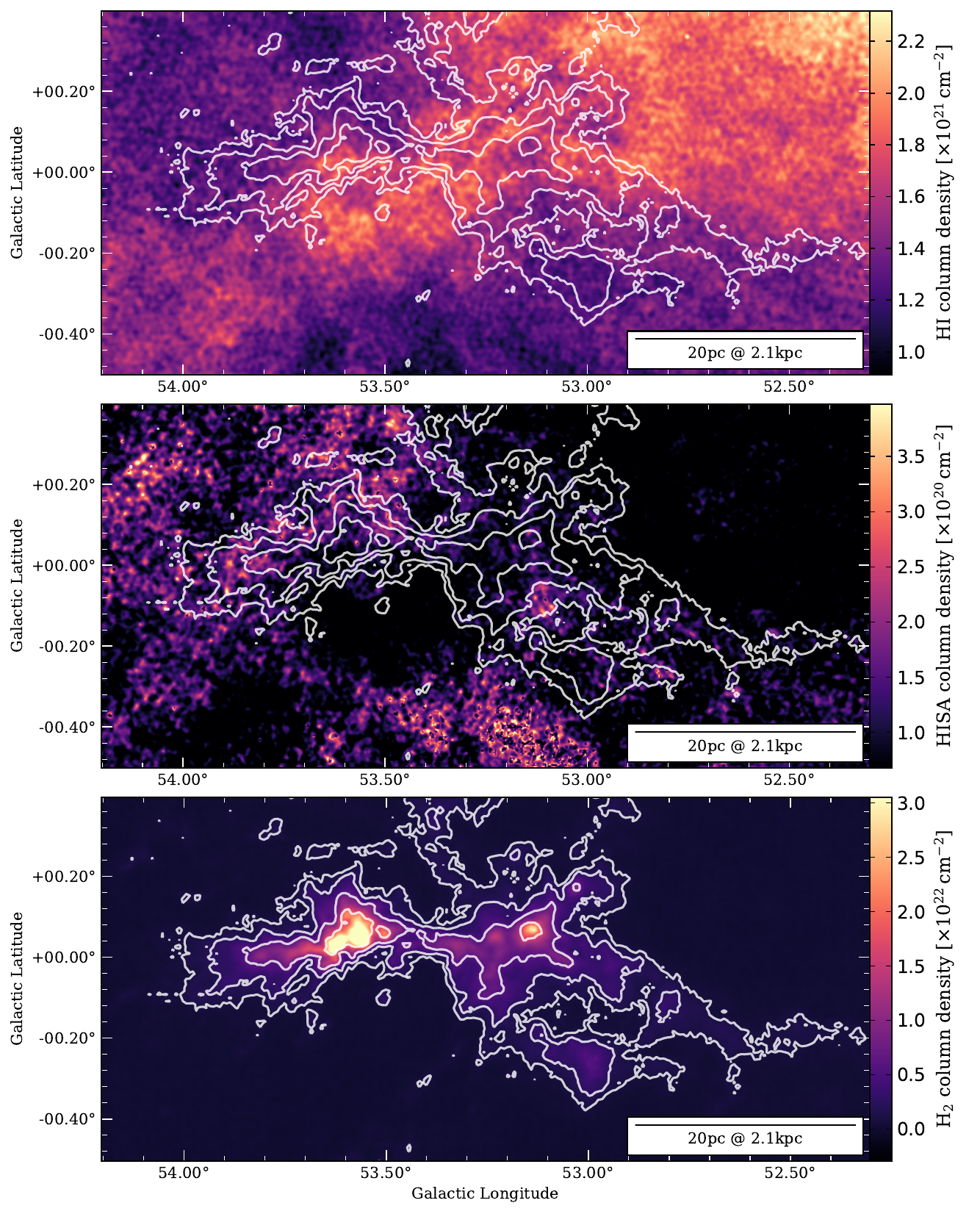}
        \caption[]{Column density toward GMF54. These maps show the column densities of atomic hydrogen traced by \ion{H}{i} emission, the cold hydrogen gas traced by HISA, and molecular hydrogen traced by \element[][13]{CO} emission, respectively. The column densities are integrated over the velocity range of the filament region given in Table~\ref{tab:overview}. The white contours in both panels show the integrated MWISP \element[][13]{CO} emission at the levels 2.5, 5.0, 10.0, and 20.0$\rm\,K\,km\,s^{-1}$. \textit{Top panel:} \ion{H}{i} column density traced by \ion{H}{i} emission. \textit{Middle panel:} HISA column density. \textit{Bottom panel:} $\rm H_2$ column density traced by \element[][13]{CO}.}
        \label{fig:coldens6}
   \end{figure*}

\end{appendix}

\end{document}